\documentclass[aps,prl,reprint,longbibliography]{revtex4-2}
\pdfoutput=1 
\usepackage{amsmath,amsthm,amsfonts,amssymb}
\usepackage{float}
\usepackage{array}
\usepackage{multirow}
\usepackage{marginnote}
\usepackage{color}
\usepackage{upgreek}
\usepackage{url}
\ProvideTextCommand{\DJ}{OT1}{\raisebox{0.25ex}{-}\kern-0.4em D}
\usepackage{rotating}
\usepackage[normalem]{ulem}
\usepackage{graphicx}
\usepackage{dcolumn}
\usepackage{bm}
\graphicspath{{./figures/}}
\usepackage{hyperref}

\newcommand\redsout{\bgroup\markoverwith{\textcolor{red}{\rule[0.5ex]{2pt}{0.4pt}}}\ULon}

\begin{document}
\title{Ultra-stretchable and Self-Healable Vitrimers with Tuneable Damping \\and Mechanical Response}

\author{Jiaxin Zhao$^{1,2}$}
\author{Nicholas J. Warren$^{3,4}$}
\author{Richard Mandle$^{1,2}$}
\author{Peter Hine$^1$}
\author{Daniel J. Read$^5$}
\author{Andrew J Wilson$^{2,6}$}
\author{Johan Mattsson$^1$}
\email{k.j.l.mattsson@leeds.ac.uk}
\affiliation{$^1$School of Physics and Astronomy, University of Leeds, Leeds LS2\,9JT, United Kingdom}
\affiliation{$^2$School of Chemistry, University of Leeds, Leeds LS2\,9JT, United Kingdom}
\affiliation{$^3$School of Chemical and Process Engineering, University of Leeds, Leeds LS2\,9JT, United Kingdom}
\affiliation{$^4$School of Chemical, Materials and Biological Engineering, University of Sheffield, Sheffield S10\,2TN, United Kingdom}
\affiliation{$^5$School of Mathematics, University of Leeds, Leeds LS2\,9JT, United Kingdom}
\affiliation{$^6$School of Chemistry, University of Birmingham, Birmingham B15\,2TT, United Kingdom}

\date{\today}%

\begin{abstract}
Vitrimers are a relatively new class of polymer materials with unique properties offered by cross-links that can undergo associative exchange dynamics. We here present a new class of vitrimers based on poly(methyl acrylate) with cross-links utilising dioxaborolane metathesis. These vitrimers demonstrate a combination of ultra-stretchability (up to $\sim$ 80 times their own length), mechanical toughness ($\sim$ 40 MJ/m$^3$), and thermal stability up to $T\sim$ 250 °C; moreover, the vitrimers demonstrate excellent mechanical damping characterised by a loss factor ($\textup{tan}(\delta)$) with a maximum of $\sim$ 2--3 and an effective value $>$0.3 across five decades in frequency (0.001--100 Hz), or correspondingly across a $T$-range of $\sim$ 35 °C near room temperature (for a probe frequency of 1 Hz). The vitrimers can be successfully re-processed using both a thermo-mechanical and a chemical processing route, and can for low crosslink density self-heal at room temperature, making them suitable for sustainable applications. The material properties are directly tuneable by variation of both the amount of cross-linker and by the degree of curing. Thus, this class of vitrimers are promising for applications where stretchability combined with mechanical toughness and/or a high mechanical dissipation is required. 
\end{abstract}

\maketitle

\color{black}

\section{Introduction}

Elastomers that combine considerable stretchability with high mechanical strength and toughness are important for a wide range of applications including actuators, sensors, soft robotics, wearable electronics, and materials for energy storage. \cite{oh2016intrinsically,shi2018highly,cheng2019stick} Standard elastomers, such as vulcanised rubbers, can readily achieve tensile strains of several 100\% combined with tensile strengths of $\gtrsim$10 MPa. \cite{leber2019stretchable,huang2016strain,mazurek2019tailor} However, elastomers that can sustain strains of several 1000\% combined with a high tensile strength (a high mechanical toughness) are challenging to produce. A variety of approaches have been attempted often focused on combining an elastic stress-bearing network with a network of sacrificial bonds that can dissipate the applied mechanical energy by bond fracture. Reported polymer systems include: block-copolymer-based thermoplastic elastomers, \cite{tong2000synthesis} interpenetrating double polymer networks, \cite{zhang2018extremely,li2022highly} elastomers based on supramolecular interactions, \cite{li2016highly,jin2023quadruple,Verjans2024JMCB} or polymeric networks based on dynamic covalent crosslinks, so called covalent adaptable networks (CANs) \cite{sun2012highly,cao2019robust,lyu2020extremely,kong2022ultra,ji2023novel}. However, elastomers that show very high stretchability ($\sim$ 1000--10000\%), typically also exhibit significant strain-softening upon yielding, leading to a rapidly decreasing mechanical strength upon deformation. \cite{li2016highly,sun2012highly,lyu2020extremely,daniel2016solvent,cao2017transparent,zhang2018exploring,zhang2019superstretchable,li2022superstretchable} Thus, it remains an on-going challenge to produce elastomers that combine high stretchability, mechanical strength and toughness. 

Here, we demonstrate that a relatively new type of dynamically cross-linked networks, so called \emph{vitrimers}, can achieve these requirements, \cite{montarnal2011silica,capelot2012metal,capelot2012catalytic}, while also providing other attractive properties such as effective self-healing, excellent mechanical damping and both thermo-mechanical and chemical re-processability. 

Polymers with permanent crosslinks -- thermosets -- show good mechanical and thermal stability, but poor flexibility, malleability and reprocessing capability. Conversely, polymers that lack permanent crosslinks -- thermoplastics -- offer reprocessability but lower stability. CANs \cite{denissen2016vitrimers} combine the advantageous properties of thermosets and thermoplastics, resulting in both mechanical stability and effective materials reprocessing. \cite{kloxin2013covalent} The subset of CANs for which the crosslink bond exchanges are associative, are termed vitrimers. \cite{montarnal2011silica,capelot2012metal,capelot2012catalytic} Vitrimers keep their crosslink density fixed during bond exchange, which results in excellent mechanical and thermal stability. At high temperatures, these materials are typically malleable and thus reprocessable, \cite{kloxin2013covalent} whereas at low temperatures, they show behaviour akin to permanently crosslinked thermosets with good mechanical and thermal stability. Vitrimers have been produced based on different strategies for the covalent exchange, such as: transesterification, \cite{montarnal2011silica} transcarbonation, \cite{snyder2018reprocessable} olefin metathesis, \cite{lu2012olefin}, transimination, \cite{hajj2020network} silyl ether metathesis, \cite{tretbar2019direct} or dioxaborolane metathesis. \cite{rottger2017high} 

The unique properties of vitrimers suggest that they should be ideal materials to combine properties such as stretchability and material strength with efficient re-processing ability. \cite{montarnal2011silica,rottger2017high,lyu2020extremely} Moreover, the dynamic nature of the bond exchange should be important in achieving other important properties such as good self-healing \cite{cash2015room}, material welding \cite{shi2023welding} capabilities, and high mechanical dissipation. \cite{cheng2024hyperbranched,shi2024dynamic} Thus, vitrimers are promising and exciting candidates for the next generation of polymer-based applications, \cite{zheng2021dynamic} including synthetic polymers acting as artificial skin, muscles, or cartilage that combine high stretching and self-healing ability, with large mechanical strength and high mechanical dissipation. \cite{oh2016intrinsically,shi2018highly,cheng2019stick,kang2018tough,xun2021tough,zheng2021dynamic} 


\begin{figure*}
\begin{center}{\includegraphics[width=1\textwidth]{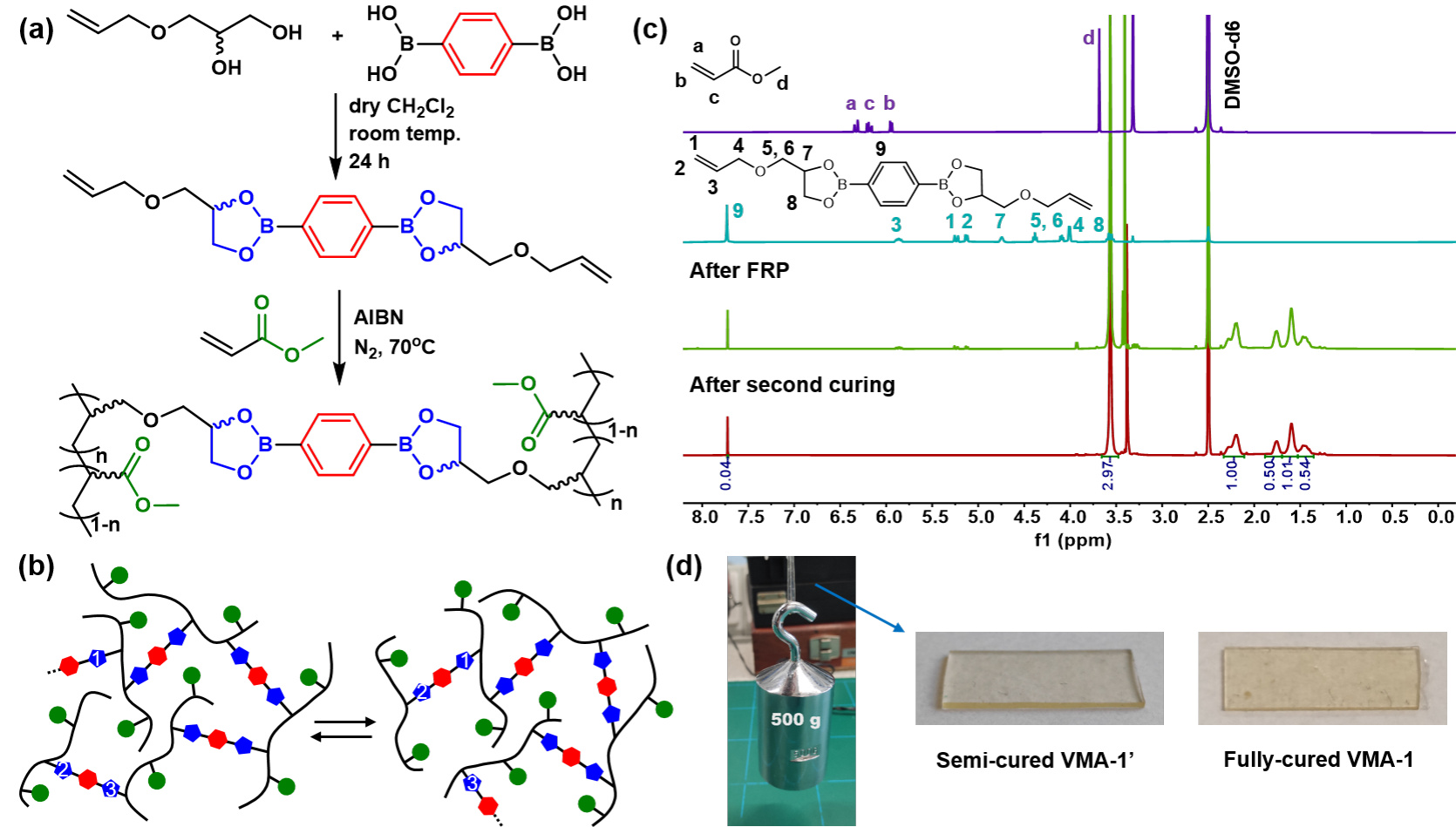}}\end{center}
  \caption{a) Schematic of the PMA-based vitrimer synthesis. b) A cartoon illustrating the associative bond exchange within the vitrimers. c) $^{1}$H NMR spectra of the raw material (monomer and crosslinker) used in the vitrimer synthesis, the synthesis product after free radical polymerization (FRP), and the synthesis product after secondary curing. d) Photographs of a semi-cured VMA-1' and a fully cured VMA-1 sample. A photograph of the VMA-1'sample lifting a 500 g weight (cross section size: 4.0 mm $\times$ 0.8 mm) is also shown.}
  \label{fig:Fig1}
\end{figure*}


Here, we present a new ultra-stretchable vitrimer system based on poly(methyl acrylate) (PMA), where the PMA-based vitrimers are prepared by free radical polymerization (FRP) of the methyl acrylate monomer and a diallyl dioxaborolane dynamic crosslinker. The associative crosslink exchange reaction, based on dioxaborolane metathesis, shows excellent thermal stability and a relatively low activation energy ($\sim$16 kJ/mol), which allows rapid bond exchange without added solvent, catalyst or free diols. \cite{rottger2017high} Long chain-length PMAs are viscous fluids at room temperature, characterised by glass-transition temperatures $T_{\textup{g}} \sim$ 15 °C, and are often used in coatings and adhesives, as liquid crystal elastomer backbone components, or as components in block co-polymers to improve material flexibility. \cite{pu2012polyacrylates,fortunato2020highly,mistry2020isotropic} Here, PMA was chosen as the matrix polymer both due to its lack of crystallinity and due to its flexible nature, as reflected in a relatively low glass transition temperature, which ensures a high segmental mobility. 

Our vitrimers demonstrate stretchability of up to 8000$\%$, combined with a high tensile strength ($\sim$ 0.6 MPa) and toughness ($\sim$ 38.5 MJ/m$^{3}$). They also demonstrate room-temperature self-healing and a high mechanical loss factor with tan($\delta$) $>$ 1 over three orders of magnitude, and tan($\delta$) $>$ 0.3 over five orders of magnitude, in frequency ($10^{-3}$--$10^2$ Hz), as well as excellent re-processing through either a thermomechanical (hot-press) or a chemical route based on hydrolysis of the boronic ester. Thus, our PMA-based vitrimers are promising materials for a wide range of application areas. 

\section{Results and Discussion}
\subsection{Vitrimer synthesis and characterization}

\begin{figure*}
\begin{center}{\includegraphics[width=1\textwidth]{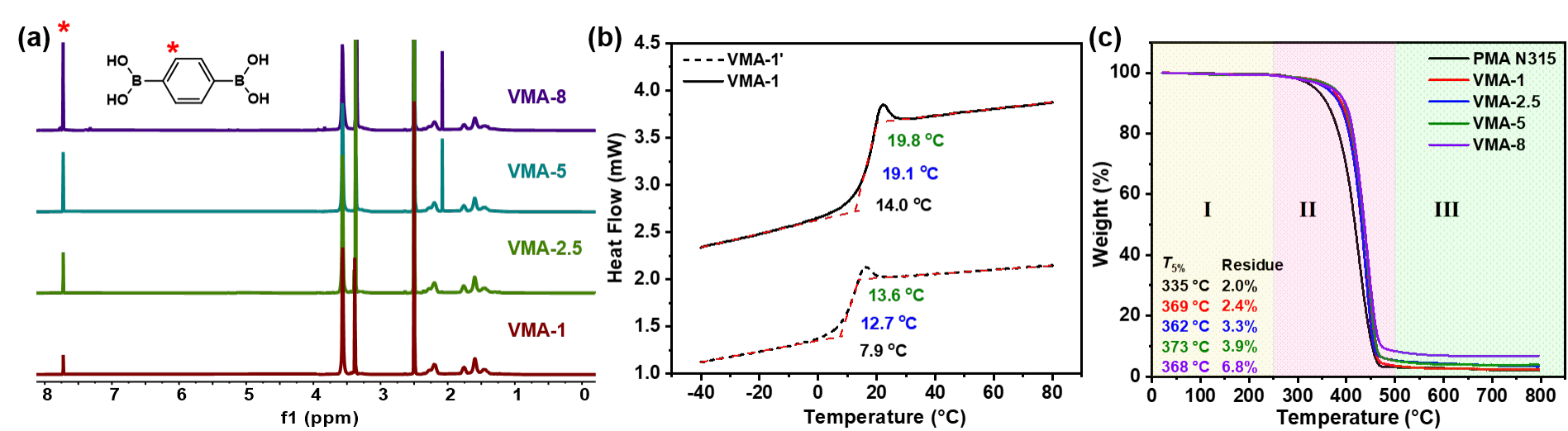}}\end{center}
  \caption{a) $^{1}$H NMR spectra of fully cured vitrimers (VMA-$x$) with different crosslink densities ($x=1, 2.5, 5$ and $8$ mol\%) after hydrolysis. b) DSC results showing the glass-transitions for the semi-cured (VMA-1') and fully cured (VMA-1) vitrimer samples as measured during heating for a rate of 10 K/min; the onset, midpoint and offset temperatures are shown. c) TGA results for the fully cured vitrimer samples together with a linear PMA for comparision.}
  \label{fig:Fig2}
\end{figure*}

A set of poly(methyl acrylate) based vitrimers (VMAs) of varying crosslink densities were produced, as shown schematically in Figure \ref{fig:Fig1}(a); the dynamic crosslinker was first synthesized, \cite{robinson2021chemical} and subsequently copolymerized with the methyl acrylate monomer using FRP to produce the vitrimers. The schematic in Figure \ref{fig:Fig1}(b) illustrates how the associative bond exchange results in topological rearrangements, which in turn lead to malleability, even though the crosslink density is maintained. After synthesis, the samples were dried in an oven and then hot-pressed at $T=120$ °C, to produce vitrimers of nominal crosslink density $x=1, 2.5, 5$ and $8$ mol\% (by stochiometry). $^1$H NMR spectra of the reactants (monomer and crosslinker) and the synthesis product are shown in Figure \ref{fig:Fig1}(c). The successful polymerization of methyl acrylate is demonstrated by the disappearance of signals due to the monomer double bond (peaks a-c in Figure \ref{fig:Fig1}(c)). The NMR spectra demonstrate a 50\%--60\% incorporation of the crosslinker after the initial FRP step (peaks labelled 1-3 in Figure \ref{fig:Fig1}(c)). The lower (than 100\%) incorporation is due to the lower reactivity of the double bond in the allyl compared to the acrylate moiety. \cite{tamezawa2012peculiar} We refer to these first-cured samples as semi-cured and denote them as VMA-$x$' (the apostrophe marks that they are semi-cured and $x$ denotes the nominal molar cross-link density). To produce fully cured vitrimers, samples were further heated to $T=220$ °C for 2 hours under vacuum. After this process, no peaks due to the allyl double bond are observed in NMR, confirming that all vinyl bonds were fully incorporated; we thus refer to these samples as fully cured and denote them as VMA-$x$. $^1$H NMR studies on the fully cured vitrimers (shown in Figure \ref{fig:Fig2}(a)) demonstrate that the resonance at 7.72 ppm, assigned to the phenyl ring of p-phenylene diboronic acid after hydrolysis (see inset), increases systematically with the nominal crosslink density, as expected for full incorporation of the crosslinker. 

To characterise the produced chain-length distribution of the vitrimer samples, a set of samples were produced for which the crosslinks were removed by hydrolysis and characterised by gel permeation chromatography (GPC) (see the Experimental Section for details). The GPC traces and results are shown in Figure S1 and Table S1, demonstrating that the pure polymers all have number-averaged molecular weights $M_{\textup{n}}$ close to 20 kDa and broad dispersities $\DJ>$ 4. The entanglement molecular weight for pure linear PMA is $M_{\textrm{e}}\approx$ 11 kDa, and the critical molecular weight is $M_{\textrm{c}}\approx$ 20 kDa\cite{fetters2007chain}. Thus, based on our GPC results, we expect the vitrimers to be weakly entangled and characterised by a broad distribution of chain-lengths. Representative photos of the produced vitrimers are included in Figure \ref{fig:Fig1}(d), which shows the $x=1$ mol\% semicured (VMA-1') and fully-cured (VMA-1) samples, respectively, together with a photo illustrating that a strip of VMA-1' (width 4.0 × thickness 0.8 mm) can readily lift a weight of 500 g, thus demonstrating the mechanically robust nature of the produced vitrimers. 

Differential scanning calorimetry (DSC) was performed to further characterise the vitrimers. The temperature ($T$) dependent heat flow recorded for a heating rate of 10 °C/min is shown in Figure \ref{fig:Fig2}(b) for two representative vitrimer samples (VMA-1' and VMA-1). Both the glass transition temperature ($T_{\textup{g}}$) and the glass transition breadth ($\Delta T_{\textup{g}}$) increase upon the additional curing (from VMA-1' to VMA-1). DSC data for all vitrimer samples are shown in Figure S2 and the corresponding analysis results are shown in Table S2. Consistent with the results shown in Figure \ref{fig:Fig2}(b), both $T_{\textup{g}}$ and $\Delta T_{\textup{g}}$ increase upon additional curing for all cross-link densities. Generally, an increase in the crosslink density (due to an increase in the nominal crosslink density or due to additional curing) restricts the polymer mobility, thus resulting in increased $T_{\textup{g}}$ values, as shown in Figure S2. Similar behaviour has previously been observed for thermosets, \cite{zheng2022understanding} supramolecular polymers, \cite{sordo2015design} and vitrimers. \cite{hajj2020network,chen2021crucial} However, we note that for our semicured vitrimers, dangling or `free' crosslinkers might act as plasticizers that can at least partially counteract the $T_{\textup{g}}$ increase, as shown in Figure S2. 


Thermogravimetric analysis (TGA) was performed to determine the vitrimer thermal stability. Data for all vitrimer samples, together with data for a linear PMA with a degree of polymerization of $N=315$ ($M_{\textup{n}}=27000$ g/mol), are shown in Figure \ref{fig:Fig2}(c). The TGA data can be roughly divided into three regimes: Regime I: stable with no significant weight loss ($T\leq$250 °C); Regime II: undergoing thermal degradation, including main-chain scission, \cite{bertini2005investigation} as previously demonstrated for PMA ($250<T<500$ °C); Regime III: degradation is practically complete ($>500$ °C). The temperature $T_{\textup{5\%}}$, denoting a 5\% thermal weight loss, is increased by $\sim$ 30 °C between the linear PMA and the PMA-based vitrimers.

\subsection{Vitrimer Reprocessability}
\begin{figure*}
\begin{center}{\includegraphics[width=1\textwidth]{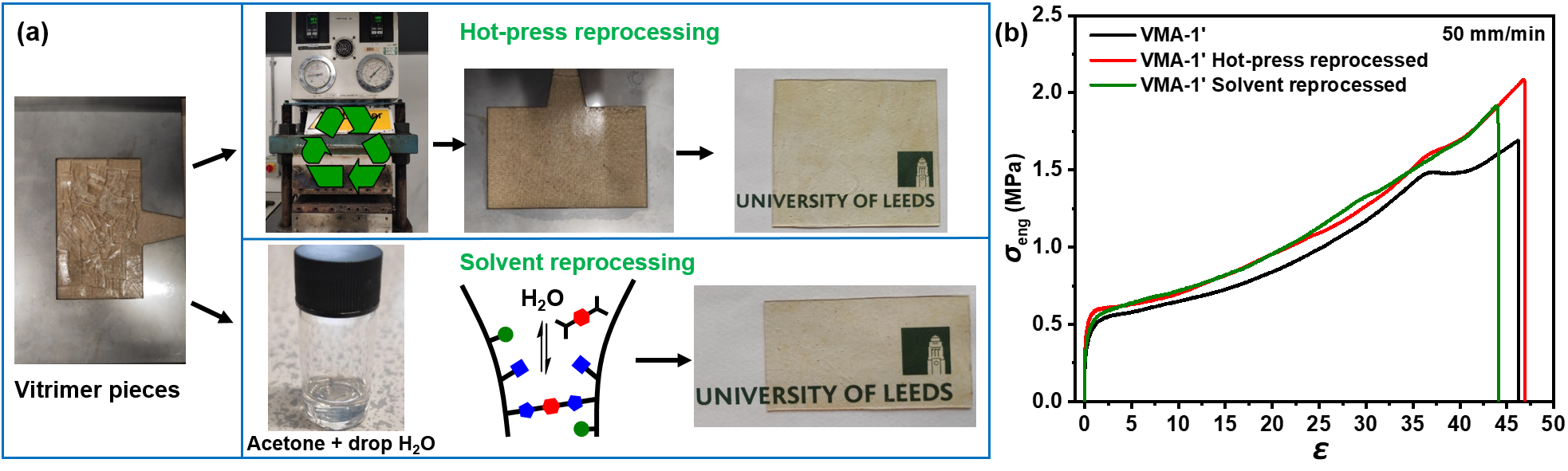}}\end{center}
  \caption{a) Illustration of the thermo-mechanical (hot-press) and solvent reprocessing routes, respectively. b) Tensile stress-strain data for the semi-cured VMA-1' vitrimer sample both before and after reprocessing through either of the two routes.}
  \label{fig:Fig3}
\end{figure*}

A strong advantage of vitrimers is the thermal reprocessability afforded by the dynamic nature of the cross-links \cite{montarnal2011silica,rottger2017high,chen2023exceptionally}. Alternative chemical reprocessing routes, either by depolymerization or by degradation (e.g. through high temperature or acidic conditions), have also been demonstrated for vitrimers; \cite{snyder2018reprocessable,yue2023one} however, these routes breaks down the vitrimer structure and reprocessing thus involves partially repeating the synthesis. In contrast, our vitrimers can be effectively re-processed using either a thermo-mechanical or a chemical route. For thermal reprocessing, the vitrimer samples were cut into pieces and placed into a hot-press mould, as shown in Figure \ref{fig:Fig3}(a), where they were kept for 15 minutes at $T=120$ °C to facilitate dynamic cross-link exchange, and thus reprocessability. In contrast, the solvent-based reprocessing route is based on boronic ester hydrolysis, where the solvent needs to contain a sufficient amount of water for the hydrolysis (we used acetone mixed with a small amount of water; $\textrm{vol/vol}$ = 200:1). The presence of the solvent initially leads to swelling of the vitrimers, followed by dissolution, where water acts to chemically ``unzip" the cross-links (see the sketch in Figure \ref{fig:Fig3}(a)); NMR characterisation demonstrated that the vitrimer samples could be successfully hydrolysed using this approach, as shown in Figure 2(a). Following the dissolution process, the vitrimers were recovered by solvent removal in a vacuum oven at 80 °C, after which the dry samples were hot-pressed into samples of suitable dimensions. To investigate how successful each reprocessing route is, tensile stress-strain measurements were performed both before and after reprocessing. As shown in Figure \ref{fig:Fig3}(b) for VMA-1', both the tensile strengths and elongation at break are largely recovered for samples undergoing either of the two reprocessing routes ($\sim$2 MPa and 4500\%), and similar behaviour is observed for the corresponding fully cured sample (VMA-1), as shown in Figure S3.  


\subsection{Vitrimer Network Response and Relaxation}

\begin{figure*}
  \begin{center}{\includegraphics[width=1\textwidth]{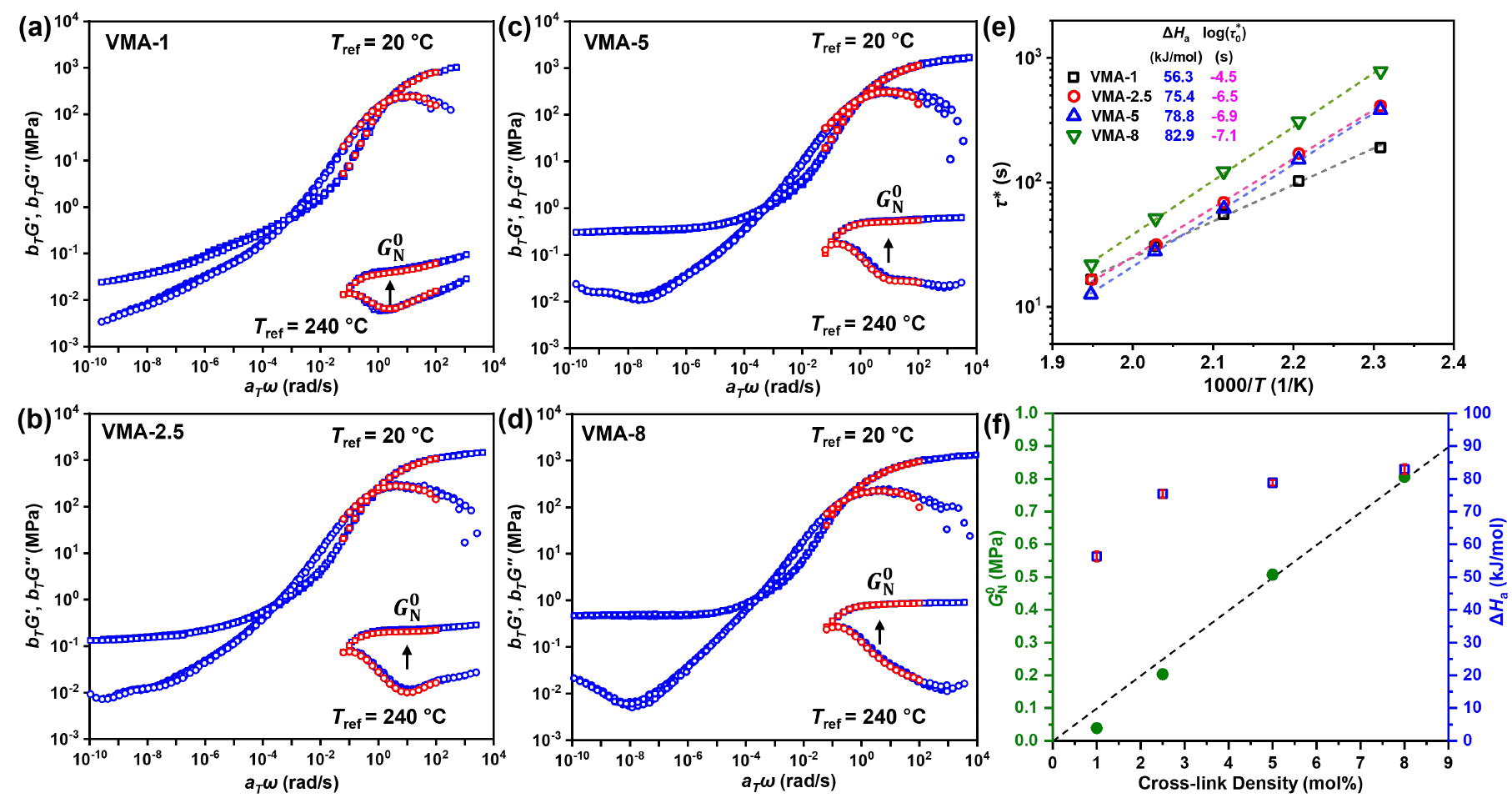}}\end{center}
  \caption{a) SAOS mastercurves formed by use of TTS for the fully cured vitrimer samples (VMA-$x$) with varying cross-link densities ($x=1, 2, 2.5$ and $8$ mol\%). The low- and high-$T$ mastercurves are shown separately as they refer to different reference temperatures, as shown in the figure. The data set corresponding to the reference temperature is marked in red squares. The arrow mark the frequencies for which the plateau moduli $G_N^0$ are determined. e) Arrhenius plot of the relaxation time $\tau^{\ast}$ characterising the slow high-temperature relaxation. f) Plot of $G_N^0$ (circles) and the activation energy of the slow high-$T$ relaxation (squares) versus the nominal cross-link density. The dashed line is a guide to the eye.}
  \label{fig:Fig4}
\end{figure*}

To evaluate the rheological response of the crosslinked network, we performed small-amplitude oscillatory shear (SAOS) rheology. The SAOS measurements were carried out from $T\sim$15 °C to 240 °C, and master curves for the storage and loss shear moduli ($G^{'}$ and $G^{''}$) were produced based on the time-temperature superposition (TTS) principle, as shown in Figure \ref{fig:Fig4} and described in detail in the Experimental Section. TTS could be accurately performed both at low and high $T$, where the shift parameters obtained for each $T$-range refer to different relaxation behaviour. For the low $T$ data, we choose a reference temperature $T_{\textrm{ref}}=$20 °C for which the structural $\alpha$ relaxation response (crossover of $G^{'}$ and $G^{''}$) lies within the experimental window. For the high $T$ data, we instead choose a reference temperature of $T_{\textrm{ref}}=$240 °C for which the onset of the high-$T$ relaxation behaviour lies within the experimental window.   

For each of the four vitrimer samples (VMA-1 to VMA-8; see Figure \ref{fig:Fig4}(a-d)) a glassy response is observed at the highest frequencies corresponding to a peak in $G^{''}$ and the establishment of a glassy plateau in $G^{'}$; for lower frequencies, a spectrum of Rouse modes are observed, as indicated by $G^{'}\sim G^{''}\sim \omega^{1/2}$, followed by a plateau in $G^{'}$ and a corresponding minimum in $G^{''}$. We define the plateau moduli ($G_{\textup{N}}^{0}$) as the $G^{'}$ value corresponding to the minima in $G^{''}$, following a commonly used approach in the literature, \cite{dealy2018structure}, as shown by the arrows in Figure \ref{fig:Fig4}(a-d). As the nominal cross-link density increases, the plateau becomes flatter and thus more well-defined and correspondingly $G_{\textup{N}}^{0}$ increases in an approximately linear manner, as shown in Figure \ref{fig:Fig4}(f) (green circles). We note that particularly for the lowest crosslink density sample (VMA-1), the `plateau' behaviour is poorly defined demonstrating the presence of a broad spectrum of relaxation modes across $\sim$8 orders of magnitude in frequency; this is consistent with the behaviour of a weakly cross-linked (and weakly entangled) system, which contains a broad distribution of dangling polymer strands, or free polymer chains, thus resulting in a very broad spectrum of relaxation frequencies. Finally, for the lowest frequencies ($T_{\textrm{ref}}=$240 °C) an additional relaxation regime is clearly observed, as shown by the crossover of the storage and loss moduli. 

The plateau modulus $G_{\textup{N}}^{0}$ can generally be attributed both to the covalent dynamic crosslinks and to topological crosslinks due to chain entanglements. The average effective molecular weight between crosslinks ($M_{\textup{x}}$) can be estimated from $G_{\textup{N}}^{0}=\rho RT/M_{\textup{x}}$, \cite{rubinstein2003polymer} where $\rho$ is the mass density, which we estimate using the literature value $\rho=1.09$ g/cm$^{3}$ for linear poly(methyl acrylate) at the relevant reference temperature of 160 °C \cite{pfefferkorn2010pressure}. The resulting $M_{\textup{x}}$ for VMA-$x$ ($x=$1, 2.5, 5 and 8) are: 109, 19, 8, and 5 kg/mol, respectively. For comparison, the entanglement molecular weight ($M_{\textup{e}}$) for linear poly(methyl acrylate) (PMA) is 11 kg/mol, \cite{fetters2007chain}, so that $M_{\textup{x}}$ is larger than $M_{\textup{e}}$ at low crosslink density, and smaller than $M_{\textup{e}}$ at high crosslink density. For context, it is also important to compare both $M_{\textup{x}}$ and $M_{\textup{e}}$ with the length of the highly polydisperse primary chains given in Table S1. For VMA-1 and VMA-2.5 we find $M_{\textup{e}} \approx M_{\textup{n}} < M_{\textup{x}} < M_{\textup{w}}$; for VMA-5 and VMA-8 we find $M_{\textup{x}} < M_{\textup{e}} \approx M_{\textup{n}} \ll  M_{\textup{w}}$. In all cases, due to the high polydispersity, we thus expect a significant number of chains that have only a small number of dynamic crosslinks, and/or which are shorter than the entanglement molecular weight. 

Moreover, in Figure S4 we compare the estimated effective crosslink densities $\mu_p$, as determined from the plateau moduli, with the corresponding nominal crosslink densities $\mu_c$ determined from the chemical composition and the mass density for the fully cured vitrimers. We determine $\mu_p$ using both the classical affine and phantom network models \cite{rubinstein2003polymer}, respectively. Thus, the molar volumetric number density of stress-bearing chain strands $\nu_p$ was determined using $G_N^0=a\cdot\nu_p \cdot RT$, where $a=1$ for the affine network model, $a=(1-2/f)$ for the phantom network model, and $f$ is the cross-link functionality. We here use $f=4$ as an approximation that is consistent with the dynamic cross-linker chemistry \cite{ricarte2023time}, and for either model the cross-link density $\mu_p$ can be determined as $\mu_p=2\nu_p/f$. We find that for the affine network model the ratio $\mu_p/\mu_c$ lies within the range $\sim$ 0.04--0.12, and for the phantom network model within the range $\sim$ 0.08--0.25, for the investigated vitrimer samples. Thus, a significant difference between the estimated and the nominal crosslink density is consistently observed.


Both with regards to the difference between $\mu_p$ and $\mu_c$, and the small influence of entanglements in determining the `rubbery' plateau, we note that: (i) since the reactivity of the crosslinker is much lower than that of the acrylate monomer, we expect a degree of heterogeneity in the cross-link distribution, and the formation of ``effective cross-links'' (each containing several nominal cross-links) is not surprising; (ii) the calculation of the nominal cross-link density does not take account of the possibility of cross-link formation within the same chain; such `loops' are not generally stress-bearing and will thus not contribute to the plateau modulus; (iii) it is highly likely, especially at low crosslink density, that many of the crosslinks lie within branched structures that are either ``free'' chains wholly unconnected to the network, or ``side branches'' connected to the network at one point only: such structures can relax their internal configuration, and so relax their contribution to the stress, without breaking of crosslinks and thus do not contribute to the plateau; (iv) finally, relaxation of such free chains or side branches will release entanglements through the well known process of dynamic dilution \cite{dealy2018structure}, thus progressively increasing the effective entanglement molecular weight and reducing the influence of entanglements on the plateau.

To further characterise the rheological response, we define two characteristic relaxation times from the crossover frequencies between the storage and loss moduli. The high-frequency (low-$T$) relaxation time is defined as $\tau_{\alpha}=1/(a_T^{\alpha}\omega_{c}^{\alpha})$, and the low frequeny (high-$T$) relaxation time as $\tau^{\ast}=1/(a_T^{\ast}\omega_{c}^{\ast})$, where near $T_{\textup{g}}$ the former corresponds to the structural $\alpha$ relaxation, and the latter corresponds to the relaxation due to dynamic cross-link exchange. $\omega_{c}^{\alpha}$ denotes the characteristic angular frequency of the low-$T$ and $\omega_{c}^{\ast}$ of the high-$T$ relaxation (each determined at their corresponding reference temperatures), and $a_T^{\alpha}$ and $a_T^{\ast}$ are the two horizontal shift parameters, respectively. $a_T^{\alpha}(T)$ for the four fully cured vitrimer samples are shown in an Arrhenius plot in Figure S5 demonstrating only very small differences with a variation of the cross-link density. The corresponding $\tau_{\alpha}(T)$ are shown in Figure S6 together with the $T_{\textup{g}}$ values determined from DSC, showing a slowing down for increasing crosslink density consistent with the observed increase in $T_{\textup{g}}$. 

The relaxation times $\tau^{\ast}(T)$ are shown in an Arrhenius plot in Figure \ref{fig:Fig4}(e). For each of the four samples, $\tau^{\ast}$ demonstrate Arrhenius behaviour: $\tau_i=\tau_{0,i}\exp{(\Delta H_i/RT)}$ ($i$ denotes a particular sample), where $R$ is the gas constant, as shown by the fits (dashed lines). The activation enthalpy ($\Delta H_a$), determined for each of the four fits, are plotted in Figure \ref{fig:Fig4}(f) (squares). We find that $\Delta H_a$ is of similar magnitude $\sim$ 80 kJ/mol for the three higher crosslink density samples, whereas a somewhat lower $\Delta H_a$ of $ \approx 60$ kJ/mol is found for the 1 mol\% sample. These values are slightly higher than those previously reported for vitrimers based on the same dynamic crosslinker, where examples include: polysiloxane vitrimers (20--57 kJ/mol), \cite{liu2021self} vitrimers prepared by a thiol-ene ``click'' reaction (50 kJ/mol), \cite{wang2021solid} polyolefin vitrimers (8--47 kJ/mol), \cite{zhao2023one} and poly(styrene-b-isoprene-b-styrene) vitrimers (44 kJ/mol). \cite{jiang2024reprocessable} 
As shown in Figure \ref{fig:Fig4}(f) the activation energy increases monotonically with increasing nominal crosslink density, even though the increase for crosslink densities above 2.5 mol\% is very small; this behaviour is consistent with reports for vitrimers based on both dioxaborolane and other exchange mechanisms. \cite{hajj2020network,ricarte2023time} 

Finally, a brief comparison with the behaviour of semi-cured vitrimers is included in Figure S7 (SI) that shows the TTS mastercurves for the semi-cured VMA-1'. At high frequencies, the $\alpha$ relaxation response is observed, followed by a plateau-like behaviour that is less distinct than the one observed for the corresponding fully cured sample (VMA-1) and the low frequency (high-$T$) relaxation is much less pronounced. The $T$-dependent $\alpha$ relaxation times $\tau_{\alpha}(T)$ are shown for both the VMA-1' and the VMA-1 sample in Figure S8 (SI). The $\alpha$ relaxation time of VMA-1' is faster than that of VMA-1, consistent with the fact that $T_{\textup{g}}$ of VMA-1' is lower than that of VMA-1. Moreover, the relaxation times for the slow network relaxations are also shown. Relatively small differences are observed in the activation enthalpies, ranging from 69 kJ/mol for VMA-1' to 56 kJ/mol for VMA-1, and the most significant observed change is the shift of $\tau^{\ast}$ towards shorter time-scales for the semi-cured sample, which is consistent with the increase in chain mobility for the less cross-linked VMA-1' sample. 

\subsection{Tensile deformation and stretchability}

\begin{figure*}
  \begin{center}{\includegraphics[width=1\textwidth]{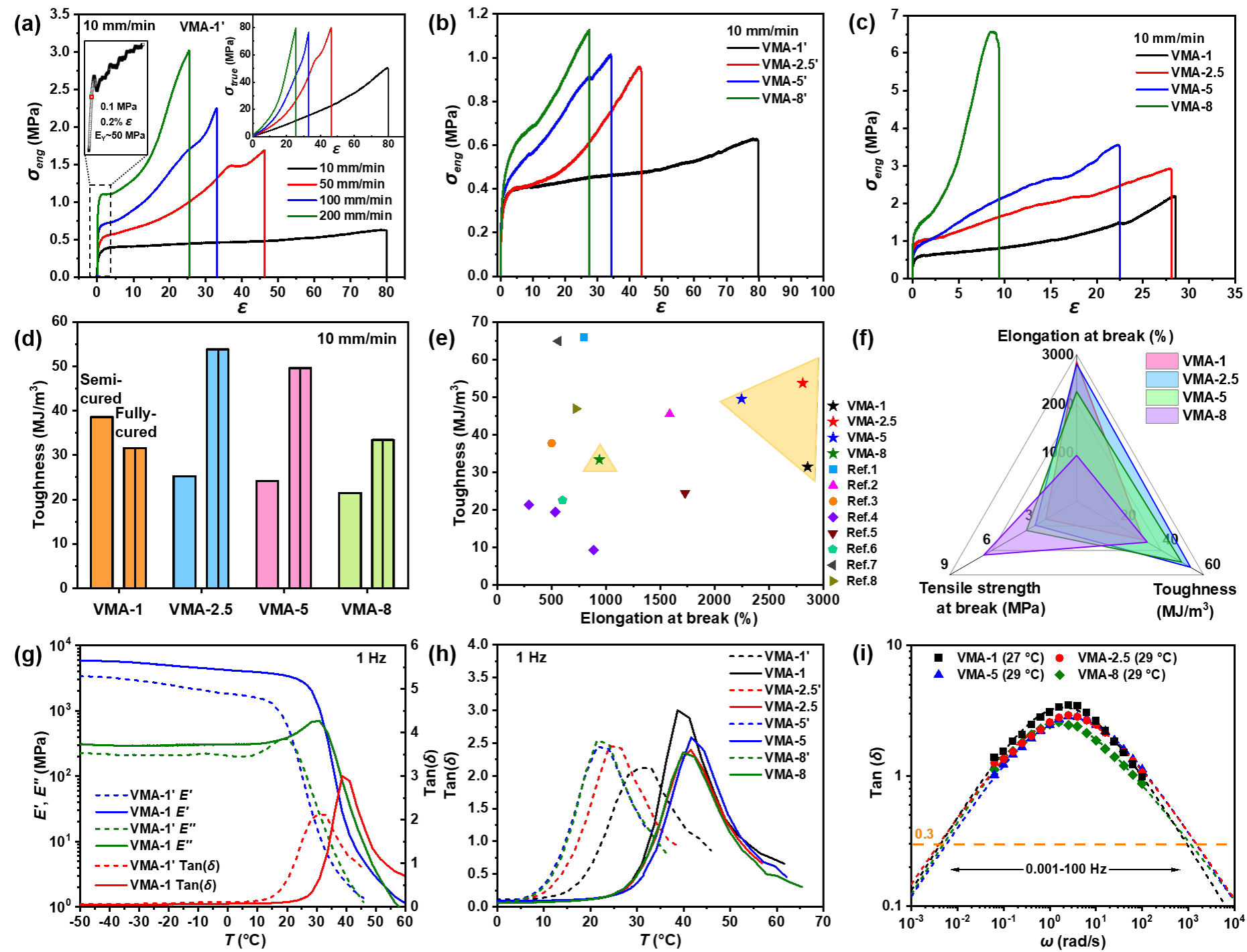}}\end{center}
  \caption{a) Tensile engineering stress vs strain data for the semi-cured VMA-1' sample for different extension speeds, as described in detail in the main text; the left inset shows a zoom-in of the data for small deformations. The right inset shows the data presented as true stress versus strain. b) Tensile engineering stress vs strain data for the semi-cured vitrimers of different crosslink densities (10 mm/min). c) Tensile engineering stress vs strain data for the fully cured vitrimers of different crosslink densities. d) Comparison of the mechanical toughness for the semi-cured and fully cured vitrimer sample of all crosslink densities. e) Comparison of the relationship between toughness and elongation at break for the fully cured vitrimer and other high-performance vitrimers or non-thermoset elastomers reported in literature \cite{zhang2020synergistic,chen2023exceptionally,lu2024high,zheng2023covalently,luo2023highly,zhou2021room,leone2022dynamically,song2019ultra}. f) Radar chart demonstrating the interrelationship of the elongation at break, tensile strength at break, and toughness of the fully cured vitrimers. g) DMA results for the storage ($E'$) and loss ($E''$) Young's moduli, and the corresponding loss tangent $\tan(\delta)$, as measured at 1 Hz. h) The $T$-dependent $\tan(\delta)$ for both the semi-cured and fully cured vitrimer samples. i) The frequency-dependent $\tan(\delta)$ for the fully cured vitrimers, together with fits to a Harvriliak-Negami expression. The horizontal dashed line indicates the limit of efficient mechanical damping.}
  \label{fig:Fig5}
\end{figure*}

The room temperature ($T$ = 20 °C) tensile stress response ($\sigma$) to an imposed strain ($\varepsilon$) was determined for four different extension speeds of 10, 50, 100 and 200 mm/min. Data for the semi-cured VMA-1' samples are shown in Figure \ref{fig:Fig5}(a) as engineering stress ($\sigma_{\textup{eng}}$) versus imposed strain. For small deformations, a linear elastic regime is observed for all applied strain rates, corresponding to a Young's modulus of $\approx$ 50 MPa, as shown in the inset to Figure \ref{fig:Fig5}(a). For larger deformations, the observed yield stress increases with applied strain rate within the range $\sim$0.4--1.1 MPa, and strain hardening is observed where the stress required for deformation rises more rapidly for increasing strain rates. Such complex strain-rate dependence is typically observed both in linear polymers \cite{haward1993strain} and in elastomers based on dynamically crosslinked networks. \cite{lyu2020extremely,zhao2021mortise,xiang2023highly,leibler1991dynamics} 

For the lowest applied strain rate, only a very weak strain hardening effect is found and a nearly horizontal plateau is observed, whereas higher rates lead to an increasing growth of the stress response. For an elastomer that is capable of large tensile strains, the engineering strain, which assumes a constant cross-sectional area, is not always representative of the tensile strength. Thus, in the inset to Figure \ref{fig:Fig5}(a) we also present the estimated true stress $\sigma_{\textrm{true}}$ ($\sigma_{\textrm{true}}=\sigma_{\textrm{eng}}(1+\varepsilon)$) by assuming a constant volume during stretching. For the three higher extension speeds, the true stress at break is effectively constant at $\sim$80 MPa, whereas at the slowest deformation, the true stress reaches only $\sim$ 50 MPa -- hence, deformation at a faster rate results in the vitrimer reaching its `true' strength at break, and thus fracture, sooner. 


Macroscopic failure of materials commonly starts with micro-cracks, where crack propagation leads to the final failure. \cite{williams1984fracture} For vitrimers, both a sufficient segmental mobility (low enough $T_{\textup{g}}$) and any associative bond kinetics fast enough to allow network relaxations on the relevant deformation time-scale \cite{leibler1991dynamics} could deter these mechanisms and thus contribute to large stretchability. Since $T_{\textup{g}}\approx13$ °C is close to room temperature, the characteristic time-scale of the structural ($\alpha$) relaxation is relatively slow, and $\tau_{\alpha}\approx$ 0.1 s (see Fig. S8). As an example for the four semi-cured samples, the imposed extensional strain rates ($\dot{\varepsilon}$) are 0.0267, 1.691, 2.251, and 3.024 s$^{-1}$, respectively. In turn, the time-scales corresponding to the inverse applied strain rates ($1/\dot{\varepsilon}$) are 37.5, 0.59, 0.49, and 0.33 s, demonstrating that the time-scale characteristic of the deformation is $\tau'\sim 1/\dot{\varepsilon}\approx \tau_{\alpha}$ for the three higher strain rates, whereas $\tau'\gg \tau_{\alpha}$ for the lowest applied rate. Thus, particularly for the slowest deformation rate, there is enough time for structural relaxation to take place during the deformation. The overall result is a negligible strain-hardening for the lowest applied strain rate and a remarkably stretchable material, for which a deformation at break of 8000\% is observed; the corresponding deformation at break is reduced to 2500\% for the faster rate of 3.024 s$^{-1}$, whereas the fracture strength is increased from $\sim$ 0.6 MPa to $\sim$3 MPa. 

We note that it is possible that the rate of the $\alpha$ relaxation and/or the bond exchange mechanism becomes significantly sped up during the deformation, particularly where the stress is focused near the tip of any micro-cracks, resulting in further contributions to the observed behaviour. The rate of cross-link exchange in the quiescent vitrimers is relatively slow, as shown by the time-scale of the network relaxation (see Fig. S8). Thus, we expect that only if this rate is significantly increased due to the applied deformation, will the dynamic nature of the crosslinks influence the stretchability. From our experiments alone, we cannot determine the role (if any) of the dynamic cross-links in the tensile deformation behaviour. However, previous studies on PMA-based elastomers with permanent crosslinks \cite{kawano2023polymers,Siavoshani2024Soft_Matter} have also demonstrated highly stretchable behaviour, with a maximum stretchability of $\sim$1200\% \cite{kawano2023polymers} observed for the lowest investigated crosslink density ($\sim 5\cdot 10^{-5}$ mol/cm$^{3}$) and strain rate (0.01 s$^{-1}$). This does not exclude a role of the dynamic cross-links, but it does illustrate the key role played by the near room temperature location of the glass transition for PMA in providing a highly stretchable material.

Consecutive tensile loading-unloading cycles were also performed on the VMA-1' sample with a waiting time of 10 min between each cycle. The results are shown for a maximum applied strain of 100\% in Fig. S9 (SI) and 1000\% in Fig. S10 (SI), respectively. No significant stress decrease is found for a repeated application of 100\% strain, even though a small residual unrecovered strain ($\lesssim10\%$) was observed right after the end of each cycle; this residual strain is similar to that previously reported ($<$ 15\%) during cyclic tensile testing of disulfide bond exchange-based PDMS vitrimers. \cite{luo2023highly}. In contrast, for large deformations (1000\% strain), the required deformation stress is increasingly reduced for successive loading-unloading cycles, and a marked unrecovered strain ($< 200\%$) is observed, which takes more than 2 hours to recover. 

The tensile response of all four semi-cured vitrimers are shown in Figure \ref{fig:Fig5} (b). Beyond the yield stress, an increasing crosslink density corresponds to an increase in the stress required to deform the sample; an increase in the fracture stress from $\sim$ 0.6 MPa to 1.1 MPa; and a reduction from 8000\% (VMA-1') to 2750\% (VMA-8') of the elongation at break. Thus, while the tensile strength increases for increasing cross-link density, the elongation at break is reduced. Similarly, as shown in Figure \ref{fig:Fig5}(c), for the fully cured vitrimer samples, the tensile strength is considerably increased for increasing crosslink density, and even though the elongation at break is reduced compared with the semi-cured samples, it remains within the range of $\sim$ 1000\% to 3000\%, and the fracture stress for VMA-8 has a high value of $\sim$ 6.5 MPa. Although there have been a few recent reports of highly stretchable elastomers, \cite{sun2012highly,li2016highly} our vitrimers uniquely combine very high stretchability with a high tensile strength, and importantly do not show the strain softening typically observed in other highly stretchable elastomers.\cite{zhang2018exploring,zhang2019superstretchable,lyu2020extremely} 

\subsection{Combination of toughness, strength and stretchability}

A measure of mechanical toughness is obtained by integration of the stress-strain data in Figures \ref{fig:Fig5}(b--c), and the results are shown in Figure \ref{fig:Fig5}(d). The toughness generally increases upon full curing, with the exception of the VMA-1 sample for which a small decrease is observed, which is directly related to the reduction in the tensile extension achievable upon full curing. In Figure \ref{fig:Fig5}(e) the calculated toughness is also plotted versus the elongation at break both for the fully cured samples (highlighted by a yellow background) and for other recently reported examples of high performance vitrimers and reprocessable elastomers in the literature; the corresponding plot for the semi-cured samples is shown in Figure S11. Our vitrimers clearly show an excellent combination of stretchability and toughness, where the specific combination is tuneable by variation of the crosslink density. In fact, the toughness of our fully cured vitrimers is comparable to the best high performance elastomers in the literature, yet surpass these materials in stretchability, as shown in Figure \ref{fig:Fig5}(e). Moreover, our semi-cured vitrimers are comparable to other reported high-performance elastomers in the literature with regards to toughness, but can significantly outperform most other elastomers in terms of stretchability, see Figure S11. 

To further illustrate the achieved materials properties, Figure \ref{fig:Fig5}(f) provides a radar chart for the fully cured samples with axes given by the elongation at break, the tensile strength at break, and the toughness, respectively. The combination of high toughness and elongation at break (particularly for VMA-2.5 and VMA-5) is noteworthy, as well as the trade-off between these properties and the tensile strength at break (as controlled by the crosslink density). The corresponding plot for the semicured samples is shown in Figure S12 (SI), demonstrating that semi-cured samples and particularly VMA-1', combine an extremely high stretchability with excellent toughness and intermediate tensile strength at break; generally, the higher crosslink vitrimers gain in tensile strength at break, while showing a reduction in toughness and elongation at break. 

\begin{figure*}
  \begin{center}{\includegraphics[width=1\textwidth]{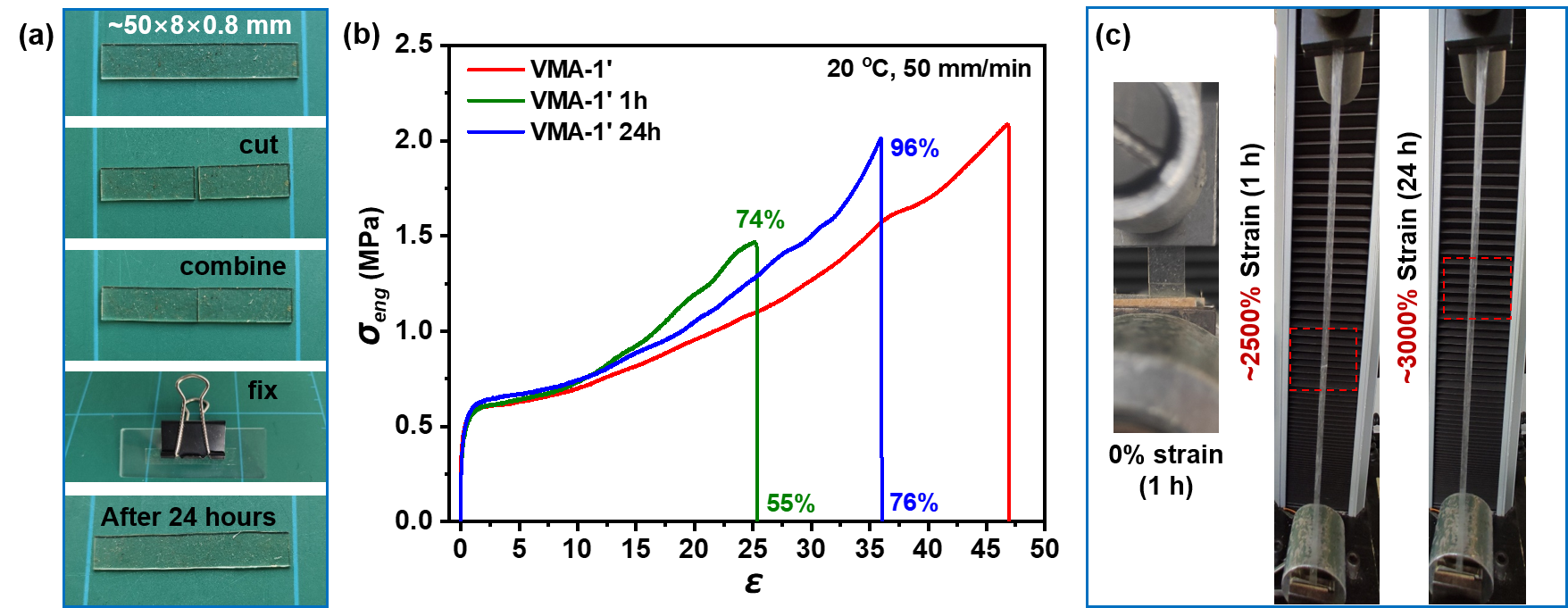}}\end{center}
  \caption{a) Illustration of the procedure used to test the room temperature self-healing properties. b) Tensile engineering stress vs strain curves for the semi-cured VMA-1' sample before cutting (red), and after self-healing for 1 hour (green) and 24 hours (blue). c) Photographs showing the sample after 1 hour without any applied strain (left), after 1 hour at a strain of $\sim$2500$\%$ (middle), and after 24 hours at a strain of $\sim$3000$\%$ (right). The small red box marks the location of the original cut, illustrating the progression of the healing process.}
  \label{fig:Fig6}
\end{figure*}

\subsection{Mechanical damping}

The mechanical response, and particularly the mechanical damping, was investigated using tensile temperature ramp experiments performed at 1 Hz using a dynamic mechanical analyser (DMA). Representative results are shown for the VMA-1' (dashed lines) and VMA-1 (solid lines) samples in Figure \ref{fig:Fig5}(g), where the storage $E^{'}$ (blue lines) and loss $E^{''}$ (green lines) are shown together with the loss tangent ($\textup{tan}(\delta)=E^{''}/E^{'}$) (red lines). The main decay in the storage modulus, corresponding to a loss peak in $E^{''}$, arises due to the structural ($\alpha$) relaxation, as observed for the probe frequency of $f=$1 Hz (corresponding to a relaxation time of $\tau_{\alpha}=1/(2\pi f)=$0.16 s; the smaller decay observed in $E^{'}$ at low temperatures, in turn corresponding to a weak peak in $E^{''}$ is due to a secondary relaxation, which persists within the glassy state. 

Comparing these temperature-sweep DMA data with the SAOS frequency-sweep data of Fig. \ref{fig:Fig4}, it is clear that the $T$-range where the ratio of loss to storage modulus, and thus $\tan(\delta)$, is the largest is found at temperatures just above the glass transition. Here, relaxations due to internal Rouse modes of the chain are superposed with a strong viscous damping due to the proximal glass transition, and the combination of these gives rise to a loss modulus significantly exceeding the storage modulus. In fact, this qualitative behaviour is seen between the glass transition and the entanglement/rubber plateau in polymeric materials in general; however, the separation of the loss modulus and the storage modulus is particularly large for polyacrylates. A plausible microscopic explanation for the effect is that dipole-dipole interactions between adjacent carbonyls (C=O $\cdot\cdot\cdot$ C=O) in polyacrylates may lead to a large relaxation modulus associated with the glass transition, \cite{xiang2023highly,sahariah2019relative} which significantly exceeds the entropic modulus of Rouse modes at the Kuhn segment length: this then leads to a greater viscous contribution of the glass transition within the Rouse relaxation region and so a higher $\tan(\delta)$, see Figure S14.

The $T$-dependent loss tangent is plotted for all semicured (dashed lines) and fully cured (solid lines) vitrimer samples in Figure \ref{fig:Fig5}(h). A maximum loss factor ($\textup{tan}(\delta)$) with a value between 2 and 3 is consistently observed. As the cross-link density increases upon full curing, the $\alpha$-relaxation slows down, as shown by the shift of the $E^{''}$ (Fig. 5g) and $\textup{tan}(\delta)$ (Fig. 5h) loss peaks to higher temperatures, which is consistent with the increase in $T_\textup{g}$ (see Table S2 in the SI). Also, increasing the cross-link density, either by addition of cross-linker during synthesis or by curing, shifts the peak in the loss tangent, and both curing and nominal cross-linking concentration can thus be used to control the $T$-position of the mechanical loss maximum. The temperature span of efficient mechanical loss ($\textup{tan}(\delta) >$ 0.3) cover 35 $\pm$5 °C, which is comparable to the $T$ ranges ($\sim40$ °C) for commercial styrene-butadiene and natural rubbers ($\textup{tan}(\delta)_\textup{max}$ $\sim$ 2.0). \cite{berki2017structure,wang2016novel} We note that the temperature range for efficient mechanical loss could be further increased by blending of semi-cured and fully cured vitrimers, thus producing materials with efficient loss factors across a $T$ range of $\sim$35 to 55 °C; such temperature ranges are comparable to those achieved for copolymer-based elastomers ($\textup{tan}(\delta)_\textup{max}$ $<$ 1.7), \cite{faghihi2011effect} however, without the inherent compatibility issues that exist for blends of different rubbers ($\textup{tan}(\delta)_\textup{max}$ $<$ 2.0). \cite{lei2019preparation} 

Since for damping applications, most (harmful) vibration sources are found in the frequency range of 0.01--100 Hz, \cite{krajnak2018health,shi2024dynamic} the frequency-dependent shear mechanical response was also measured (see Experimental Section). The results are shown in Figure \ref{fig:Fig5}(i) for the fully cured vitrimers, demonstrating a maximum mechanical loss factor tan($\delta$)$\geq$2.5, in agreement with the single-frequency (1Hz) experiments (see panel (h)). Due to the restricted frequency window of the rheometer ($\sim$ 3 decades), an empirical Havriliak-Negami expression \cite{havriliak1967complex} was used to describe the frequency-dependent loss tangent \cite{szabo2002method,madigosky2006method}, and thus provide a realistic estimate of the frequency-range for which the mechanical damping is efficient (tan($\delta$)$>$0.3, \cite{zhao2020bio}) We conclude that our vitrimers demonstrate good mechanical damping across a frequency range of 0.001--100 Hz, corresponding to more than five orders of magnitude in frequency. 

To put our results into perspective, the maximum value of the damping loss factor reported in literature is $\sim$ 1.0 by controlling the hydrogen bond properties in supramolecular networks, \cite{xu2014molecular,zhang2016unusual,hou2022bioinspired} 1.0 to 1.4 by use of microphase separation in long-fluorinated side-group polyacrylates or gels prepared by copolymerization of dual monomers with different solvent solubilities. \cite{xiang2023highly,zhang2024dielectric} It is $\sim$ 1.2 by utilising a mortise-and-tenon joint inspired rotaxane-containing mechanically interlocked supramolecular network, \cite{zhao2021mortise} around 1.0--1.5 for liquid crystal elastomers, \cite{ohzono2019enhanced,saed2021impact,farre2022dynamic} 1.5 by use of synergistic covalent and supramolecular interactions, \cite{zhang2020muscle} and close to 2.0 in polymer-fluid-based gels. \cite{huang2021ultrahigh} Thus, our vitrimers reach loss factor values at least as high as those for the best currently identified materials, and demonstrate effective loss factors across a wide ($\sim$ 5 decades) and relevant frequency range for damping, spanning a wide $T$-range around room temperature. 

\subsection{Self-healing}

At elevated temperatures where the crosslink exchange kinetics are sufficiently fast, vitrimers are typically reprocessable. The dynamic nature of the network of covalent cross-links could also facilitate efficient self-healing if the exchange dynamics are well tuned to the operational temperature range. Efficient self-healing depends on both segmental mobility and bond exchange kinetics, in addition to interfacial properties such as wetting and adhesion. \cite{stukalin2013self,kim1983theory,wool2008self} Thus, for many investigated vitrimer systems, temperatures significantly above room temperature are required to achieve self-healing on realistic time-scales. \cite{liu2018self,han2018catalyst,tang2021bio} Previous work on a polymer network with crosslink exchange reactions based on boronic transesterification demonstrated room temperature self-healing with a healing time of 4 days; it was also shown that addition of water to the fracture surface accelerates self-healing as it promotes the formation of free boronic acid and diol groups at the interface, which can in turn lead to the formation of bonds across the interface. \cite{cash2015room} These results, suggest that room temperature self-healing might be possible also in our PMA-based vitrimers, where the crosslink exchange is based on a similar mechanism. 

To investigate the ability of our vitrimers to self-heal at room temperature (20 °C), the test procedure outlined in Figure \ref{fig:Fig6}(a) was followed. The tests were focused on the semicured VMA-1' sample for which $T_{\textup{g}}$ is lower than room temperature ($\sim$20 °C), thus ensuring good segmental mobility at the fracture interface, The VMA-1' sample was cut into two parts using a razor blade, after which the two parts were manually joined. To ensure that the two parts were fixed during healing, two glass slides and a small clip were used as an effective ``splint" akin to the use of splints in the healing of bone fractures. Samples were prepared using both a 1 hour and a 24 hour healing time, respectively, and elongational stress-strain measurements were performed using an extensional speed of 50 mm/min (corresponding to a strain rate of $\sim$1.7 s$^{-1}$), as shown in Figure \ref{fig:Fig6}(b). Data recorded for the original VMA-1' sample (red) were compared with data recorded for the sample that had undergone healing for 1 hour (green) and 24 hours (blue), respectively. For the 1 hour sample, an elongation at fracture of 55\%, and a tensile strength at fracture of 74\%, of the corresponding values for the original sample were observed. After 24 hours, an elongation at fracture of 76\% and a tensile strength at fracture of 96\% were achieved. 

The photos in Figure \ref{fig:Fig6}(c) show a representative VMA-1' vitrimer sample before deformation at 0\% strain (left), after 1 hour at a strain of $\sim$ 2500$\%$ (centre), and after 24 hours at a strain of $\sim$ 3000$\%$ (right). In the left and centre photos, the cut can still be observed, but after 24 hours of room temperature self-healing it is no longer easily distinguishable, as shown in the right photo. Thus, we conclude that our semi-cured vitrimers demonstrate good self-healing properties at room temperature, without the addition of water or any other `solvent'. We do note that due to the slower segmental dynamics (ca. 20 °C higher $T_{\textup{g}}$) and slower (circa two orders of magnitude; see Figure S8) network rearrangements, the fully cured samples do not exhibit noticeable room temperature self-healing on a time-scale of 24 hours. However, fully cured vitrimers have the potential to self-heal at temperatures above room temperature.

\section{Conclusion}

We have synthesized and characterised a new vitrimer system based on poly(methyl acrylate) with dynamic crosslinks utilising the associative exchange reactions of dioxaborolane metathesis. We demonstrate that these vitrimers combine ultra-stretchability and mechanical toughness with high mechanical dissipation and self-healing. Vitrimers with nominal cross-link densities between 2 and 8 mol\% were prepared according to two different preparation routes: the first route involved a single curing stage (semi-cured) and the second an additional curing stage (fully cured); the two series of vitrimers showed complementary properties with excellent mechanical and thermal stability over a wide temperature range ($T<$250 °C), and they could each be effectively re-processed using thermal or chemical processing conditions. The produced set of vitrimers demonstrated an excellent combination of ultra-stretchability (up to 8000$\%$ for semi-cured and 3000$\%$ for fully cured vitrimers) and mechanical toughness (20--55 MJ/m$^3$), where the balance between the two is tuneable by variation of the crosslink density and the processing conditions. This combination of stretchability and toughness significantly outperforms most other elastomers in the literature, and the lowest cross-link semi-cured vitrimers also demonstrate good self-healing at room temperature. Moreover, the vitrimers show excellent mechanical damping properties with a maximum mechanical loss factor $\textup{tan}(\delta)$ value of $\sim$ 2.0--3.0, which corresponds to the values for the best currently identified damping materials, and shows a significant loss factor $\textup{tan}(\delta) >$ 0.3 across a wide (and for application relevant) frequency range of $\sim$ 5 decades (0.001--100 Hz), or correspondingly across a wide $T$-range (35 $\pm$5 °C for $f=$1 Hz) around room temperature. Both the processing conditions and the variation of the nominal cross-link density can be used to tune the $T$-range of efficient mechanical loss. Table S3 (SI) provides a comparison between the properties of our vitrimers and those of high performance elastomers in literature. 

Our vitrimers are promising candidates in polymer-based applications where stretchability combined with mechanical toughness and/or a high mechanical dissipation is required. Examples include applications in soft robotics or biomimetic synthetic materials, such as artificial skin, muscles, or cartilage, where a combination of high stretching and self-healing ability, together with a high mechanical strength and dissipation are required. The vitrimers are also suited to applications such as protective devices (helmets or clothing) or in vibration dampers or shock absorbers. Future work will focus on further developing the detailed understanding of the links between the primary polymer chain length, the detailed dynamics and the distribution of cross-link, and the resulting vitrimer physical properties, as well as to expand these investigations to the wider family of polyacrylate-based vitrimers.  

\section{Experimental Section}

\subsection{Materials}

Methyl acrylate ($\geq$99\%, contains $\leq$100 ppm monomethyl ether hydroquinone as inhibitor), dichloromethane (DCM), 1,4-dioxane (anhydrous, $\geq$99.8\%), deuterated dimethyl sulfoxide (DMSO-$d_{6}$, 99.9 atom \% D) and deuterium oxide (D$_{2}$O, 99.9 atom \% D) were purchased from Sigma Aldrich. 3-Allyloxy-1,2-Propanediol ($\geq$99\%) was purchased from Santa Cruz Biotechnology. Benzene-1 4-diboronic acid ($\geq$98\%) was purchased from Apollo Scientific Ltd. Magnesium sulfate (extra pure, dried), was purchased from Fisher Chemical. Azobisisobutyronitrile (AIBN) was purchased from VWR International Ltd. All the above reagents were used as received without further purification. Poly(methyl acrylate) (PMA-N315) was purchased from Sigma Aldrich, and this sample was supplied as a toluene solution. The solvent was first removed using a rotary evaporator, followed by three dissolution and reprecipitation cycles using THF and methanol to remove low molecular weight components to reach lower dispersity, and finally dried in a vacuum oven at 80 °C before use.

\subsection{Synthesis of diallyl dioxaborolane dynamic crosslinker}
Benzene-1,4-diboronic acid (5.074 g, 30 mmol, 1 equiv.), 3-allyloxy-1,2-diol (7.50 ml, 60 mmol, 2 equiv.), dichloromethane (50 ml) and magnesium sulphate (10 g), used to absorb the water generated during the reaction, were added together in a 250 ml round-bottomed flask and stirred at room temperature for 3 hours. The solid magnesium sulphate was filtered off at the end of the reaction and a white solid product was obtained from the solution after removing the solvent using a rotary evaporator. $^{1}$H NMR (500 MHz, CDCl$_{3}$): $\delta$ (ppm) = 7.83 (s, 4H, -Ph-), 5.88 (m, 1H, –C\textit{H}=), 5.28 (dd, \textit{J} = 17.3, 1.4 Hz, 1H, =C\textit{H}$_{2}$), 5.19 (dd, \textit{J} = 10.4, 1.4 Hz, 1H, =C\textit{H}$_{2}$), 4.73 (m, 1H, –CH$_{2}$–C(O–)\textit{H}–CH$_{2}$–), 4.43 (dd, \textit{J} = 9.1, 8.2 Hz, 1H, –C(O–)H–C\textit{H}$_{2}$–O–), 4.18 (dd, \textit{J} = 9.1, 6.6 Hz, 1H, –C(O–)H–C\textit{H}$_{2}$–O–), 4.06 (dq, \textit{J} = 5.7, 1.6 Hz, 2H, –C\textit{H}$_{2}$–CH=), 3.64 (dd, \textit{J} = 10.2, 5.1 Hz, 1H, –B–O–C\textit{H}$_{2}$–CH–), 3.57 (dd, \textit{J} = 10.2, 5.1 Hz, 1H, –B–O–C\textit{H}$_{2}$–CH–). $^{13}$C NMR (125 MHz, CDCl$_{3}$): $\delta$ (ppm) = 134.47 (–\textit{C}H=CH$_{2}$), 134.22 (–O–B(O–)C–\textit{C}H–), 130.86 (–O–B(O–)\textit{C}–), 117.60 (=\textit{C}H$_{2}$), 76.34 (-\textit{C}H(CH$_{2}$–)O–B–, 72.71 (=CH–\textit{C}H$_{2}$–), 72.03 (–B–O–\textit{C}H$_{2}$–), 68.62 (=CH–CH$_{2}$–O–\textit{C}H$_{2}$–).

\subsection{Synthesis and preparation of PMA vitrimers}

As an example, VMA-1 was prepared in a 100 ml Schlenk flask by adding previously synthesised diallyl dioxaborolane dynamic crosslinker 0.716 g (2 mmol), methyl acrylate 17.8 ml (198 mmol), initiator AIBN 0.164 g (1 mmol) and anhydrous dioxane 30 ml. Three freeze-pump-thaw cycles were performed before the reaction, and then the reaction was stirred at 70 °C under a nitrogen atmosphere until gelation. The reaction products were removed and left in a fume hood for two days to evaporate most of the solvents, and were then dried to constant weight in a vacuum oven at 80 °C. The dried product was placed in a mould and hot pressed at 120 °C using a pressure of 5 MPa for 15 minutes, then removed and naturally cooled to obtain the semi-cured VMA-1' sample. To obtain the fully cured VMA-1 sample, the VMA-1' sample was placed in a vacuum oven at 220 °C for 2 hours for secondary curing. Before testing, the fully cured vitrimers were hot pressed, as previously described, to prepare suitable sample geometries for the experiments.

\subsection{Nuclear magnetic resonance (NMR) spectroscopy}
$^{1}$H NMR spectra were recorded on a Bruker AV4 NEO 11.75 T 500 MHz spectrometer fitted with a 5 mm Bruker C/H cryoprobe. For the $^{1}$H NMR experiments, the spectral width was 9090 Hz, the relaxation delay time was 2 s, and the number of scans was 16. Around 0.5 mg VMA and 0.7 ml DMSO-$d_{6}$ were first added to an NMR tube, followed by the addition of 5 $\mu$l of D$_{2}$O to ensure hydrolysis of the boronic ester crosslinker. The samples were left for 48 hours after sample preparation and then measured to ensure that the hydrolysis was complete. Chemical shifts (in ppm) are referenced to the DMSO-$d_{6}$ solvent peak with a chemical shift of 2.50 ppm. 

\subsection{Gel permeation chromatography (GPC)}
The number average molecular weight ($M_{\textup{n}}$), weight average molecular weight ($M_{\textup{w}}$), and polymer dispersity ($\DJ$) were measured using an Agilent 1260 Infinity system GPC instrument fitted with two 5 $\mu$m Mixed-C columns plus a guard column, a refractive index detector, and a UV/Vis detector operating at 309 nm. THF (mixed with a small amount of water, vol/vol = 200:1) was used as the eluent (1 ml/min), and polystyrene standards ($M_{\textup{P}}$ ranging from 370 to 2,520,000 g/mol) were used for calibration. A polymer solution at a concentration of approximately 3 mg/ml was used for the measurements. The samples were left for 48 hours after sample preparation and then measured to ensure that hydrolysis was complete.

\subsection{Differential scanning calorimetry (DSC)}
DSC measurements were performed using a TA Instruments Q2000 instrument equipped using a refrigerated cooling system. Samples with a typical weight of 5-10 mg were placed in hermetically sealed aluminium pans (Tzero pans from TA instruments). Experiments were performed between -70 °C and 100 °C and three cycles were run using a heating and cooling scan rate of 10 °C/min under a nitrogen flow of 50 ml/min. 

\subsection{Thermogravimetric analysis (TGA)}
TGA measurements were carried out using a TA Instruments Q50 TGA instrument under a nitrogen atmosphere from room temperature to 800 °C at a constant rate of 10 °C/min with a sample mass of $\sim$10 mg. 

\subsection{Tensile Test}
Tensile tests were performed room temperature (20 °C) using an Instron 5564 universal testing machine with a 2 kN capacity load cell. The tests were carried out on rectangular specimens ($\sim$50 (length) $\times$ 8 (width) $\times$ 0.8 (thickness) mm) for varying extension speeds ranging from 10 to 200 mm/min.

\subsection{Dynamic mechanical analysis (DMA)}

DMA experiments were performed using a TA Instruments Q850 DMA on vitrimer samples with sizes of around 10 mm (length) $\times$ 4 mm (width) $\times$ 0.5 mm (thickness). Temperature ramps from -50 to 100 °C were performed using a heating rate of 3 °C/min at a frequency of 1 Hz, using strain amplitudes within the linear viscoelastic region; a preload force of 1 N and a force track of 125\% were used. 

\subsection{Small-amplitude oscillatory shear (SAOS) rheology}

SAOS was performed over a frequency range of 0.06-126 rad/s (0.01--20 Hz) using a Rheometrics ARES strain-controlled rheometer. A 5 mm diameter parallel plate geometry and a sample thickness of about 1.5 mm was used, and while cooling down, the sample thickness was gradually reduced (due to the increase in sample density) to maintain a constant sample radius. The temperature was controlled ($\pm 1$ K) with a liquid nitrogen cooling system controlling a forced convection oven. Strain sweeps were performed prior to the frequency sweeps to confirm that for the strain amplitudes used, the measurements were performed in the linear viscoelastic region (LVR). The strain amplitude was decreased with increasing temperature to ensure that the measured torque remained within the relevant transducer range and that the measurements were performed within the LVR. 

\subsection{Time temperature superposition of SAOS rheology data}

Rheology data measured at different temperatures can often be combined into a mastercurve using so-called Time Temperature Superposition (TTS). However, TTS is only applicable when a single relaxation process controls the data, i.e. the data are rheologically simple. To investigate the relevance of TTS analysis, we plot the phase angle $\delta$ vs the complex modulus $|G^*|$ in a so-called van Gurp Palmen (vGP) plot \cite{van1998time}, as shown for the VMA-8 sample in Fig. S15 (a). This representation of the data removes the explicit time-dependence and thus demonstrates whether TTS based on a simple horizontal shift parameter $a_{T}$ is a reasonable approximation. As shown in Fig. S5, the first observation is that for high moduli (low temperatures) a horizontal shifting works reasonably well, even though it is clear that generally a vertical shift parameter is also required. The vertical shift parameter for polymers is typically expressed as $b_T = T_{\textup{ref}}\cdot\rho_{\textup{ref}}/T\cdot \rho$, where $\rho$ is the mass density. For our vitrimers, the detailed temperature variation of the density is unknown. Thus, we set $b_{T} = T_{\textup{ref}}/T$ as an approximation for the vertical shift parameter. To investigate the accuracy of this approach, we have plotted a modulus-renormalised vGP plot in Fig. S15 (b). The results confirm that a simple $T$-renormalization provides a good approximation of the vertical shift. 

Using the TA Instruments Orchestrator software, we first created mastercurves for the loss tangent ($\textup{tan}(\delta)=G^{'}/G^{''}$) data, by determining the 
horizontal shift parameters $a_T(T)$ required to overlap data for neighbouring temperatures; any vertical shifts required between data sets are removed for the loss tangent, thus facilitating an unbiased determination. Mastercurves for the shear storage and loss moduli, $G^{'}$ and $G^{''}$, were subsequently created by also applying the vertical shift parameters $b_T(T)$, as described above. The resulting master curves for the shear storage and loss moduli, $G^{'}$ and $G^{''}$ are shown in Figure \ref{fig:Fig4} for the four fully cured vitrimer samples, and in Fig. S7 (SI) for the semicured VMA-1' sample. At low-$T$ the structural $\alpha$ relaxation is the dominating mechanism, whereas at high-$T$, a mechanism related to cross-link exchange is instead controlling the behaviour. Thus, we plot the mastercurves corresponding to these low and high temperature regimes separately in Fig. \ref{fig:Fig4} and Fig. S7 (SI), respectively.  

\subsection{Supporting Information}
\noindent Supplementary Information to this article is available from.

\subsection{Acknowledgements}
\noindent This work was supported by the EPSRC EP/T011726/1 and the Leverhulme Trust [RPG-2019-169]. Jiaxin Zhao thanks the China Scholarship Council and University of Leeds for awarding him a joint scholarship (202106220064). 

\subsection{Conflict of Interest}
\noindent The authors declare no conflict of interest.

\subsection{Data Availability Statement}
\noindent The data in this paper are available in the Leeds Data Repository at https://doi.org/10.5518/1650.




\begin{thebibliography}{99}%
\makeatletter
\providecommand \@ifxundefined [1]{%
 \@ifx{#1\undefined}
}%
\providecommand \@ifnum [1]{%
 \ifnum #1\expandafter \@firstoftwo
 \else \expandafter \@secondoftwo
 \fi
}%
\providecommand \@ifx [1]{%
 \ifx #1\expandafter \@firstoftwo
 \else \expandafter \@secondoftwo
 \fi
}%
\providecommand \natexlab [1]{#1}%
\providecommand \enquote  [1]{``#1''}%
\providecommand \bibnamefont  [1]{#1}%
\providecommand \bibfnamefont [1]{#1}%
\providecommand \citenamefont [1]{#1}%
\providecommand \href@noop [0]{\@secondoftwo}%
\providecommand \href [0]{\begingroup \@sanitize@url \@href}%
\providecommand \@href[1]{\@@startlink{#1}\@@href}%
\providecommand \@@href[1]{\endgroup#1\@@endlink}%
\providecommand \@sanitize@url [0]{\catcode `\\12\catcode `\$12\catcode `\&12\catcode `\#12\catcode `\^12\catcode `\_12\catcode `\%12\relax}%
\providecommand \@@startlink[1]{}%
\providecommand \@@endlink[0]{}%
\providecommand \url  [0]{\begingroup\@sanitize@url \@url }%
\providecommand \@url [1]{\endgroup\@href {#1}{\urlprefix }}%
\providecommand \urlprefix  [0]{URL }%
\providecommand \Eprint [0]{\href }%
\providecommand \doibase [0]{https://doi.org/}%
\providecommand \selectlanguage [0]{\@gobble}%
\providecommand \bibinfo  [0]{\@secondoftwo}%
\providecommand \bibfield  [0]{\@secondoftwo}%
\providecommand \translation [1]{[#1]}%
\providecommand \BibitemOpen [0]{}%
\providecommand \bibitemStop [0]{}%
\providecommand \bibitemNoStop [0]{.\EOS\space}%
\providecommand \EOS [0]{\spacefactor3000\relax}%
\providecommand \BibitemShut  [1]{\csname bibitem#1\endcsname}%
\let\auto@bib@innerbib\@empty
\bibitem [{\citenamefont {Oh}\ \emph {et~al.}(2016)\citenamefont {Oh}, \citenamefont {Rondeau-Gagn{\'e}}, \citenamefont {Chiu}, \citenamefont {Chortos}, \citenamefont {Lissel}, \citenamefont {Wang}, \citenamefont {Schroeder}, \citenamefont {Kurosawa}, \citenamefont {Lopez}, \citenamefont {Katsumata} \emph {et~al.}}]{oh2016intrinsically}%
  \BibitemOpen
  \bibfield  {author} {\bibinfo {author} {\bibfnamefont {J.~Y.}\ \bibnamefont {Oh}}, \bibinfo {author} {\bibfnamefont {S.}~\bibnamefont {Rondeau-Gagn{\'e}}}, \bibinfo {author} {\bibfnamefont {Y.-C.}\ \bibnamefont {Chiu}}, \bibinfo {author} {\bibfnamefont {A.}~\bibnamefont {Chortos}}, \bibinfo {author} {\bibfnamefont {F.}~\bibnamefont {Lissel}}, \bibinfo {author} {\bibfnamefont {G.-J.~N.}\ \bibnamefont {Wang}}, \bibinfo {author} {\bibfnamefont {B.~C.}\ \bibnamefont {Schroeder}}, \bibinfo {author} {\bibfnamefont {T.}~\bibnamefont {Kurosawa}}, \bibinfo {author} {\bibfnamefont {J.}~\bibnamefont {Lopez}}, \bibinfo {author} {\bibfnamefont {T.}~\bibnamefont {Katsumata}}, \emph {et~al.},\ }\bibfield  {title} {\bibinfo {title} {Intrinsically stretchable and healable semiconducting polymer for organic transistors},\ }\href@noop {} {\bibfield  {journal} {\bibinfo  {journal} {Nature}\ }\textbf {\bibinfo {volume} {539}},\ \bibinfo {pages} {411} (\bibinfo {year} {2016})}\BibitemShut {NoStop}%
\bibitem [{\citenamefont {Shi}\ \emph {et~al.}(2018)\citenamefont {Shi}, \citenamefont {Zhu}, \citenamefont {Gao}, \citenamefont {Zhang}, \citenamefont {Wei}, \citenamefont {Liu},\ and\ \citenamefont {Ding}}]{shi2018highly}%
  \BibitemOpen
  \bibfield  {author} {\bibinfo {author} {\bibfnamefont {L.}~\bibnamefont {Shi}}, \bibinfo {author} {\bibfnamefont {T.}~\bibnamefont {Zhu}}, \bibinfo {author} {\bibfnamefont {G.}~\bibnamefont {Gao}}, \bibinfo {author} {\bibfnamefont {X.}~\bibnamefont {Zhang}}, \bibinfo {author} {\bibfnamefont {W.}~\bibnamefont {Wei}}, \bibinfo {author} {\bibfnamefont {W.}~\bibnamefont {Liu}},\ and\ \bibinfo {author} {\bibfnamefont {S.}~\bibnamefont {Ding}},\ }\bibfield  {title} {\bibinfo {title} {Highly stretchable and transparent ionic conducting elastomers},\ }\href@noop {} {\bibfield  {journal} {\bibinfo  {journal} {Nature communications}\ }\textbf {\bibinfo {volume} {9}},\ \bibinfo {pages} {2630} (\bibinfo {year} {2018})}\BibitemShut {NoStop}%
\bibitem [{\citenamefont {Cheng}\ \emph {et~al.}(2019)\citenamefont {Cheng}, \citenamefont {Narang}, \citenamefont {Yang}, \citenamefont {Suo},\ and\ \citenamefont {Howe}}]{cheng2019stick}%
  \BibitemOpen
  \bibfield  {author} {\bibinfo {author} {\bibfnamefont {S.}~\bibnamefont {Cheng}}, \bibinfo {author} {\bibfnamefont {Y.~S.}\ \bibnamefont {Narang}}, \bibinfo {author} {\bibfnamefont {C.}~\bibnamefont {Yang}}, \bibinfo {author} {\bibfnamefont {Z.}~\bibnamefont {Suo}},\ and\ \bibinfo {author} {\bibfnamefont {R.~D.}\ \bibnamefont {Howe}},\ }\bibfield  {title} {\bibinfo {title} {Stick-on large-strain sensors for soft robots},\ }\href@noop {} {\bibfield  {journal} {\bibinfo  {journal} {Advanced Materials Interfaces}\ }\textbf {\bibinfo {volume} {6}},\ \bibinfo {pages} {1900985} (\bibinfo {year} {2019})}\BibitemShut {NoStop}%
\bibitem [{\citenamefont {Leber}\ \emph {et~al.}(2019)\citenamefont {Leber}, \citenamefont {Cholst}, \citenamefont {Sandt}, \citenamefont {Vogel},\ and\ \citenamefont {Kolle}}]{leber2019stretchable}%
  \BibitemOpen
  \bibfield  {author} {\bibinfo {author} {\bibfnamefont {A.}~\bibnamefont {Leber}}, \bibinfo {author} {\bibfnamefont {B.}~\bibnamefont {Cholst}}, \bibinfo {author} {\bibfnamefont {J.}~\bibnamefont {Sandt}}, \bibinfo {author} {\bibfnamefont {N.}~\bibnamefont {Vogel}},\ and\ \bibinfo {author} {\bibfnamefont {M.}~\bibnamefont {Kolle}},\ }\bibfield  {title} {\bibinfo {title} {Stretchable thermoplastic elastomer optical fibers for sensing of extreme deformations},\ }\href@noop {} {\bibfield  {journal} {\bibinfo  {journal} {Advanced Functional Materials}\ }\textbf {\bibinfo {volume} {29}},\ \bibinfo {pages} {1802629} (\bibinfo {year} {2019})}\BibitemShut {NoStop}%
\bibitem [{\citenamefont {Huang}\ \emph {et~al.}(2016)\citenamefont {Huang}, \citenamefont {Tunnicliffe}, \citenamefont {Zhuang}, \citenamefont {Ren}, \citenamefont {Yan},\ and\ \citenamefont {Busfield}}]{huang2016strain}%
  \BibitemOpen
  \bibfield  {author} {\bibinfo {author} {\bibfnamefont {M.}~\bibnamefont {Huang}}, \bibinfo {author} {\bibfnamefont {L.~B.}\ \bibnamefont {Tunnicliffe}}, \bibinfo {author} {\bibfnamefont {J.}~\bibnamefont {Zhuang}}, \bibinfo {author} {\bibfnamefont {W.}~\bibnamefont {Ren}}, \bibinfo {author} {\bibfnamefont {H.}~\bibnamefont {Yan}},\ and\ \bibinfo {author} {\bibfnamefont {J.~J.}\ \bibnamefont {Busfield}},\ }\bibfield  {title} {\bibinfo {title} {Strain-dependent dielectric behavior of carbon black reinforced natural rubber},\ }\href@noop {} {\bibfield  {journal} {\bibinfo  {journal} {Macromolecules}\ }\textbf {\bibinfo {volume} {49}},\ \bibinfo {pages} {2339} (\bibinfo {year} {2016})}\BibitemShut {NoStop}%
\bibitem [{\citenamefont {Mazurek}\ \emph {et~al.}(2019)\citenamefont {Mazurek}, \citenamefont {Vudayagiri},\ and\ \citenamefont {Skov}}]{mazurek2019tailor}%
  \BibitemOpen
  \bibfield  {author} {\bibinfo {author} {\bibfnamefont {P.}~\bibnamefont {Mazurek}}, \bibinfo {author} {\bibfnamefont {S.}~\bibnamefont {Vudayagiri}},\ and\ \bibinfo {author} {\bibfnamefont {A.~L.}\ \bibnamefont {Skov}},\ }\bibfield  {title} {\bibinfo {title} {How to tailor flexible silicone elastomers with mechanical integrity: A tutorial review},\ }\href@noop {} {\bibfield  {journal} {\bibinfo  {journal} {Chemical Society Reviews}\ }\textbf {\bibinfo {volume} {48}},\ \bibinfo {pages} {1448} (\bibinfo {year} {2019})}\BibitemShut {NoStop}%
\bibitem [{\citenamefont {Tong}\ and\ \citenamefont {J{\'e}r{\^o}me}(2000)}]{tong2000synthesis}%
  \BibitemOpen
  \bibfield  {author} {\bibinfo {author} {\bibfnamefont {J.}~\bibnamefont {Tong}}\ and\ \bibinfo {author} {\bibfnamefont {R.}~\bibnamefont {J{\'e}r{\^o}me}},\ }\bibfield  {title} {\bibinfo {title} {Synthesis of poly (methyl methacrylate)-b-poly (n-butyl acrylate)-b-poly (methyl methacrylate) triblocks and their potential as thermoplastic elastomers},\ }\href@noop {} {\bibfield  {journal} {\bibinfo  {journal} {Polymer}\ }\textbf {\bibinfo {volume} {41}},\ \bibinfo {pages} {2499} (\bibinfo {year} {2000})}\BibitemShut {NoStop}%
\bibitem [{\citenamefont {Zhang}\ \emph {et~al.}(2018{\natexlab{a}})\citenamefont {Zhang}, \citenamefont {Niu},\ and\ \citenamefont {Zhang}}]{zhang2018extremely}%
  \BibitemOpen
  \bibfield  {author} {\bibinfo {author} {\bibfnamefont {H.}~\bibnamefont {Zhang}}, \bibinfo {author} {\bibfnamefont {W.}~\bibnamefont {Niu}},\ and\ \bibinfo {author} {\bibfnamefont {S.}~\bibnamefont {Zhang}},\ }\bibfield  {title} {\bibinfo {title} {Extremely stretchable, stable, and durable strain sensors based on double-network organogels},\ }\href@noop {} {\bibfield  {journal} {\bibinfo  {journal} {ACS applied materials \& interfaces}\ }\textbf {\bibinfo {volume} {10}},\ \bibinfo {pages} {32640} (\bibinfo {year} {2018}{\natexlab{a}})}\BibitemShut {NoStop}%
\bibitem [{\citenamefont {Li}\ \emph {et~al.}(2022{\natexlab{a}})\citenamefont {Li}, \citenamefont {Huang}, \citenamefont {Deng}, \citenamefont {Guo}, \citenamefont {Cai}, \citenamefont {Zhang},\ and\ \citenamefont {Guo}}]{li2022highly}%
  \BibitemOpen
  \bibfield  {author} {\bibinfo {author} {\bibfnamefont {G.}~\bibnamefont {Li}}, \bibinfo {author} {\bibfnamefont {K.}~\bibnamefont {Huang}}, \bibinfo {author} {\bibfnamefont {J.}~\bibnamefont {Deng}}, \bibinfo {author} {\bibfnamefont {M.}~\bibnamefont {Guo}}, \bibinfo {author} {\bibfnamefont {M.}~\bibnamefont {Cai}}, \bibinfo {author} {\bibfnamefont {Y.}~\bibnamefont {Zhang}},\ and\ \bibinfo {author} {\bibfnamefont {C.~F.}\ \bibnamefont {Guo}},\ }\bibfield  {title} {\bibinfo {title} {Highly conducting and stretchable double-network hydrogel for soft bioelectronics},\ }\href@noop {} {\bibfield  {journal} {\bibinfo  {journal} {Advanced Materials}\ }\textbf {\bibinfo {volume} {34}},\ \bibinfo {pages} {2200261} (\bibinfo {year} {2022}{\natexlab{a}})}\BibitemShut {NoStop}%
\bibitem [{\citenamefont {Li}\ \emph {et~al.}(2016)\citenamefont {Li}, \citenamefont {Wang}, \citenamefont {Keplinger}, \citenamefont {Zuo}, \citenamefont {Jin}, \citenamefont {Sun}, \citenamefont {Zheng}, \citenamefont {Cao}, \citenamefont {Lissel}, \citenamefont {Linder} \emph {et~al.}}]{li2016highly}%
  \BibitemOpen
  \bibfield  {author} {\bibinfo {author} {\bibfnamefont {C.-H.}\ \bibnamefont {Li}}, \bibinfo {author} {\bibfnamefont {C.}~\bibnamefont {Wang}}, \bibinfo {author} {\bibfnamefont {C.}~\bibnamefont {Keplinger}}, \bibinfo {author} {\bibfnamefont {J.-L.}\ \bibnamefont {Zuo}}, \bibinfo {author} {\bibfnamefont {L.}~\bibnamefont {Jin}}, \bibinfo {author} {\bibfnamefont {Y.}~\bibnamefont {Sun}}, \bibinfo {author} {\bibfnamefont {P.}~\bibnamefont {Zheng}}, \bibinfo {author} {\bibfnamefont {Y.}~\bibnamefont {Cao}}, \bibinfo {author} {\bibfnamefont {F.}~\bibnamefont {Lissel}}, \bibinfo {author} {\bibfnamefont {C.}~\bibnamefont {Linder}}, \emph {et~al.},\ }\bibfield  {title} {\bibinfo {title} {A highly stretchable autonomous self-healing elastomer},\ }\href@noop {} {\bibfield  {journal} {\bibinfo  {journal} {Nature chemistry}\ }\textbf {\bibinfo {volume} {8}},\ \bibinfo {pages} {618} (\bibinfo {year} {2016})}\BibitemShut {NoStop}%
\bibitem [{\citenamefont {Jin}\ \emph {et~al.}(2023)\citenamefont {Jin}, \citenamefont {Du}, \citenamefont {Tang}, \citenamefont {Zhao}, \citenamefont {Peng}, \citenamefont {Li}, \citenamefont {Zhang}, \citenamefont {Zhu}, \citenamefont {Huang}, \citenamefont {Kong} \emph {et~al.}}]{jin2023quadruple}%
  \BibitemOpen
  \bibfield  {author} {\bibinfo {author} {\bibfnamefont {Q.}~\bibnamefont {Jin}}, \bibinfo {author} {\bibfnamefont {R.}~\bibnamefont {Du}}, \bibinfo {author} {\bibfnamefont {H.}~\bibnamefont {Tang}}, \bibinfo {author} {\bibfnamefont {Y.}~\bibnamefont {Zhao}}, \bibinfo {author} {\bibfnamefont {W.}~\bibnamefont {Peng}}, \bibinfo {author} {\bibfnamefont {Y.}~\bibnamefont {Li}}, \bibinfo {author} {\bibfnamefont {J.}~\bibnamefont {Zhang}}, \bibinfo {author} {\bibfnamefont {T.}~\bibnamefont {Zhu}}, \bibinfo {author} {\bibfnamefont {X.}~\bibnamefont {Huang}}, \bibinfo {author} {\bibfnamefont {D.}~\bibnamefont {Kong}}, \emph {et~al.},\ }\bibfield  {title} {\bibinfo {title} {Quadruple h-bonding and polyrotaxanes dual cross-linking supramolecular elastomer for high toughness and self-healing conductors},\ }\href@noop {} {\bibfield  {journal} {\bibinfo  {journal} {Angewandte Chemie International Edition}\ }\textbf {\bibinfo {volume} {62}},\ \bibinfo {pages} {e202305282} (\bibinfo {year} {2023})}\BibitemShut {NoStop}%
\bibitem [{\citenamefont {Verjans}\ \emph {et~al.}(2024)\citenamefont {Verjans}, \citenamefont {Alexis}, \citenamefont {Sedlacik}, \citenamefont {Aksakal}, \citenamefont {van Ruymbeke},\ and\ \citenamefont {Hoogenboom}}]{Verjans2024JMCB}%
  \BibitemOpen
  \bibfield  {author} {\bibinfo {author} {\bibfnamefont {J.}~\bibnamefont {Verjans}}, \bibinfo {author} {\bibfnamefont {A.}~\bibnamefont {Alexis}}, \bibinfo {author} {\bibfnamefont {T.}~\bibnamefont {Sedlacik}}, \bibinfo {author} {\bibfnamefont {R.}~\bibnamefont {Aksakal}}, \bibinfo {author} {\bibfnamefont {E.}~\bibnamefont {van Ruymbeke}},\ and\ \bibinfo {author} {\bibfnamefont {R.}~\bibnamefont {Hoogenboom}},\ }\bibfield  {title} {\bibinfo {title} {Physically crosslinked polyacrylates by quadruple hydrogen bonding side chains},\ }\href@noop {} {\bibfield  {journal} {\bibinfo  {journal} {Journal of Materials Chemistry B}\ }\textbf {\bibinfo {volume} {12}},\ \bibinfo {pages} {12378} (\bibinfo {year} {2024})}\BibitemShut {NoStop}%
\bibitem [{\citenamefont {Sun}\ \emph {et~al.}(2012)\citenamefont {Sun}, \citenamefont {Zhao}, \citenamefont {Illeperuma}, \citenamefont {Chaudhuri}, \citenamefont {Oh}, \citenamefont {Mooney}, \citenamefont {Vlassak},\ and\ \citenamefont {Suo}}]{sun2012highly}%
  \BibitemOpen
  \bibfield  {author} {\bibinfo {author} {\bibfnamefont {J.-Y.}\ \bibnamefont {Sun}}, \bibinfo {author} {\bibfnamefont {X.}~\bibnamefont {Zhao}}, \bibinfo {author} {\bibfnamefont {W.~R.}\ \bibnamefont {Illeperuma}}, \bibinfo {author} {\bibfnamefont {O.}~\bibnamefont {Chaudhuri}}, \bibinfo {author} {\bibfnamefont {K.~H.}\ \bibnamefont {Oh}}, \bibinfo {author} {\bibfnamefont {D.~J.}\ \bibnamefont {Mooney}}, \bibinfo {author} {\bibfnamefont {J.~J.}\ \bibnamefont {Vlassak}},\ and\ \bibinfo {author} {\bibfnamefont {Z.}~\bibnamefont {Suo}},\ }\bibfield  {title} {\bibinfo {title} {Highly stretchable and tough hydrogels},\ }\href@noop {} {\bibfield  {journal} {\bibinfo  {journal} {Nature}\ }\textbf {\bibinfo {volume} {489}},\ \bibinfo {pages} {133} (\bibinfo {year} {2012})}\BibitemShut {NoStop}%
\bibitem [{\citenamefont {Cao}\ \emph {et~al.}(2019)\citenamefont {Cao}, \citenamefont {Fan}, \citenamefont {Huang},\ and\ \citenamefont {Chen}}]{cao2019robust}%
  \BibitemOpen
  \bibfield  {author} {\bibinfo {author} {\bibfnamefont {L.}~\bibnamefont {Cao}}, \bibinfo {author} {\bibfnamefont {J.}~\bibnamefont {Fan}}, \bibinfo {author} {\bibfnamefont {J.}~\bibnamefont {Huang}},\ and\ \bibinfo {author} {\bibfnamefont {Y.}~\bibnamefont {Chen}},\ }\bibfield  {title} {\bibinfo {title} {A robust and stretchable cross-linked rubber network with recyclable and self-healable capabilities based on dynamic covalent bonds},\ }\href@noop {} {\bibfield  {journal} {\bibinfo  {journal} {Journal of materials chemistry A}\ }\textbf {\bibinfo {volume} {7}},\ \bibinfo {pages} {4922} (\bibinfo {year} {2019})}\BibitemShut {NoStop}%
\bibitem [{\citenamefont {Lyu}\ and\ \citenamefont {Wu}(2020)}]{lyu2020extremely}%
  \BibitemOpen
  \bibfield  {author} {\bibinfo {author} {\bibfnamefont {Z.}~\bibnamefont {Lyu}}\ and\ \bibinfo {author} {\bibfnamefont {T.}~\bibnamefont {Wu}},\ }\bibfield  {title} {\bibinfo {title} {Extremely stretchable vitrimers},\ }\href@noop {} {\bibfield  {journal} {\bibinfo  {journal} {Macromolecular Rapid Communications}\ }\textbf {\bibinfo {volume} {41}},\ \bibinfo {pages} {2000265} (\bibinfo {year} {2020})}\BibitemShut {NoStop}%
\bibitem [{\citenamefont {Kong}\ \emph {et~al.}(2022)\citenamefont {Kong}, \citenamefont {Yang}, \citenamefont {Wang}, \citenamefont {Cheng}, \citenamefont {Yan}, \citenamefont {Huang}, \citenamefont {Ning}, \citenamefont {Zeng}, \citenamefont {Cai},\ and\ \citenamefont {Wang}}]{kong2022ultra}%
  \BibitemOpen
  \bibfield  {author} {\bibinfo {author} {\bibfnamefont {W.}~\bibnamefont {Kong}}, \bibinfo {author} {\bibfnamefont {Y.}~\bibnamefont {Yang}}, \bibinfo {author} {\bibfnamefont {Y.}~\bibnamefont {Wang}}, \bibinfo {author} {\bibfnamefont {H.}~\bibnamefont {Cheng}}, \bibinfo {author} {\bibfnamefont {P.}~\bibnamefont {Yan}}, \bibinfo {author} {\bibfnamefont {L.}~\bibnamefont {Huang}}, \bibinfo {author} {\bibfnamefont {J.}~\bibnamefont {Ning}}, \bibinfo {author} {\bibfnamefont {F.}~\bibnamefont {Zeng}}, \bibinfo {author} {\bibfnamefont {X.}~\bibnamefont {Cai}},\ and\ \bibinfo {author} {\bibfnamefont {M.}~\bibnamefont {Wang}},\ }\bibfield  {title} {\bibinfo {title} {An ultra-low hysteresis, self-healing and stretchable conductor based on dynamic disulfide covalent adaptable networks},\ }\href@noop {} {\bibfield  {journal} {\bibinfo  {journal} {Journal of Materials Chemistry A}\ }\textbf {\bibinfo {volume} {10}},\ \bibinfo {pages} {2012} (\bibinfo {year} {2022})}\BibitemShut {NoStop}%
\bibitem [{\citenamefont {Ji}\ \emph {et~al.}(2023)\citenamefont {Ji}, \citenamefont {Luo}, \citenamefont {Zhang}, \citenamefont {Sun}, \citenamefont {Wang}, \citenamefont {Qin}, \citenamefont {Zhuo},\ and\ \citenamefont {Dai}}]{ji2023novel}%
  \BibitemOpen
  \bibfield  {author} {\bibinfo {author} {\bibfnamefont {N.}~\bibnamefont {Ji}}, \bibinfo {author} {\bibfnamefont {J.}~\bibnamefont {Luo}}, \bibinfo {author} {\bibfnamefont {W.}~\bibnamefont {Zhang}}, \bibinfo {author} {\bibfnamefont {J.}~\bibnamefont {Sun}}, \bibinfo {author} {\bibfnamefont {J.}~\bibnamefont {Wang}}, \bibinfo {author} {\bibfnamefont {C.}~\bibnamefont {Qin}}, \bibinfo {author} {\bibfnamefont {Q.}~\bibnamefont {Zhuo}},\ and\ \bibinfo {author} {\bibfnamefont {L.}~\bibnamefont {Dai}},\ }\bibfield  {title} {\bibinfo {title} {A novel polyvinyl alcohol-based hydrogel with ultra-fast self-healing ability and excellent stretchability based on multi dynamic covalent bond cross-linking},\ }\href@noop {} {\bibfield  {journal} {\bibinfo  {journal} {Macromolecular Materials and Engineering}\ }\textbf {\bibinfo {volume} {308}},\ \bibinfo {pages} {2200525} (\bibinfo {year} {2023})}\BibitemShut {NoStop}%
\bibitem [{\citenamefont {Daniel}\ \emph {et~al.}(2016)\citenamefont {Daniel}, \citenamefont {Burdy{\'n}ska}, \citenamefont {Vatankhah-Varnoosfaderani}, \citenamefont {Matyjaszewski}, \citenamefont {Paturej}, \citenamefont {Rubinstein}, \citenamefont {Dobrynin},\ and\ \citenamefont {Sheiko}}]{daniel2016solvent}%
  \BibitemOpen
  \bibfield  {author} {\bibinfo {author} {\bibfnamefont {W.~F.}\ \bibnamefont {Daniel}}, \bibinfo {author} {\bibfnamefont {J.}~\bibnamefont {Burdy{\'n}ska}}, \bibinfo {author} {\bibfnamefont {M.}~\bibnamefont {Vatankhah-Varnoosfaderani}}, \bibinfo {author} {\bibfnamefont {K.}~\bibnamefont {Matyjaszewski}}, \bibinfo {author} {\bibfnamefont {J.}~\bibnamefont {Paturej}}, \bibinfo {author} {\bibfnamefont {M.}~\bibnamefont {Rubinstein}}, \bibinfo {author} {\bibfnamefont {A.~V.}\ \bibnamefont {Dobrynin}},\ and\ \bibinfo {author} {\bibfnamefont {S.~S.}\ \bibnamefont {Sheiko}},\ }\bibfield  {title} {\bibinfo {title} {Solvent-free, supersoft and superelastic bottlebrush melts and networks},\ }\href@noop {} {\bibfield  {journal} {\bibinfo  {journal} {Nature materials}\ }\textbf {\bibinfo {volume} {15}},\ \bibinfo {pages} {183} (\bibinfo {year} {2016})}\BibitemShut {NoStop}%
\bibitem [{\citenamefont {Cao}\ \emph {et~al.}(2017)\citenamefont {Cao}, \citenamefont {Morrissey}, \citenamefont {Acome}, \citenamefont {Allec}, \citenamefont {Wong}, \citenamefont {Keplinger},\ and\ \citenamefont {Wang}}]{cao2017transparent}%
  \BibitemOpen
  \bibfield  {author} {\bibinfo {author} {\bibfnamefont {Y.}~\bibnamefont {Cao}}, \bibinfo {author} {\bibfnamefont {T.~G.}\ \bibnamefont {Morrissey}}, \bibinfo {author} {\bibfnamefont {E.}~\bibnamefont {Acome}}, \bibinfo {author} {\bibfnamefont {S.~I.}\ \bibnamefont {Allec}}, \bibinfo {author} {\bibfnamefont {B.~M.}\ \bibnamefont {Wong}}, \bibinfo {author} {\bibfnamefont {C.}~\bibnamefont {Keplinger}},\ and\ \bibinfo {author} {\bibfnamefont {C.}~\bibnamefont {Wang}},\ }\bibfield  {title} {\bibinfo {title} {A transparent, self-healing, highly stretchable ionic conductor},\ }\href@noop {} {\bibfield  {journal} {\bibinfo  {journal} {Advanced Materials}\ }\textbf {\bibinfo {volume} {29}},\ \bibinfo {pages} {1605099} (\bibinfo {year} {2017})}\BibitemShut {NoStop}%
\bibitem [{\citenamefont {Zhang}\ \emph {et~al.}(2018{\natexlab{b}})\citenamefont {Zhang}, \citenamefont {Shi}, \citenamefont {Qu}, \citenamefont {Long}, \citenamefont {Feringa},\ and\ \citenamefont {Tian}}]{zhang2018exploring}%
  \BibitemOpen
  \bibfield  {author} {\bibinfo {author} {\bibfnamefont {Q.}~\bibnamefont {Zhang}}, \bibinfo {author} {\bibfnamefont {C.-Y.}\ \bibnamefont {Shi}}, \bibinfo {author} {\bibfnamefont {D.-H.}\ \bibnamefont {Qu}}, \bibinfo {author} {\bibfnamefont {Y.-T.}\ \bibnamefont {Long}}, \bibinfo {author} {\bibfnamefont {B.~L.}\ \bibnamefont {Feringa}},\ and\ \bibinfo {author} {\bibfnamefont {H.}~\bibnamefont {Tian}},\ }\bibfield  {title} {\bibinfo {title} {Exploring a naturally tailored small molecule for stretchable, self-healing, and adhesive supramolecular polymers},\ }\href@noop {} {\bibfield  {journal} {\bibinfo  {journal} {Science advances}\ }\textbf {\bibinfo {volume} {4}},\ \bibinfo {pages} {eaat8192} (\bibinfo {year} {2018}{\natexlab{b}})}\BibitemShut {NoStop}%
\bibitem [{\citenamefont {Zhang}\ \emph {et~al.}(2019)\citenamefont {Zhang}, \citenamefont {Wu}, \citenamefont {Yang}, \citenamefont {Wang}, \citenamefont {Yu}, \citenamefont {Lai}, \citenamefont {Shi}, \citenamefont {Wang}, \citenamefont {Cui}, \citenamefont {Xiang} \emph {et~al.}}]{zhang2019superstretchable}%
  \BibitemOpen
  \bibfield  {author} {\bibinfo {author} {\bibfnamefont {H.}~\bibnamefont {Zhang}}, \bibinfo {author} {\bibfnamefont {Y.}~\bibnamefont {Wu}}, \bibinfo {author} {\bibfnamefont {J.}~\bibnamefont {Yang}}, \bibinfo {author} {\bibfnamefont {D.}~\bibnamefont {Wang}}, \bibinfo {author} {\bibfnamefont {P.}~\bibnamefont {Yu}}, \bibinfo {author} {\bibfnamefont {C.~T.}\ \bibnamefont {Lai}}, \bibinfo {author} {\bibfnamefont {A.-C.}\ \bibnamefont {Shi}}, \bibinfo {author} {\bibfnamefont {J.}~\bibnamefont {Wang}}, \bibinfo {author} {\bibfnamefont {S.}~\bibnamefont {Cui}}, \bibinfo {author} {\bibfnamefont {J.}~\bibnamefont {Xiang}}, \emph {et~al.},\ }\bibfield  {title} {\bibinfo {title} {Superstretchable dynamic polymer networks},\ }\href@noop {} {\bibfield  {journal} {\bibinfo  {journal} {Advanced Materials}\ }\textbf {\bibinfo {volume} {31}},\ \bibinfo {pages} {1904029} (\bibinfo {year} {2019})}\BibitemShut {NoStop}%
\bibitem [{\citenamefont {Li}\ \emph {et~al.}(2022{\natexlab{b}})\citenamefont {Li}, \citenamefont {Chen}, \citenamefont {Li}, \citenamefont {Dai}, \citenamefont {Jin}, \citenamefont {Zhang}, \citenamefont {Feng}, \citenamefont {Yan}, \citenamefont {Cao},\ and\ \citenamefont {Wang}}]{li2022superstretchable}%
  \BibitemOpen
  \bibfield  {author} {\bibinfo {author} {\bibfnamefont {M.}~\bibnamefont {Li}}, \bibinfo {author} {\bibfnamefont {L.}~\bibnamefont {Chen}}, \bibinfo {author} {\bibfnamefont {Y.}~\bibnamefont {Li}}, \bibinfo {author} {\bibfnamefont {X.}~\bibnamefont {Dai}}, \bibinfo {author} {\bibfnamefont {Z.}~\bibnamefont {Jin}}, \bibinfo {author} {\bibfnamefont {Y.}~\bibnamefont {Zhang}}, \bibinfo {author} {\bibfnamefont {W.}~\bibnamefont {Feng}}, \bibinfo {author} {\bibfnamefont {L.-T.}\ \bibnamefont {Yan}}, \bibinfo {author} {\bibfnamefont {Y.}~\bibnamefont {Cao}},\ and\ \bibinfo {author} {\bibfnamefont {C.}~\bibnamefont {Wang}},\ }\bibfield  {title} {\bibinfo {title} {Superstretchable, yet stiff, fatigue-resistant ligament-like elastomers},\ }\href@noop {} {\bibfield  {journal} {\bibinfo  {journal} {Nature Communications}\ }\textbf {\bibinfo {volume} {13}},\ \bibinfo {pages} {2279} (\bibinfo {year} {2022}{\natexlab{b}})}\BibitemShut {NoStop}%
\bibitem [{\citenamefont {Montarnal}\ \emph {et~al.}(2011)\citenamefont {Montarnal}, \citenamefont {Capelot}, \citenamefont {Tournilhac},\ and\ \citenamefont {Leibler}}]{montarnal2011silica}%
  \BibitemOpen
  \bibfield  {author} {\bibinfo {author} {\bibfnamefont {D.}~\bibnamefont {Montarnal}}, \bibinfo {author} {\bibfnamefont {M.}~\bibnamefont {Capelot}}, \bibinfo {author} {\bibfnamefont {F.}~\bibnamefont {Tournilhac}},\ and\ \bibinfo {author} {\bibfnamefont {L.}~\bibnamefont {Leibler}},\ }\bibfield  {title} {\bibinfo {title} {Silica-like malleable materials from permanent organic networks},\ }\href@noop {} {\bibfield  {journal} {\bibinfo  {journal} {Science}\ }\textbf {\bibinfo {volume} {334}},\ \bibinfo {pages} {965} (\bibinfo {year} {2011})}\BibitemShut {NoStop}%
\bibitem [{\citenamefont {Capelot}\ \emph {et~al.}(2012{\natexlab{a}})\citenamefont {Capelot}, \citenamefont {Montarnal}, \citenamefont {Tournilhac},\ and\ \citenamefont {Leibler}}]{capelot2012metal}%
  \BibitemOpen
  \bibfield  {author} {\bibinfo {author} {\bibfnamefont {M.}~\bibnamefont {Capelot}}, \bibinfo {author} {\bibfnamefont {D.}~\bibnamefont {Montarnal}}, \bibinfo {author} {\bibfnamefont {F.}~\bibnamefont {Tournilhac}},\ and\ \bibinfo {author} {\bibfnamefont {L.}~\bibnamefont {Leibler}},\ }\bibfield  {title} {\bibinfo {title} {Metal-catalyzed transesterification for healing and assembling of thermosets},\ }\href@noop {} {\bibfield  {journal} {\bibinfo  {journal} {Journal of the american chemical society}\ }\textbf {\bibinfo {volume} {134}},\ \bibinfo {pages} {7664} (\bibinfo {year} {2012}{\natexlab{a}})}\BibitemShut {NoStop}%
\bibitem [{\citenamefont {Capelot}\ \emph {et~al.}(2012{\natexlab{b}})\citenamefont {Capelot}, \citenamefont {Unterlass}, \citenamefont {Tournilhac},\ and\ \citenamefont {Leibler}}]{capelot2012catalytic}%
  \BibitemOpen
  \bibfield  {author} {\bibinfo {author} {\bibfnamefont {M.}~\bibnamefont {Capelot}}, \bibinfo {author} {\bibfnamefont {M.~M.}\ \bibnamefont {Unterlass}}, \bibinfo {author} {\bibfnamefont {F.}~\bibnamefont {Tournilhac}},\ and\ \bibinfo {author} {\bibfnamefont {L.}~\bibnamefont {Leibler}},\ }\bibfield  {title} {\bibinfo {title} {Catalytic control of the vitrimer glass transition},\ }\href@noop {} {\bibfield  {journal} {\bibinfo  {journal} {ACS Macro Letters}\ }\textbf {\bibinfo {volume} {1}},\ \bibinfo {pages} {789} (\bibinfo {year} {2012}{\natexlab{b}})}\BibitemShut {NoStop}%
\bibitem [{\citenamefont {Denissen}\ \emph {et~al.}(2016)\citenamefont {Denissen}, \citenamefont {Winne},\ and\ \citenamefont {Du~Prez}}]{denissen2016vitrimers}%
  \BibitemOpen
  \bibfield  {author} {\bibinfo {author} {\bibfnamefont {W.}~\bibnamefont {Denissen}}, \bibinfo {author} {\bibfnamefont {J.~M.}\ \bibnamefont {Winne}},\ and\ \bibinfo {author} {\bibfnamefont {F.~E.}\ \bibnamefont {Du~Prez}},\ }\bibfield  {title} {\bibinfo {title} {Vitrimers: permanent organic networks with glass-like fluidity},\ }\href@noop {} {\bibfield  {journal} {\bibinfo  {journal} {Chemical science}\ }\textbf {\bibinfo {volume} {7}},\ \bibinfo {pages} {30} (\bibinfo {year} {2016})}\BibitemShut {NoStop}%
\bibitem [{\citenamefont {Kloxin}\ and\ \citenamefont {Bowman}(2013)}]{kloxin2013covalent}%
  \BibitemOpen
  \bibfield  {author} {\bibinfo {author} {\bibfnamefont {C.~J.}\ \bibnamefont {Kloxin}}\ and\ \bibinfo {author} {\bibfnamefont {C.~N.}\ \bibnamefont {Bowman}},\ }\bibfield  {title} {\bibinfo {title} {Covalent adaptable networks: smart, reconfigurable and responsive network systems},\ }\href@noop {} {\bibfield  {journal} {\bibinfo  {journal} {Chemical Society Reviews}\ }\textbf {\bibinfo {volume} {42}},\ \bibinfo {pages} {7161} (\bibinfo {year} {2013})}\BibitemShut {NoStop}%
\bibitem [{\citenamefont {Snyder}\ \emph {et~al.}(2018)\citenamefont {Snyder}, \citenamefont {Fortman}, \citenamefont {De~Hoe}, \citenamefont {Hillmyer},\ and\ \citenamefont {Dichtel}}]{snyder2018reprocessable}%
  \BibitemOpen
  \bibfield  {author} {\bibinfo {author} {\bibfnamefont {R.~L.}\ \bibnamefont {Snyder}}, \bibinfo {author} {\bibfnamefont {D.~J.}\ \bibnamefont {Fortman}}, \bibinfo {author} {\bibfnamefont {G.~X.}\ \bibnamefont {De~Hoe}}, \bibinfo {author} {\bibfnamefont {M.~A.}\ \bibnamefont {Hillmyer}},\ and\ \bibinfo {author} {\bibfnamefont {W.~R.}\ \bibnamefont {Dichtel}},\ }\bibfield  {title} {\bibinfo {title} {Reprocessable acid-degradable polycarbonate vitrimers},\ }\href@noop {} {\bibfield  {journal} {\bibinfo  {journal} {Macromolecules}\ }\textbf {\bibinfo {volume} {51}},\ \bibinfo {pages} {389} (\bibinfo {year} {2018})}\BibitemShut {NoStop}%
\bibitem [{\citenamefont {Lu}\ and\ \citenamefont {Guan}(2012)}]{lu2012olefin}%
  \BibitemOpen
  \bibfield  {author} {\bibinfo {author} {\bibfnamefont {Y.-X.}\ \bibnamefont {Lu}}\ and\ \bibinfo {author} {\bibfnamefont {Z.}~\bibnamefont {Guan}},\ }\bibfield  {title} {\bibinfo {title} {Olefin metathesis for effective polymer healing via dynamic exchange of strong carbon--carbon double bonds},\ }\href@noop {} {\bibfield  {journal} {\bibinfo  {journal} {Journal of the American Chemical Society}\ }\textbf {\bibinfo {volume} {134}},\ \bibinfo {pages} {14226} (\bibinfo {year} {2012})}\BibitemShut {NoStop}%
\bibitem [{\citenamefont {Hajj}\ \emph {et~al.}(2020)\citenamefont {Hajj}, \citenamefont {Duval}, \citenamefont {Dhers},\ and\ \citenamefont {Av{\'e}rous}}]{hajj2020network}%
  \BibitemOpen
  \bibfield  {author} {\bibinfo {author} {\bibfnamefont {R.}~\bibnamefont {Hajj}}, \bibinfo {author} {\bibfnamefont {A.}~\bibnamefont {Duval}}, \bibinfo {author} {\bibfnamefont {S.}~\bibnamefont {Dhers}},\ and\ \bibinfo {author} {\bibfnamefont {L.}~\bibnamefont {Av{\'e}rous}},\ }\bibfield  {title} {\bibinfo {title} {Network design to control polyimine vitrimer properties: physical versus chemical approach},\ }\href@noop {} {\bibfield  {journal} {\bibinfo  {journal} {Macromolecules}\ }\textbf {\bibinfo {volume} {53}},\ \bibinfo {pages} {3796} (\bibinfo {year} {2020})}\BibitemShut {NoStop}%
\bibitem [{\citenamefont {Tretbar}\ \emph {et~al.}(2019)\citenamefont {Tretbar}, \citenamefont {Neal},\ and\ \citenamefont {Guan}}]{tretbar2019direct}%
  \BibitemOpen
  \bibfield  {author} {\bibinfo {author} {\bibfnamefont {C.~A.}\ \bibnamefont {Tretbar}}, \bibinfo {author} {\bibfnamefont {J.~A.}\ \bibnamefont {Neal}},\ and\ \bibinfo {author} {\bibfnamefont {Z.}~\bibnamefont {Guan}},\ }\bibfield  {title} {\bibinfo {title} {Direct silyl ether metathesis for vitrimers with exceptional thermal stability},\ }\href@noop {} {\bibfield  {journal} {\bibinfo  {journal} {Journal of the American Chemical Society}\ }\textbf {\bibinfo {volume} {141}},\ \bibinfo {pages} {16595} (\bibinfo {year} {2019})}\BibitemShut {NoStop}%
\bibitem [{\citenamefont {R{\"o}ttger}\ \emph {et~al.}(2017)\citenamefont {R{\"o}ttger}, \citenamefont {Domenech}, \citenamefont {van Der~Weegen}, \citenamefont {Breuillac}, \citenamefont {Nicola{\"y}},\ and\ \citenamefont {Leibler}}]{rottger2017high}%
  \BibitemOpen
  \bibfield  {author} {\bibinfo {author} {\bibfnamefont {M.}~\bibnamefont {R{\"o}ttger}}, \bibinfo {author} {\bibfnamefont {T.}~\bibnamefont {Domenech}}, \bibinfo {author} {\bibfnamefont {R.}~\bibnamefont {van Der~Weegen}}, \bibinfo {author} {\bibfnamefont {A.}~\bibnamefont {Breuillac}}, \bibinfo {author} {\bibfnamefont {R.}~\bibnamefont {Nicola{\"y}}},\ and\ \bibinfo {author} {\bibfnamefont {L.}~\bibnamefont {Leibler}},\ }\bibfield  {title} {\bibinfo {title} {High-performance vitrimers from commodity thermoplastics through dioxaborolane metathesis},\ }\href@noop {} {\bibfield  {journal} {\bibinfo  {journal} {Science}\ }\textbf {\bibinfo {volume} {356}},\ \bibinfo {pages} {62} (\bibinfo {year} {2017})}\BibitemShut {NoStop}%
\bibitem [{\citenamefont {Cash}\ \emph {et~al.}(2015)\citenamefont {Cash}, \citenamefont {Kubo}, \citenamefont {Bapat},\ and\ \citenamefont {Sumerlin}}]{cash2015room}%
  \BibitemOpen
  \bibfield  {author} {\bibinfo {author} {\bibfnamefont {J.~J.}\ \bibnamefont {Cash}}, \bibinfo {author} {\bibfnamefont {T.}~\bibnamefont {Kubo}}, \bibinfo {author} {\bibfnamefont {A.~P.}\ \bibnamefont {Bapat}},\ and\ \bibinfo {author} {\bibfnamefont {B.~S.}\ \bibnamefont {Sumerlin}},\ }\bibfield  {title} {\bibinfo {title} {Room-temperature self-healing polymers based on dynamic-covalent boronic esters},\ }\href@noop {} {\bibfield  {journal} {\bibinfo  {journal} {Macromolecules}\ }\textbf {\bibinfo {volume} {48}},\ \bibinfo {pages} {2098} (\bibinfo {year} {2015})}\BibitemShut {NoStop}%
\bibitem [{\citenamefont {Shi}\ \emph {et~al.}(2023)\citenamefont {Shi}, \citenamefont {Jin}, \citenamefont {Chen}, \citenamefont {An},\ and\ \citenamefont {Wang}}]{shi2023welding}%
  \BibitemOpen
  \bibfield  {author} {\bibinfo {author} {\bibfnamefont {Q.}~\bibnamefont {Shi}}, \bibinfo {author} {\bibfnamefont {C.}~\bibnamefont {Jin}}, \bibinfo {author} {\bibfnamefont {Z.}~\bibnamefont {Chen}}, \bibinfo {author} {\bibfnamefont {L.}~\bibnamefont {An}},\ and\ \bibinfo {author} {\bibfnamefont {T.}~\bibnamefont {Wang}},\ }\bibfield  {title} {\bibinfo {title} {On the welding of vitrimers: Chemistry, mechanics and applications},\ }\href@noop {} {\bibfield  {journal} {\bibinfo  {journal} {Advanced Functional Materials}\ }\textbf {\bibinfo {volume} {33}},\ \bibinfo {pages} {2300288} (\bibinfo {year} {2023})}\BibitemShut {NoStop}%
\bibitem [{\citenamefont {Cheng}\ \emph {et~al.}(2024)\citenamefont {Cheng}, \citenamefont {Zhao}, \citenamefont {Xiong}, \citenamefont {Liu}, \citenamefont {Yan},\ and\ \citenamefont {Yu}}]{cheng2024hyperbranched}%
  \BibitemOpen
  \bibfield  {author} {\bibinfo {author} {\bibfnamefont {L.}~\bibnamefont {Cheng}}, \bibinfo {author} {\bibfnamefont {J.}~\bibnamefont {Zhao}}, \bibinfo {author} {\bibfnamefont {Z.}~\bibnamefont {Xiong}}, \bibinfo {author} {\bibfnamefont {S.}~\bibnamefont {Liu}}, \bibinfo {author} {\bibfnamefont {X.}~\bibnamefont {Yan}},\ and\ \bibinfo {author} {\bibfnamefont {W.}~\bibnamefont {Yu}},\ }\bibfield  {title} {\bibinfo {title} {Hyperbranched vitrimer for ultrahigh energy dissipation},\ }\href@noop {} {\bibfield  {journal} {\bibinfo  {journal} {Angewandte Chemie}\ ,\ \bibinfo {pages} {e202406937}} (\bibinfo {year} {2024})}\BibitemShut {NoStop}%
\bibitem [{\citenamefont {Shi}\ \emph {et~al.}(2024)\citenamefont {Shi}, \citenamefont {Zhou}, \citenamefont {He}, \citenamefont {Huang},\ and\ \citenamefont {Liu}}]{shi2024dynamic}%
  \BibitemOpen
  \bibfield  {author} {\bibinfo {author} {\bibfnamefont {W.}~\bibnamefont {Shi}}, \bibinfo {author} {\bibfnamefont {T.}~\bibnamefont {Zhou}}, \bibinfo {author} {\bibfnamefont {B.}~\bibnamefont {He}}, \bibinfo {author} {\bibfnamefont {J.}~\bibnamefont {Huang}},\ and\ \bibinfo {author} {\bibfnamefont {M.}~\bibnamefont {Liu}},\ }\bibfield  {title} {\bibinfo {title} {Dynamic-bond-mediated chain reptation enhances energy dissipation of elastomers},\ }\href@noop {} {\bibfield  {journal} {\bibinfo  {journal} {Angewandte Chemie International Edition}\ }\textbf {\bibinfo {volume} {63}},\ \bibinfo {pages} {e202401845} (\bibinfo {year} {2024})}\BibitemShut {NoStop}%
\bibitem [{\citenamefont {Zheng}\ \emph {et~al.}(2021)\citenamefont {Zheng}, \citenamefont {Xu}, \citenamefont {Zhao},\ and\ \citenamefont {Xie}}]{zheng2021dynamic}%
  \BibitemOpen
  \bibfield  {author} {\bibinfo {author} {\bibfnamefont {N.}~\bibnamefont {Zheng}}, \bibinfo {author} {\bibfnamefont {Y.}~\bibnamefont {Xu}}, \bibinfo {author} {\bibfnamefont {Q.}~\bibnamefont {Zhao}},\ and\ \bibinfo {author} {\bibfnamefont {T.}~\bibnamefont {Xie}},\ }\bibfield  {title} {\bibinfo {title} {Dynamic covalent polymer networks: a molecular platform for designing functions beyond chemical recycling and self-healing},\ }\href@noop {} {\bibfield  {journal} {\bibinfo  {journal} {Chemical Reviews}\ }\textbf {\bibinfo {volume} {121}},\ \bibinfo {pages} {1716} (\bibinfo {year} {2021})}\BibitemShut {NoStop}%
\bibitem [{\citenamefont {Kang}\ \emph {et~al.}(2018)\citenamefont {Kang}, \citenamefont {Son}, \citenamefont {Wang}, \citenamefont {Liu}, \citenamefont {Lopez}, \citenamefont {Kim}, \citenamefont {Oh}, \citenamefont {Katsumata}, \citenamefont {Mun}, \citenamefont {Lee} \emph {et~al.}}]{kang2018tough}%
  \BibitemOpen
  \bibfield  {author} {\bibinfo {author} {\bibfnamefont {J.}~\bibnamefont {Kang}}, \bibinfo {author} {\bibfnamefont {D.}~\bibnamefont {Son}}, \bibinfo {author} {\bibfnamefont {G.-J.~N.}\ \bibnamefont {Wang}}, \bibinfo {author} {\bibfnamefont {Y.}~\bibnamefont {Liu}}, \bibinfo {author} {\bibfnamefont {J.}~\bibnamefont {Lopez}}, \bibinfo {author} {\bibfnamefont {Y.}~\bibnamefont {Kim}}, \bibinfo {author} {\bibfnamefont {J.~Y.}\ \bibnamefont {Oh}}, \bibinfo {author} {\bibfnamefont {T.}~\bibnamefont {Katsumata}}, \bibinfo {author} {\bibfnamefont {J.}~\bibnamefont {Mun}}, \bibinfo {author} {\bibfnamefont {Y.}~\bibnamefont {Lee}}, \emph {et~al.},\ }\bibfield  {title} {\bibinfo {title} {Tough and water-insensitive self-healing elastomer for robust electronic skin},\ }\href@noop {} {\bibfield  {journal} {\bibinfo  {journal} {Advanced Materials}\ }\textbf {\bibinfo {volume} {30}},\ \bibinfo {pages} {1706846} (\bibinfo {year} {2018})}\BibitemShut {NoStop}%
\bibitem [{\citenamefont {Xun}\ \emph {et~al.}(2021)\citenamefont {Xun}, \citenamefont {Zhao}, \citenamefont {Li}, \citenamefont {Zhao}, \citenamefont {Ouyang}, \citenamefont {Zhang}, \citenamefont {Kang}, \citenamefont {Liao},\ and\ \citenamefont {Zhang}}]{xun2021tough}%
  \BibitemOpen
  \bibfield  {author} {\bibinfo {author} {\bibfnamefont {X.}~\bibnamefont {Xun}}, \bibinfo {author} {\bibfnamefont {X.}~\bibnamefont {Zhao}}, \bibinfo {author} {\bibfnamefont {Q.}~\bibnamefont {Li}}, \bibinfo {author} {\bibfnamefont {B.}~\bibnamefont {Zhao}}, \bibinfo {author} {\bibfnamefont {T.}~\bibnamefont {Ouyang}}, \bibinfo {author} {\bibfnamefont {Z.}~\bibnamefont {Zhang}}, \bibinfo {author} {\bibfnamefont {Z.}~\bibnamefont {Kang}}, \bibinfo {author} {\bibfnamefont {Q.}~\bibnamefont {Liao}},\ and\ \bibinfo {author} {\bibfnamefont {Y.}~\bibnamefont {Zhang}},\ }\bibfield  {title} {\bibinfo {title} {Tough and degradable self-healing elastomer from synergistic soft--hard segments design for biomechano-robust artificial skin},\ }\href@noop {} {\bibfield  {journal} {\bibinfo  {journal} {ACS nano}\ }\textbf {\bibinfo {volume} {15}},\ \bibinfo {pages} {20656} (\bibinfo {year} {2021})}\BibitemShut {NoStop}%
\bibitem [{\citenamefont {Pu}\ \emph {et~al.}(2012)\citenamefont {Pu}, \citenamefont {Dubay}, \citenamefont {Zhang}, \citenamefont {Severtson},\ and\ \citenamefont {Houtman}}]{pu2012polyacrylates}%
  \BibitemOpen
  \bibfield  {author} {\bibinfo {author} {\bibfnamefont {G.}~\bibnamefont {Pu}}, \bibinfo {author} {\bibfnamefont {M.~R.}\ \bibnamefont {Dubay}}, \bibinfo {author} {\bibfnamefont {J.}~\bibnamefont {Zhang}}, \bibinfo {author} {\bibfnamefont {S.~J.}\ \bibnamefont {Severtson}},\ and\ \bibinfo {author} {\bibfnamefont {C.~J.}\ \bibnamefont {Houtman}},\ }\bibfield  {title} {\bibinfo {title} {Polyacrylates with high biomass contents for pressure-sensitive adhesives prepared via mini-emulsion polymerization},\ }\href@noop {} {\bibfield  {journal} {\bibinfo  {journal} {Industrial \& engineering chemistry research}\ }\textbf {\bibinfo {volume} {51}},\ \bibinfo {pages} {12145} (\bibinfo {year} {2012})}\BibitemShut {NoStop}%
\bibitem [{\citenamefont {Fortunato}\ \emph {et~al.}(2020)\citenamefont {Fortunato}, \citenamefont {Tatsi}, \citenamefont {Rigatelli}, \citenamefont {Turri},\ and\ \citenamefont {Griffini}}]{fortunato2020highly}%
  \BibitemOpen
  \bibfield  {author} {\bibinfo {author} {\bibfnamefont {G.}~\bibnamefont {Fortunato}}, \bibinfo {author} {\bibfnamefont {E.}~\bibnamefont {Tatsi}}, \bibinfo {author} {\bibfnamefont {B.}~\bibnamefont {Rigatelli}}, \bibinfo {author} {\bibfnamefont {S.}~\bibnamefont {Turri}},\ and\ \bibinfo {author} {\bibfnamefont {G.}~\bibnamefont {Griffini}},\ }\bibfield  {title} {\bibinfo {title} {Highly transparent and colorless self-healing polyacrylate coatings based on diels--alder chemistry},\ }\href@noop {} {\bibfield  {journal} {\bibinfo  {journal} {Macromolecular Materials and Engineering}\ }\textbf {\bibinfo {volume} {305}},\ \bibinfo {pages} {1900652} (\bibinfo {year} {2020})}\BibitemShut {NoStop}%
\bibitem [{\citenamefont {Mistry}\ \emph {et~al.}(2020)\citenamefont {Mistry}, \citenamefont {Nikkhou}, \citenamefont {Raistrick}, \citenamefont {Hussain}, \citenamefont {Jull}, \citenamefont {Baker},\ and\ \citenamefont {Gleeson}}]{mistry2020isotropic}%
  \BibitemOpen
  \bibfield  {author} {\bibinfo {author} {\bibfnamefont {D.}~\bibnamefont {Mistry}}, \bibinfo {author} {\bibfnamefont {M.}~\bibnamefont {Nikkhou}}, \bibinfo {author} {\bibfnamefont {T.}~\bibnamefont {Raistrick}}, \bibinfo {author} {\bibfnamefont {M.}~\bibnamefont {Hussain}}, \bibinfo {author} {\bibfnamefont {E.~I.}\ \bibnamefont {Jull}}, \bibinfo {author} {\bibfnamefont {D.~L.}\ \bibnamefont {Baker}},\ and\ \bibinfo {author} {\bibfnamefont {H.~F.}\ \bibnamefont {Gleeson}},\ }\bibfield  {title} {\bibinfo {title} {Isotropic liquid crystal elastomers as exceptional photoelastic strain sensors},\ }\href@noop {} {\bibfield  {journal} {\bibinfo  {journal} {Macromolecules}\ }\textbf {\bibinfo {volume} {53}},\ \bibinfo {pages} {3709} (\bibinfo {year} {2020})}\BibitemShut {NoStop}%
\bibitem [{\citenamefont {Robinson}\ \emph {et~al.}(2021)\citenamefont {Robinson}, \citenamefont {Self}, \citenamefont {Fusi}, \citenamefont {Bates}, \citenamefont {Read~de Alaniz}, \citenamefont {Hawker}, \citenamefont {Bates},\ and\ \citenamefont {Sample}}]{robinson2021chemical}%
  \BibitemOpen
  \bibfield  {author} {\bibinfo {author} {\bibfnamefont {L.~L.}\ \bibnamefont {Robinson}}, \bibinfo {author} {\bibfnamefont {J.~L.}\ \bibnamefont {Self}}, \bibinfo {author} {\bibfnamefont {A.~D.}\ \bibnamefont {Fusi}}, \bibinfo {author} {\bibfnamefont {M.~W.}\ \bibnamefont {Bates}}, \bibinfo {author} {\bibfnamefont {J.}~\bibnamefont {Read~de Alaniz}}, \bibinfo {author} {\bibfnamefont {C.~J.}\ \bibnamefont {Hawker}}, \bibinfo {author} {\bibfnamefont {C.~M.}\ \bibnamefont {Bates}},\ and\ \bibinfo {author} {\bibfnamefont {C.~S.}\ \bibnamefont {Sample}},\ }\bibfield  {title} {\bibinfo {title} {Chemical and mechanical tunability of 3d-printed dynamic covalent networks based on boronate esters},\ }\href@noop {} {\bibfield  {journal} {\bibinfo  {journal} {ACS Macro Letters}\ }\textbf {\bibinfo {volume} {10}},\ \bibinfo {pages} {857} (\bibinfo {year} {2021})}\BibitemShut {NoStop}%
\bibitem [{\citenamefont {Tamezawa}\ \emph {et~al.}(2012)\citenamefont {Tamezawa}, \citenamefont {Kumagai}, \citenamefont {Aota}, \citenamefont {Matsumoto}, \citenamefont {Totsuka},\ and\ \citenamefont {Fujie}}]{tamezawa2012peculiar}%
  \BibitemOpen
  \bibfield  {author} {\bibinfo {author} {\bibfnamefont {H.}~\bibnamefont {Tamezawa}}, \bibinfo {author} {\bibfnamefont {T.}~\bibnamefont {Kumagai}}, \bibinfo {author} {\bibfnamefont {H.}~\bibnamefont {Aota}}, \bibinfo {author} {\bibfnamefont {A.}~\bibnamefont {Matsumoto}}, \bibinfo {author} {\bibfnamefont {T.}~\bibnamefont {Totsuka}},\ and\ \bibinfo {author} {\bibfnamefont {H.}~\bibnamefont {Fujie}},\ }\bibfield  {title} {\bibinfo {title} {Peculiar initiation behavior of azo-initiators in allyl polymerization},\ }\href@noop {} {\bibfield  {journal} {\bibinfo  {journal} {Journal of Polymer Science Part A: Polymer Chemistry}\ }\textbf {\bibinfo {volume} {50}},\ \bibinfo {pages} {2732} (\bibinfo {year} {2012})}\BibitemShut {NoStop}%
\bibitem [{\citenamefont {Fetters}\ \emph {et~al.}(2007)\citenamefont {Fetters}, \citenamefont {Lohse},\ and\ \citenamefont {Colby}}]{fetters2007chain}%
  \BibitemOpen
  \bibfield  {author} {\bibinfo {author} {\bibfnamefont {L.}~\bibnamefont {Fetters}}, \bibinfo {author} {\bibfnamefont {D.}~\bibnamefont {Lohse}},\ and\ \bibinfo {author} {\bibfnamefont {R.}~\bibnamefont {Colby}},\ }\bibfield  {title} {\bibinfo {title} {Chain dimensions and entanglement spacings},\ }in\ \href@noop {} {\emph {\bibinfo {booktitle} {Physical properties of polymers handbook}}}\ (\bibinfo  {publisher} {Springer},\ \bibinfo {year} {2007})\ pp.\ \bibinfo {pages} {447--454}\BibitemShut {NoStop}%
\bibitem [{\citenamefont {Zheng}\ \emph {et~al.}(2022)\citenamefont {Zheng}, \citenamefont {Guo}, \citenamefont {Douglas},\ and\ \citenamefont {Xia}}]{zheng2022understanding}%
  \BibitemOpen
  \bibfield  {author} {\bibinfo {author} {\bibfnamefont {X.}~\bibnamefont {Zheng}}, \bibinfo {author} {\bibfnamefont {Y.}~\bibnamefont {Guo}}, \bibinfo {author} {\bibfnamefont {J.~F.}\ \bibnamefont {Douglas}},\ and\ \bibinfo {author} {\bibfnamefont {W.}~\bibnamefont {Xia}},\ }\bibfield  {title} {\bibinfo {title} {Understanding the role of cross-link density in the segmental dynamics and elastic properties of cross-linked thermosets},\ }\href@noop {} {\bibfield  {journal} {\bibinfo  {journal} {The Journal of Chemical Physics}\ }\textbf {\bibinfo {volume} {157}} (\bibinfo {year} {2022})}\BibitemShut {NoStop}%
\bibitem [{\citenamefont {Sordo}\ \emph {et~al.}(2015)\citenamefont {Sordo}, \citenamefont {Mougnier}, \citenamefont {Loureiro}, \citenamefont {Tournilhac},\ and\ \citenamefont {Michaud}}]{sordo2015design}%
  \BibitemOpen
  \bibfield  {author} {\bibinfo {author} {\bibfnamefont {F.}~\bibnamefont {Sordo}}, \bibinfo {author} {\bibfnamefont {S.-J.}\ \bibnamefont {Mougnier}}, \bibinfo {author} {\bibfnamefont {N.}~\bibnamefont {Loureiro}}, \bibinfo {author} {\bibfnamefont {F.}~\bibnamefont {Tournilhac}},\ and\ \bibinfo {author} {\bibfnamefont {V.}~\bibnamefont {Michaud}},\ }\bibfield  {title} {\bibinfo {title} {Design of self-healing supramolecular rubbers with a tunable number of chemical cross-links},\ }\href@noop {} {\bibfield  {journal} {\bibinfo  {journal} {Macromolecules}\ }\textbf {\bibinfo {volume} {48}},\ \bibinfo {pages} {4394} (\bibinfo {year} {2015})}\BibitemShut {NoStop}%
\bibitem [{\citenamefont {Chen}\ \emph {et~al.}(2021)\citenamefont {Chen}, \citenamefont {Si}, \citenamefont {Zhang}, \citenamefont {Zhou}, \citenamefont {Wu}, \citenamefont {Song}, \citenamefont {Kang},\ and\ \citenamefont {Zhao}}]{chen2021crucial}%
  \BibitemOpen
  \bibfield  {author} {\bibinfo {author} {\bibfnamefont {M.}~\bibnamefont {Chen}}, \bibinfo {author} {\bibfnamefont {H.}~\bibnamefont {Si}}, \bibinfo {author} {\bibfnamefont {H.}~\bibnamefont {Zhang}}, \bibinfo {author} {\bibfnamefont {L.}~\bibnamefont {Zhou}}, \bibinfo {author} {\bibfnamefont {Y.}~\bibnamefont {Wu}}, \bibinfo {author} {\bibfnamefont {L.}~\bibnamefont {Song}}, \bibinfo {author} {\bibfnamefont {M.}~\bibnamefont {Kang}},\ and\ \bibinfo {author} {\bibfnamefont {X.-L.}\ \bibnamefont {Zhao}},\ }\bibfield  {title} {\bibinfo {title} {The crucial role in controlling the dynamic properties of polyester-based epoxy vitrimers: the density of exchangeable ester bonds ($\upsilon$)},\ }\href@noop {} {\bibfield  {journal} {\bibinfo  {journal} {Macromolecules}\ }\textbf {\bibinfo {volume} {54}},\ \bibinfo {pages} {10110} (\bibinfo {year} {2021})}\BibitemShut {NoStop}%
\bibitem [{\citenamefont {Bertini}\ \emph {et~al.}(2005)\citenamefont {Bertini}, \citenamefont {Audisio},\ and\ \citenamefont {Zuev}}]{bertini2005investigation}%
  \BibitemOpen
  \bibfield  {author} {\bibinfo {author} {\bibfnamefont {F.}~\bibnamefont {Bertini}}, \bibinfo {author} {\bibfnamefont {G.}~\bibnamefont {Audisio}},\ and\ \bibinfo {author} {\bibfnamefont {V.~V.}\ \bibnamefont {Zuev}},\ }\bibfield  {title} {\bibinfo {title} {Investigation on the thermal degradation of poly-n-alkyl acrylates and poly-n-alkyl methacrylates (c1--c12)},\ }\href@noop {} {\bibfield  {journal} {\bibinfo  {journal} {Polymer degradation and stability}\ }\textbf {\bibinfo {volume} {89}},\ \bibinfo {pages} {233} (\bibinfo {year} {2005})}\BibitemShut {NoStop}%
\bibitem [{\citenamefont {Chen}\ \emph {et~al.}(2023)\citenamefont {Chen}, \citenamefont {Zhao}, \citenamefont {Li}, \citenamefont {Sokolov}, \citenamefont {Tian}, \citenamefont {Advincula},\ and\ \citenamefont {Cao}}]{chen2023exceptionally}%
  \BibitemOpen
  \bibfield  {author} {\bibinfo {author} {\bibfnamefont {Q.}~\bibnamefont {Chen}}, \bibinfo {author} {\bibfnamefont {X.}~\bibnamefont {Zhao}}, \bibinfo {author} {\bibfnamefont {B.}~\bibnamefont {Li}}, \bibinfo {author} {\bibfnamefont {A.~P.}\ \bibnamefont {Sokolov}}, \bibinfo {author} {\bibfnamefont {M.}~\bibnamefont {Tian}}, \bibinfo {author} {\bibfnamefont {R.~C.}\ \bibnamefont {Advincula}},\ and\ \bibinfo {author} {\bibfnamefont {P.-F.}\ \bibnamefont {Cao}},\ }\bibfield  {title} {\bibinfo {title} {Exceptionally recyclable, extremely tough, vitrimer-like polydimethylsiloxane elastomers via rational network design},\ }\href@noop {} {\bibfield  {journal} {\bibinfo  {journal} {Matter}\ }\textbf {\bibinfo {volume} {6}},\ \bibinfo {pages} {3378} (\bibinfo {year} {2023})}\BibitemShut {NoStop}%
\bibitem [{\citenamefont {Yue}\ \emph {et~al.}(2023)\citenamefont {Yue}, \citenamefont {Su}, \citenamefont {Li}, \citenamefont {Yu}, \citenamefont {Montgomery}, \citenamefont {Sun}, \citenamefont {Finn}, \citenamefont {Gutekunst}, \citenamefont {Ramprasad},\ and\ \citenamefont {Qi}}]{yue2023one}%
  \BibitemOpen
  \bibfield  {author} {\bibinfo {author} {\bibfnamefont {L.}~\bibnamefont {Yue}}, \bibinfo {author} {\bibfnamefont {Y.-L.}\ \bibnamefont {Su}}, \bibinfo {author} {\bibfnamefont {M.}~\bibnamefont {Li}}, \bibinfo {author} {\bibfnamefont {L.}~\bibnamefont {Yu}}, \bibinfo {author} {\bibfnamefont {S.~M.}\ \bibnamefont {Montgomery}}, \bibinfo {author} {\bibfnamefont {X.}~\bibnamefont {Sun}}, \bibinfo {author} {\bibfnamefont {M.}~\bibnamefont {Finn}}, \bibinfo {author} {\bibfnamefont {W.~R.}\ \bibnamefont {Gutekunst}}, \bibinfo {author} {\bibfnamefont {R.}~\bibnamefont {Ramprasad}},\ and\ \bibinfo {author} {\bibfnamefont {H.~J.}\ \bibnamefont {Qi}},\ }\bibfield  {title} {\bibinfo {title} {One-pot synthesis of depolymerizable $\delta$-lactone based vitrimers},\ }\href@noop {} {\bibfield  {journal} {\bibinfo  {journal} {Advanced Materials}\ }\textbf {\bibinfo {volume} {35}},\ \bibinfo {pages} {2300954} (\bibinfo {year} {2023})}\BibitemShut {NoStop}%
\bibitem [{\citenamefont {Dealy}\ \emph {et~al.}(2018)\citenamefont {Dealy}, \citenamefont {Read},\ and\ \citenamefont {Larson}}]{dealy2018structure}%
  \BibitemOpen
  \bibfield  {author} {\bibinfo {author} {\bibfnamefont {J.~M.}\ \bibnamefont {Dealy}}, \bibinfo {author} {\bibfnamefont {D.~J.}\ \bibnamefont {Read}},\ and\ \bibinfo {author} {\bibfnamefont {R.~G.}\ \bibnamefont {Larson}},\ }\href@noop {} {\emph {\bibinfo {title} {Structure and rheology of molten polymers: from structure to flow behavior and back again}}}\ (\bibinfo  {publisher} {Carl Hanser Verlag GmbH Co KG},\ \bibinfo {year} {2018})\BibitemShut {NoStop}%
\bibitem [{\citenamefont {Rubinstein}\ and\ \citenamefont {Colby}(2003)}]{rubinstein2003polymer}%
  \BibitemOpen
  \bibfield  {author} {\bibinfo {author} {\bibfnamefont {M.}~\bibnamefont {Rubinstein}}\ and\ \bibinfo {author} {\bibfnamefont {R.~H.}\ \bibnamefont {Colby}},\ }\href@noop {} {\emph {\bibinfo {title} {Polymer physics}}}\ (\bibinfo  {publisher} {Oxford university press},\ \bibinfo {year} {2003})\BibitemShut {NoStop}%
\bibitem [{\citenamefont {Pfefferkorn}\ \emph {et~al.}(2010)\citenamefont {Pfefferkorn}, \citenamefont {Sonntag}, \citenamefont {Kyeremateng}, \citenamefont {Funke}, \citenamefont {Kammer},\ and\ \citenamefont {Kressler}}]{pfefferkorn2010pressure}%
  \BibitemOpen
  \bibfield  {author} {\bibinfo {author} {\bibfnamefont {D.}~\bibnamefont {Pfefferkorn}}, \bibinfo {author} {\bibfnamefont {S.}~\bibnamefont {Sonntag}}, \bibinfo {author} {\bibfnamefont {S.~O.}\ \bibnamefont {Kyeremateng}}, \bibinfo {author} {\bibfnamefont {Z.}~\bibnamefont {Funke}}, \bibinfo {author} {\bibfnamefont {H.-W.}\ \bibnamefont {Kammer}},\ and\ \bibinfo {author} {\bibfnamefont {J.}~\bibnamefont {Kressler}},\ }\bibfield  {title} {\bibinfo {title} {Pressure--volume--temperature data and surface tension of blends of poly (ethylene oxide) and poly (methyl acrylate) in the melt},\ }\href@noop {} {\bibfield  {journal} {\bibinfo  {journal} {Journal of Polymer Science Part B: Polymer Physics}\ }\textbf {\bibinfo {volume} {48}},\ \bibinfo {pages} {1893} (\bibinfo {year} {2010})}\BibitemShut {NoStop}%
\bibitem [{\citenamefont {Ricarte}\ \emph {et~al.}(2023)\citenamefont {Ricarte}, \citenamefont {Shanbhag}, \citenamefont {Ezzeddine}, \citenamefont {Barzycki},\ and\ \citenamefont {Fay}}]{ricarte2023time}%
  \BibitemOpen
  \bibfield  {author} {\bibinfo {author} {\bibfnamefont {R.~G.}\ \bibnamefont {Ricarte}}, \bibinfo {author} {\bibfnamefont {S.}~\bibnamefont {Shanbhag}}, \bibinfo {author} {\bibfnamefont {D.}~\bibnamefont {Ezzeddine}}, \bibinfo {author} {\bibfnamefont {D.}~\bibnamefont {Barzycki}},\ and\ \bibinfo {author} {\bibfnamefont {K.}~\bibnamefont {Fay}},\ }\bibfield  {title} {\bibinfo {title} {Time--temperature superposition of polybutadiene vitrimers},\ }\href@noop {} {\bibfield  {journal} {\bibinfo  {journal} {Macromolecules}\ }\textbf {\bibinfo {volume} {56}},\ \bibinfo {pages} {6806} (\bibinfo {year} {2023})}\BibitemShut {NoStop}%
\bibitem [{\citenamefont {Liu}\ \emph {et~al.}(2021)\citenamefont {Liu}, \citenamefont {Xiao}, \citenamefont {Liu}, \citenamefont {Xiang}, \citenamefont {Rong},\ and\ \citenamefont {Zhang}}]{liu2021self}%
  \BibitemOpen
  \bibfield  {author} {\bibinfo {author} {\bibfnamefont {Z.}~\bibnamefont {Liu}}, \bibinfo {author} {\bibfnamefont {D.}~\bibnamefont {Xiao}}, \bibinfo {author} {\bibfnamefont {G.}~\bibnamefont {Liu}}, \bibinfo {author} {\bibfnamefont {H.}~\bibnamefont {Xiang}}, \bibinfo {author} {\bibfnamefont {M.}~\bibnamefont {Rong}},\ and\ \bibinfo {author} {\bibfnamefont {M.}~\bibnamefont {Zhang}},\ }\bibfield  {title} {\bibinfo {title} {Self-healing and reprocessing of transparent uv-cured polysiloxane elastomer},\ }\href@noop {} {\bibfield  {journal} {\bibinfo  {journal} {Progress in Organic Coatings}\ }\textbf {\bibinfo {volume} {159}},\ \bibinfo {pages} {106450} (\bibinfo {year} {2021})}\BibitemShut {NoStop}%
\bibitem [{\citenamefont {Wang}\ \emph {et~al.}(2021)\citenamefont {Wang}, \citenamefont {Xue}, \citenamefont {Zhou},\ and\ \citenamefont {Cui}}]{wang2021solid}%
  \BibitemOpen
  \bibfield  {author} {\bibinfo {author} {\bibfnamefont {S.}~\bibnamefont {Wang}}, \bibinfo {author} {\bibfnamefont {L.-L.}\ \bibnamefont {Xue}}, \bibinfo {author} {\bibfnamefont {X.-Z.}\ \bibnamefont {Zhou}},\ and\ \bibinfo {author} {\bibfnamefont {J.-X.}\ \bibnamefont {Cui}},\ }\bibfield  {title} {\bibinfo {title} {“solid-liquid” vitrimers based on dynamic boronic ester networks},\ }\href@noop {} {\bibfield  {journal} {\bibinfo  {journal} {Chinese Journal of Polymer Science}\ }\textbf {\bibinfo {volume} {39}},\ \bibinfo {pages} {1292} (\bibinfo {year} {2021})}\BibitemShut {NoStop}%
\bibitem [{\citenamefont {Zhao}\ \emph {et~al.}(2023)\citenamefont {Zhao}, \citenamefont {Li}, \citenamefont {Ma}, \citenamefont {Wang}, \citenamefont {Jiang}, \citenamefont {Yan}, \citenamefont {Hao}, \citenamefont {Qin}, \citenamefont {Shi},\ and\ \citenamefont {Zhang}}]{zhao2023one}%
  \BibitemOpen
  \bibfield  {author} {\bibinfo {author} {\bibfnamefont {Y.}~\bibnamefont {Zhao}}, \bibinfo {author} {\bibfnamefont {J.}~\bibnamefont {Li}}, \bibinfo {author} {\bibfnamefont {Y.}~\bibnamefont {Ma}}, \bibinfo {author} {\bibfnamefont {Y.}~\bibnamefont {Wang}}, \bibinfo {author} {\bibfnamefont {C.}~\bibnamefont {Jiang}}, \bibinfo {author} {\bibfnamefont {H.}~\bibnamefont {Yan}}, \bibinfo {author} {\bibfnamefont {R.}~\bibnamefont {Hao}}, \bibinfo {author} {\bibfnamefont {J.}~\bibnamefont {Qin}}, \bibinfo {author} {\bibfnamefont {X.}~\bibnamefont {Shi}},\ and\ \bibinfo {author} {\bibfnamefont {G.}~\bibnamefont {Zhang}},\ }\bibfield  {title} {\bibinfo {title} {One-step reactive processing of vitrimeric thermoplastic polyolefin elastomer with greatly improved thermo-mechanical property},\ }\href@noop {} {\bibfield  {journal} {\bibinfo  {journal} {Polymer}\ }\textbf {\bibinfo {volume} {282}},\ \bibinfo {pages} {126185} (\bibinfo {year} {2023})}\BibitemShut {NoStop}%
\bibitem [{\citenamefont {Jiang}\ \emph {et~al.}(2024)\citenamefont {Jiang}, \citenamefont {Wang}, \citenamefont {Zhang}, \citenamefont {Bi}, \citenamefont {Wu}, \citenamefont {Wang}, \citenamefont {Zhu}, \citenamefont {Qin}, \citenamefont {Zhao}, \citenamefont {Shi} \emph {et~al.}}]{jiang2024reprocessable}%
  \BibitemOpen
  \bibfield  {author} {\bibinfo {author} {\bibfnamefont {C.}~\bibnamefont {Jiang}}, \bibinfo {author} {\bibfnamefont {C.}~\bibnamefont {Wang}}, \bibinfo {author} {\bibfnamefont {S.}~\bibnamefont {Zhang}}, \bibinfo {author} {\bibfnamefont {H.}~\bibnamefont {Bi}}, \bibinfo {author} {\bibfnamefont {Y.}~\bibnamefont {Wu}}, \bibinfo {author} {\bibfnamefont {J.}~\bibnamefont {Wang}}, \bibinfo {author} {\bibfnamefont {Y.}~\bibnamefont {Zhu}}, \bibinfo {author} {\bibfnamefont {J.}~\bibnamefont {Qin}}, \bibinfo {author} {\bibfnamefont {Y.}~\bibnamefont {Zhao}}, \bibinfo {author} {\bibfnamefont {X.}~\bibnamefont {Shi}}, \emph {et~al.},\ }\bibfield  {title} {\bibinfo {title} {Reprocessable and self-healable boronic-ester based poly (styrene-b-isoprene-b-styrene) vitrimeric elastomer with improved thermo-mechanical property and adhesive performance},\ }\href@noop {} {\bibfield  {journal} {\bibinfo  {journal} {Reactive and Functional Polymers}\ }\textbf {\bibinfo {volume} {198}},\ \bibinfo {pages} {105893} (\bibinfo
  {year} {2024})}\BibitemShut {NoStop}%
\bibitem [{\citenamefont {Haward}(1993)}]{haward1993strain}%
  \BibitemOpen
  \bibfield  {author} {\bibinfo {author} {\bibfnamefont {R.}~\bibnamefont {Haward}},\ }\bibfield  {title} {\bibinfo {title} {Strain hardening of thermoplastics},\ }\href@noop {} {\bibfield  {journal} {\bibinfo  {journal} {Macromolecules}\ }\textbf {\bibinfo {volume} {26}},\ \bibinfo {pages} {5860} (\bibinfo {year} {1993})}\BibitemShut {NoStop}%
\bibitem [{\citenamefont {Zhao}\ \emph {et~al.}(2021)\citenamefont {Zhao}, \citenamefont {Zhang}, \citenamefont {Zhao}, \citenamefont {Liu}, \citenamefont {Liu}, \citenamefont {Li}, \citenamefont {Zhang}, \citenamefont {Bai}, \citenamefont {Yang},\ and\ \citenamefont {Yan}}]{zhao2021mortise}%
  \BibitemOpen
  \bibfield  {author} {\bibinfo {author} {\bibfnamefont {D.}~\bibnamefont {Zhao}}, \bibinfo {author} {\bibfnamefont {Z.}~\bibnamefont {Zhang}}, \bibinfo {author} {\bibfnamefont {J.}~\bibnamefont {Zhao}}, \bibinfo {author} {\bibfnamefont {K.}~\bibnamefont {Liu}}, \bibinfo {author} {\bibfnamefont {Y.}~\bibnamefont {Liu}}, \bibinfo {author} {\bibfnamefont {G.}~\bibnamefont {Li}}, \bibinfo {author} {\bibfnamefont {X.}~\bibnamefont {Zhang}}, \bibinfo {author} {\bibfnamefont {R.}~\bibnamefont {Bai}}, \bibinfo {author} {\bibfnamefont {X.}~\bibnamefont {Yang}},\ and\ \bibinfo {author} {\bibfnamefont {X.}~\bibnamefont {Yan}},\ }\bibfield  {title} {\bibinfo {title} {A mortise-and-tenon joint inspired mechanically interlocked network},\ }\href@noop {} {\bibfield  {journal} {\bibinfo  {journal} {Angewandte Chemie International Edition}\ }\textbf {\bibinfo {volume} {60}},\ \bibinfo {pages} {16224} (\bibinfo {year} {2021})}\BibitemShut {NoStop}%
\bibitem [{\citenamefont {Xiang}\ \emph {et~al.}(2023)\citenamefont {Xiang}, \citenamefont {Li}, \citenamefont {Wu}, \citenamefont {Sun},\ and\ \citenamefont {Wu}}]{xiang2023highly}%
  \BibitemOpen
  \bibfield  {author} {\bibinfo {author} {\bibfnamefont {H.}~\bibnamefont {Xiang}}, \bibinfo {author} {\bibfnamefont {X.}~\bibnamefont {Li}}, \bibinfo {author} {\bibfnamefont {B.}~\bibnamefont {Wu}}, \bibinfo {author} {\bibfnamefont {S.}~\bibnamefont {Sun}},\ and\ \bibinfo {author} {\bibfnamefont {P.}~\bibnamefont {Wu}},\ }\bibfield  {title} {\bibinfo {title} {Highly damping and self-healable ionic elastomer from dynamic phase separation of sticky fluorinated polymers},\ }\href@noop {} {\bibfield  {journal} {\bibinfo  {journal} {Advanced Materials}\ }\textbf {\bibinfo {volume} {35}},\ \bibinfo {pages} {2209581} (\bibinfo {year} {2023})}\BibitemShut {NoStop}%
\bibitem [{\citenamefont {Leibler}\ \emph {et~al.}(1991)\citenamefont {Leibler}, \citenamefont {Rubinstein},\ and\ \citenamefont {Colby}}]{leibler1991dynamics}%
  \BibitemOpen
  \bibfield  {author} {\bibinfo {author} {\bibfnamefont {L.}~\bibnamefont {Leibler}}, \bibinfo {author} {\bibfnamefont {M.}~\bibnamefont {Rubinstein}},\ and\ \bibinfo {author} {\bibfnamefont {R.~H.}\ \bibnamefont {Colby}},\ }\bibfield  {title} {\bibinfo {title} {Dynamics of reversible networks},\ }\href@noop {} {\bibfield  {journal} {\bibinfo  {journal} {Macromolecules}\ }\textbf {\bibinfo {volume} {24}},\ \bibinfo {pages} {4701} (\bibinfo {year} {1991})}\BibitemShut {NoStop}%
\bibitem [{\citenamefont {Zhang}\ \emph {et~al.}(2020{\natexlab{a}})\citenamefont {Zhang}, \citenamefont {Cheng}, \citenamefont {Zhao}, \citenamefont {Wang}, \citenamefont {Liu}, \citenamefont {Yu},\ and\ \citenamefont {Yan}}]{zhang2020synergistic}%
  \BibitemOpen
  \bibfield  {author} {\bibinfo {author} {\bibfnamefont {Z.}~\bibnamefont {Zhang}}, \bibinfo {author} {\bibfnamefont {L.}~\bibnamefont {Cheng}}, \bibinfo {author} {\bibfnamefont {J.}~\bibnamefont {Zhao}}, \bibinfo {author} {\bibfnamefont {L.}~\bibnamefont {Wang}}, \bibinfo {author} {\bibfnamefont {K.}~\bibnamefont {Liu}}, \bibinfo {author} {\bibfnamefont {W.}~\bibnamefont {Yu}},\ and\ \bibinfo {author} {\bibfnamefont {X.}~\bibnamefont {Yan}},\ }\bibfield  {title} {\bibinfo {title} {Synergistic covalent and supramolecular polymers for mechanically robust but dynamic materials},\ }\href@noop {} {\bibfield  {journal} {\bibinfo  {journal} {Angewandte Chemie}\ }\textbf {\bibinfo {volume} {132}},\ \bibinfo {pages} {12237} (\bibinfo {year} {2020}{\natexlab{a}})}\BibitemShut {NoStop}%
\bibitem [{\citenamefont {Lu}\ \emph {et~al.}(2024)\citenamefont {Lu}, \citenamefont {Xu}, \citenamefont {Li}, \citenamefont {Xing}, \citenamefont {Zhou},\ and\ \citenamefont {Zhang}}]{lu2024high}%
  \BibitemOpen
  \bibfield  {author} {\bibinfo {author} {\bibfnamefont {L.}~\bibnamefont {Lu}}, \bibinfo {author} {\bibfnamefont {J.}~\bibnamefont {Xu}}, \bibinfo {author} {\bibfnamefont {J.}~\bibnamefont {Li}}, \bibinfo {author} {\bibfnamefont {Y.}~\bibnamefont {Xing}}, \bibinfo {author} {\bibfnamefont {Z.}~\bibnamefont {Zhou}},\ and\ \bibinfo {author} {\bibfnamefont {F.}~\bibnamefont {Zhang}},\ }\bibfield  {title} {\bibinfo {title} {High-toughness and intrinsically self-healing cross-linked polyurea elastomers with dynamic sextuple h-bonds},\ }\href@noop {} {\bibfield  {journal} {\bibinfo  {journal} {Macromolecules}\ } (\bibinfo {year} {2024})}\BibitemShut {NoStop}%
\bibitem [{\citenamefont {Zheng}\ \emph {et~al.}(2023)\citenamefont {Zheng}, \citenamefont {Liu}, \citenamefont {Si},\ and\ \citenamefont {Chen}}]{zheng2023covalently}%
  \BibitemOpen
  \bibfield  {author} {\bibinfo {author} {\bibfnamefont {S.}~\bibnamefont {Zheng}}, \bibinfo {author} {\bibfnamefont {Y.}~\bibnamefont {Liu}}, \bibinfo {author} {\bibfnamefont {G.}~\bibnamefont {Si}},\ and\ \bibinfo {author} {\bibfnamefont {M.}~\bibnamefont {Chen}},\ }\bibfield  {title} {\bibinfo {title} {Covalently crosslinked networks from telechelic polycyclooctene with reinforced properties},\ }\href@noop {} {\bibfield  {journal} {\bibinfo  {journal} {Chinese Journal of Chemistry}\ }\textbf {\bibinfo {volume} {41}},\ \bibinfo {pages} {2002} (\bibinfo {year} {2023})}\BibitemShut {NoStop}%
\bibitem [{\citenamefont {Luo}\ \emph {et~al.}(2023)\citenamefont {Luo}, \citenamefont {Zhao}, \citenamefont {Ju}, \citenamefont {Chen}, \citenamefont {Zhao}, \citenamefont {Demchuk}, \citenamefont {Li}, \citenamefont {Bocharova}, \citenamefont {Carrillo}, \citenamefont {Keum} \emph {et~al.}}]{luo2023highly}%
  \BibitemOpen
  \bibfield  {author} {\bibinfo {author} {\bibfnamefont {J.}~\bibnamefont {Luo}}, \bibinfo {author} {\bibfnamefont {X.}~\bibnamefont {Zhao}}, \bibinfo {author} {\bibfnamefont {H.}~\bibnamefont {Ju}}, \bibinfo {author} {\bibfnamefont {X.}~\bibnamefont {Chen}}, \bibinfo {author} {\bibfnamefont {S.}~\bibnamefont {Zhao}}, \bibinfo {author} {\bibfnamefont {Z.}~\bibnamefont {Demchuk}}, \bibinfo {author} {\bibfnamefont {B.}~\bibnamefont {Li}}, \bibinfo {author} {\bibfnamefont {V.}~\bibnamefont {Bocharova}}, \bibinfo {author} {\bibfnamefont {J.-M.~Y.}\ \bibnamefont {Carrillo}}, \bibinfo {author} {\bibfnamefont {J.~K.}\ \bibnamefont {Keum}}, \emph {et~al.},\ }\bibfield  {title} {\bibinfo {title} {Highly recyclable and tough elastic vitrimers from a defined polydimethylsiloxane network},\ }\href@noop {} {\bibfield  {journal} {\bibinfo  {journal} {Angewandte Chemie International Edition}\ }\textbf {\bibinfo {volume} {62}},\ \bibinfo {pages} {e202310989} (\bibinfo {year} {2023})}\BibitemShut {NoStop}%
\bibitem [{\citenamefont {Zhou}\ \emph {et~al.}(2021)\citenamefont {Zhou}, \citenamefont {Chen}, \citenamefont {Xu}, \citenamefont {Chen}, \citenamefont {Xu}, \citenamefont {Zeng},\ and\ \citenamefont {Zhang}}]{zhou2021room}%
  \BibitemOpen
  \bibfield  {author} {\bibinfo {author} {\bibfnamefont {Z.}~\bibnamefont {Zhou}}, \bibinfo {author} {\bibfnamefont {S.}~\bibnamefont {Chen}}, \bibinfo {author} {\bibfnamefont {X.}~\bibnamefont {Xu}}, \bibinfo {author} {\bibfnamefont {Y.}~\bibnamefont {Chen}}, \bibinfo {author} {\bibfnamefont {L.}~\bibnamefont {Xu}}, \bibinfo {author} {\bibfnamefont {Y.}~\bibnamefont {Zeng}},\ and\ \bibinfo {author} {\bibfnamefont {F.}~\bibnamefont {Zhang}},\ }\bibfield  {title} {\bibinfo {title} {Room temperature self-healing crosslinked elastomer constructed by dynamic urea bond and hydrogen bond},\ }\href@noop {} {\bibfield  {journal} {\bibinfo  {journal} {Progress in Organic Coatings}\ }\textbf {\bibinfo {volume} {154}},\ \bibinfo {pages} {106213} (\bibinfo {year} {2021})}\BibitemShut {NoStop}%
\bibitem [{\citenamefont {Leone}\ \emph {et~al.}(2022)\citenamefont {Leone}, \citenamefont {Palucci}, \citenamefont {Zanchin}, \citenamefont {Vignali}, \citenamefont {Ricci},\ and\ \citenamefont {Bertini}}]{leone2022dynamically}%
  \BibitemOpen
  \bibfield  {author} {\bibinfo {author} {\bibfnamefont {G.}~\bibnamefont {Leone}}, \bibinfo {author} {\bibfnamefont {B.}~\bibnamefont {Palucci}}, \bibinfo {author} {\bibfnamefont {G.}~\bibnamefont {Zanchin}}, \bibinfo {author} {\bibfnamefont {A.}~\bibnamefont {Vignali}}, \bibinfo {author} {\bibfnamefont {G.}~\bibnamefont {Ricci}},\ and\ \bibinfo {author} {\bibfnamefont {F.}~\bibnamefont {Bertini}},\ }\bibfield  {title} {\bibinfo {title} {Dynamically cross-linked polyolefins via hydrogen bonds: Tough yet soft thermoplastic elastomers with high elastic recovery},\ }\href@noop {} {\bibfield  {journal} {\bibinfo  {journal} {ACS Applied Polymer Materials}\ }\textbf {\bibinfo {volume} {4}},\ \bibinfo {pages} {3770} (\bibinfo {year} {2022})}\BibitemShut {NoStop}%
\bibitem [{\citenamefont {Song}\ \emph {et~al.}(2019)\citenamefont {Song}, \citenamefont {Zhu}, \citenamefont {Yuan}, \citenamefont {Zhou}, \citenamefont {Zhang}, \citenamefont {Wang},\ and\ \citenamefont {Tang}}]{song2019ultra}%
  \BibitemOpen
  \bibfield  {author} {\bibinfo {author} {\bibfnamefont {L.}~\bibnamefont {Song}}, \bibinfo {author} {\bibfnamefont {T.}~\bibnamefont {Zhu}}, \bibinfo {author} {\bibfnamefont {L.}~\bibnamefont {Yuan}}, \bibinfo {author} {\bibfnamefont {J.}~\bibnamefont {Zhou}}, \bibinfo {author} {\bibfnamefont {Y.}~\bibnamefont {Zhang}}, \bibinfo {author} {\bibfnamefont {Z.}~\bibnamefont {Wang}},\ and\ \bibinfo {author} {\bibfnamefont {C.}~\bibnamefont {Tang}},\ }\bibfield  {title} {\bibinfo {title} {Ultra-strong long-chain polyamide elastomers with programmable supramolecular interactions and oriented crystalline microstructures},\ }\href@noop {} {\bibfield  {journal} {\bibinfo  {journal} {Nature Communications}\ }\textbf {\bibinfo {volume} {10}},\ \bibinfo {pages} {1315} (\bibinfo {year} {2019})}\BibitemShut {NoStop}%
\bibitem [{\citenamefont {Williams}(1984)}]{williams1984fracture}%
  \BibitemOpen
  \bibfield  {author} {\bibinfo {author} {\bibfnamefont {J.~G.}\ \bibnamefont {Williams}},\ }\bibfield  {title} {\bibinfo {title} {Fracture mechanics of polymers},\ }\href@noop {} {\bibfield  {journal} {\bibinfo  {journal} {Ellis Horwood Limited, Market Cross House, Cooper St, Chichester, West Sussex, PO 19, 1 EB, UK, 1984. 302}\ } (\bibinfo {year} {1984})}\BibitemShut {NoStop}%
\bibitem [{\citenamefont {Kawano}\ \emph {et~al.}(2023)\citenamefont {Kawano}, \citenamefont {Masai}, \citenamefont {Nakagawa}, \citenamefont {Yoshie},\ and\ \citenamefont {Terao}}]{kawano2023polymers}%
  \BibitemOpen
  \bibfield  {author} {\bibinfo {author} {\bibfnamefont {Y.}~\bibnamefont {Kawano}}, \bibinfo {author} {\bibfnamefont {H.}~\bibnamefont {Masai}}, \bibinfo {author} {\bibfnamefont {S.}~\bibnamefont {Nakagawa}}, \bibinfo {author} {\bibfnamefont {N.}~\bibnamefont {Yoshie}},\ and\ \bibinfo {author} {\bibfnamefont {J.}~\bibnamefont {Terao}},\ }\bibfield  {title} {\bibinfo {title} {Effects of alkyl ester chain length on the toughness of polyacrylate-based network materials},\ }\href@noop {} {\bibfield  {journal} {\bibinfo  {journal} {Polymers}\ }\textbf {\bibinfo {volume} {15}},\ \bibinfo {pages} {2389} (\bibinfo {year} {2023})}\BibitemShut {NoStop}%
\bibitem [{\citenamefont {Siavoshani}\ \emph {et~al.}(2024)\citenamefont {Siavoshani}, \citenamefont {Fan}, \citenamefont {Yang}, \citenamefont {Liu}, \citenamefont {Wang}, \citenamefont {Liu}, \citenamefont {Xu}, \citenamefont {Wang}, \citenamefont {Lin},\ and\ \citenamefont {Wang}}]{Siavoshani2024Soft_Matter}%
  \BibitemOpen
  \bibfield  {author} {\bibinfo {author} {\bibfnamefont {A.}~\bibnamefont {Siavoshani}}, \bibinfo {author} {\bibfnamefont {Z.}~\bibnamefont {Fan}}, \bibinfo {author} {\bibfnamefont {M.}~\bibnamefont {Yang}}, \bibinfo {author} {\bibfnamefont {S.}~\bibnamefont {Liu}}, \bibinfo {author} {\bibfnamefont {M.-C.}\ \bibnamefont {Wang}}, \bibinfo {author} {\bibfnamefont {J.}~\bibnamefont {Liu}}, \bibinfo {author} {\bibfnamefont {W.}~\bibnamefont {Xu}}, \bibinfo {author} {\bibfnamefont {J.}~\bibnamefont {Wang}}, \bibinfo {author} {\bibfnamefont {S.}~\bibnamefont {Lin}},\ and\ \bibinfo {author} {\bibfnamefont {S.-Q.}\ \bibnamefont {Wang}},\ }\bibfield  {title} {\bibinfo {title} {How do stretch rate, temperature, and solvent exchange affect elastic network rupture?},\ }\href@noop {} {\bibfield  {journal} {\bibinfo  {journal} {Soft Matter}\ }\textbf {\bibinfo {volume} {20}},\ \bibinfo {pages} {7657} (\bibinfo {year} {2024})}\BibitemShut {NoStop}%
\bibitem [{\citenamefont {Sahariah}\ and\ \citenamefont {Sarma}(2019)}]{sahariah2019relative}%
  \BibitemOpen
  \bibfield  {author} {\bibinfo {author} {\bibfnamefont {B.}~\bibnamefont {Sahariah}}\ and\ \bibinfo {author} {\bibfnamefont {B.~K.}\ \bibnamefont {Sarma}},\ }\bibfield  {title} {\bibinfo {title} {Relative orientation of the carbonyl groups determines the nature of orbital interactions in carbonyl--carbonyl short contacts},\ }\href@noop {} {\bibfield  {journal} {\bibinfo  {journal} {Chemical Science}\ }\textbf {\bibinfo {volume} {10}},\ \bibinfo {pages} {909} (\bibinfo {year} {2019})}\BibitemShut {NoStop}%
\bibitem [{\citenamefont {Berki}\ \emph {et~al.}(2017)\citenamefont {Berki}, \citenamefont {G{\"o}bl},\ and\ \citenamefont {Karger-Kocsis}}]{berki2017structure}%
  \BibitemOpen
  \bibfield  {author} {\bibinfo {author} {\bibfnamefont {P.}~\bibnamefont {Berki}}, \bibinfo {author} {\bibfnamefont {R.}~\bibnamefont {G{\"o}bl}},\ and\ \bibinfo {author} {\bibfnamefont {J.}~\bibnamefont {Karger-Kocsis}},\ }\bibfield  {title} {\bibinfo {title} {Structure and properties of styrene-butadiene rubber (sbr) with pyrolytic and industrial carbon black},\ }\href@noop {} {\bibfield  {journal} {\bibinfo  {journal} {Polymer Testing}\ }\textbf {\bibinfo {volume} {61}},\ \bibinfo {pages} {404} (\bibinfo {year} {2017})}\BibitemShut {NoStop}%
\bibitem [{\citenamefont {Wang}\ \emph {et~al.}(2016)\citenamefont {Wang}, \citenamefont {Zhao}, \citenamefont {Yang}, \citenamefont {Nishi}, \citenamefont {Ito}, \citenamefont {Zhao},\ and\ \citenamefont {Zhang}}]{wang2016novel}%
  \BibitemOpen
  \bibfield  {author} {\bibinfo {author} {\bibfnamefont {W.}~\bibnamefont {Wang}}, \bibinfo {author} {\bibfnamefont {D.}~\bibnamefont {Zhao}}, \bibinfo {author} {\bibfnamefont {J.}~\bibnamefont {Yang}}, \bibinfo {author} {\bibfnamefont {T.}~\bibnamefont {Nishi}}, \bibinfo {author} {\bibfnamefont {K.}~\bibnamefont {Ito}}, \bibinfo {author} {\bibfnamefont {X.}~\bibnamefont {Zhao}},\ and\ \bibinfo {author} {\bibfnamefont {L.}~\bibnamefont {Zhang}},\ }\bibfield  {title} {\bibinfo {title} {Novel slide-ring material/natural rubber composites with high damping property},\ }\href@noop {} {\bibfield  {journal} {\bibinfo  {journal} {Scientific reports}\ }\textbf {\bibinfo {volume} {6}},\ \bibinfo {pages} {22810} (\bibinfo {year} {2016})}\BibitemShut {NoStop}%
\bibitem [{\citenamefont {Faghihi}\ \emph {et~al.}(2011)\citenamefont {Faghihi}, \citenamefont {Mohammadi},\ and\ \citenamefont {Hazendonk}}]{faghihi2011effect}%
  \BibitemOpen
  \bibfield  {author} {\bibinfo {author} {\bibfnamefont {F.}~\bibnamefont {Faghihi}}, \bibinfo {author} {\bibfnamefont {N.}~\bibnamefont {Mohammadi}},\ and\ \bibinfo {author} {\bibfnamefont {P.}~\bibnamefont {Hazendonk}},\ }\bibfield  {title} {\bibinfo {title} {Effect of restricted phase segregation and resultant nanostructural heterogeneity on glass transition of nonuniform acrylic random copolymers},\ }\href@noop {} {\bibfield  {journal} {\bibinfo  {journal} {Macromolecules}\ }\textbf {\bibinfo {volume} {44}},\ \bibinfo {pages} {2154} (\bibinfo {year} {2011})}\BibitemShut {NoStop}%
\bibitem [{\citenamefont {Lei}\ \emph {et~al.}(2019)\citenamefont {Lei}, \citenamefont {Zhang}, \citenamefont {Kuang},\ and\ \citenamefont {Yang}}]{lei2019preparation}%
  \BibitemOpen
  \bibfield  {author} {\bibinfo {author} {\bibfnamefont {T.}~\bibnamefont {Lei}}, \bibinfo {author} {\bibfnamefont {Y.-W.}\ \bibnamefont {Zhang}}, \bibinfo {author} {\bibfnamefont {D.-L.}\ \bibnamefont {Kuang}},\ and\ \bibinfo {author} {\bibfnamefont {Y.-R.}\ \bibnamefont {Yang}},\ }\bibfield  {title} {\bibinfo {title} {Preparation and properties of rubber blends for high-damping-isolation bearings},\ }\href@noop {} {\bibfield  {journal} {\bibinfo  {journal} {Polymers}\ }\textbf {\bibinfo {volume} {11}},\ \bibinfo {pages} {1374} (\bibinfo {year} {2019})}\BibitemShut {NoStop}%
\bibitem [{\citenamefont {Krajnak}(2018)}]{krajnak2018health}%
  \BibitemOpen
  \bibfield  {author} {\bibinfo {author} {\bibfnamefont {K.}~\bibnamefont {Krajnak}},\ }\bibfield  {title} {\bibinfo {title} {Health effects associated with occupational exposure to hand-arm or whole body vibration},\ }\href@noop {} {\bibfield  {journal} {\bibinfo  {journal} {Journal of Toxicology and Environmental Health, Part B}\ }\textbf {\bibinfo {volume} {21}},\ \bibinfo {pages} {320} (\bibinfo {year} {2018})}\BibitemShut {NoStop}%
\bibitem [{\citenamefont {Havriliak}\ and\ \citenamefont {Negami}(1967)}]{havriliak1967complex}%
  \BibitemOpen
  \bibfield  {author} {\bibinfo {author} {\bibfnamefont {S.}~\bibnamefont {Havriliak}}\ and\ \bibinfo {author} {\bibfnamefont {S.}~\bibnamefont {Negami}},\ }\bibfield  {title} {\bibinfo {title} {A complex plane representation of dielectric and mechanical relaxation processes in some polymers},\ }\href@noop {} {\bibfield  {journal} {\bibinfo  {journal} {Polymer}\ }\textbf {\bibinfo {volume} {8}},\ \bibinfo {pages} {161} (\bibinfo {year} {1967})}\BibitemShut {NoStop}%
\bibitem [{\citenamefont {Szabo}\ and\ \citenamefont {Keough}(2002)}]{szabo2002method}%
  \BibitemOpen
  \bibfield  {author} {\bibinfo {author} {\bibfnamefont {J.~P.}\ \bibnamefont {Szabo}}\ and\ \bibinfo {author} {\bibfnamefont {I.~A.}\ \bibnamefont {Keough}},\ }\bibfield  {title} {\bibinfo {title} {Method for analysis of dynamic mechanical thermal analysis data using the havriliak--negami model},\ }\href@noop {} {\bibfield  {journal} {\bibinfo  {journal} {Thermochimica acta}\ }\textbf {\bibinfo {volume} {392}},\ \bibinfo {pages} {1} (\bibinfo {year} {2002})}\BibitemShut {NoStop}%
\bibitem [{\citenamefont {Madigosky}\ \emph {et~al.}(2006)\citenamefont {Madigosky}, \citenamefont {Lee},\ and\ \citenamefont {Niemiec}}]{madigosky2006method}%
  \BibitemOpen
  \bibfield  {author} {\bibinfo {author} {\bibfnamefont {W.~M.}\ \bibnamefont {Madigosky}}, \bibinfo {author} {\bibfnamefont {G.~F.}\ \bibnamefont {Lee}},\ and\ \bibinfo {author} {\bibfnamefont {J.~M.}\ \bibnamefont {Niemiec}},\ }\bibfield  {title} {\bibinfo {title} {A method for modeling polymer viscoelastic data and the temperature shift function},\ }\href@noop {} {\bibfield  {journal} {\bibinfo  {journal} {The Journal of the Acoustical Society of America}\ }\textbf {\bibinfo {volume} {119}},\ \bibinfo {pages} {3760} (\bibinfo {year} {2006})}\BibitemShut {NoStop}%
\bibitem [{\citenamefont {Zhao}\ \emph {et~al.}(2020)\citenamefont {Zhao}, \citenamefont {Shou}, \citenamefont {Liang}, \citenamefont {Hu}, \citenamefont {Yu},\ and\ \citenamefont {Zhang}}]{zhao2020bio}%
  \BibitemOpen
  \bibfield  {author} {\bibinfo {author} {\bibfnamefont {X.}~\bibnamefont {Zhao}}, \bibinfo {author} {\bibfnamefont {T.}~\bibnamefont {Shou}}, \bibinfo {author} {\bibfnamefont {R.}~\bibnamefont {Liang}}, \bibinfo {author} {\bibfnamefont {S.}~\bibnamefont {Hu}}, \bibinfo {author} {\bibfnamefont {P.}~\bibnamefont {Yu}},\ and\ \bibinfo {author} {\bibfnamefont {L.}~\bibnamefont {Zhang}},\ }\bibfield  {title} {\bibinfo {title} {Bio-based thermoplastic polyurethane derived from polylactic acid with high-damping performance},\ }\href@noop {} {\bibfield  {journal} {\bibinfo  {journal} {Industrial crops and products}\ }\textbf {\bibinfo {volume} {154}},\ \bibinfo {pages} {112619} (\bibinfo {year} {2020})}\BibitemShut {NoStop}%
\bibitem [{\citenamefont {Xu}\ \emph {et~al.}(2014)\citenamefont {Xu}, \citenamefont {Zhang}, \citenamefont {Zhang}, \citenamefont {Hu}, \citenamefont {Wu},\ and\ \citenamefont {Guo}}]{xu2014molecular}%
  \BibitemOpen
  \bibfield  {author} {\bibinfo {author} {\bibfnamefont {K.}~\bibnamefont {Xu}}, \bibinfo {author} {\bibfnamefont {F.}~\bibnamefont {Zhang}}, \bibinfo {author} {\bibfnamefont {X.}~\bibnamefont {Zhang}}, \bibinfo {author} {\bibfnamefont {Q.}~\bibnamefont {Hu}}, \bibinfo {author} {\bibfnamefont {H.}~\bibnamefont {Wu}},\ and\ \bibinfo {author} {\bibfnamefont {S.}~\bibnamefont {Guo}},\ }\bibfield  {title} {\bibinfo {title} {Molecular insights into hydrogen bonds in polyurethane/hindered phenol hybrids: evolution and relationship with damping properties},\ }\href@noop {} {\bibfield  {journal} {\bibinfo  {journal} {Journal of Materials Chemistry A}\ }\textbf {\bibinfo {volume} {2}},\ \bibinfo {pages} {8545} (\bibinfo {year} {2014})}\BibitemShut {NoStop}%
\bibitem [{\citenamefont {Zhang}\ \emph {et~al.}(2016)\citenamefont {Zhang}, \citenamefont {Zhao}, \citenamefont {Zou}, \citenamefont {Luo},\ and\ \citenamefont {Xie}}]{zhang2016unusual}%
  \BibitemOpen
  \bibfield  {author} {\bibinfo {author} {\bibfnamefont {G.}~\bibnamefont {Zhang}}, \bibinfo {author} {\bibfnamefont {Q.}~\bibnamefont {Zhao}}, \bibinfo {author} {\bibfnamefont {W.}~\bibnamefont {Zou}}, \bibinfo {author} {\bibfnamefont {Y.}~\bibnamefont {Luo}},\ and\ \bibinfo {author} {\bibfnamefont {T.}~\bibnamefont {Xie}},\ }\bibfield  {title} {\bibinfo {title} {Unusual aspects of supramolecular networks: plasticity to elasticity, ultrasoft shape memory, and dynamic mechanical properties},\ }\href@noop {} {\bibfield  {journal} {\bibinfo  {journal} {Advanced Functional Materials}\ }\textbf {\bibinfo {volume} {26}},\ \bibinfo {pages} {931} (\bibinfo {year} {2016})}\BibitemShut {NoStop}%
\bibitem [{\citenamefont {Hou}\ \emph {et~al.}(2022)\citenamefont {Hou}, \citenamefont {Peng}, \citenamefont {Li}, \citenamefont {Wu}, \citenamefont {Zhang}, \citenamefont {Li}, \citenamefont {Zhou},\ and\ \citenamefont {Wu}}]{hou2022bioinspired}%
  \BibitemOpen
  \bibfield  {author} {\bibinfo {author} {\bibfnamefont {Y.}~\bibnamefont {Hou}}, \bibinfo {author} {\bibfnamefont {Y.}~\bibnamefont {Peng}}, \bibinfo {author} {\bibfnamefont {P.}~\bibnamefont {Li}}, \bibinfo {author} {\bibfnamefont {Q.}~\bibnamefont {Wu}}, \bibinfo {author} {\bibfnamefont {J.}~\bibnamefont {Zhang}}, \bibinfo {author} {\bibfnamefont {W.}~\bibnamefont {Li}}, \bibinfo {author} {\bibfnamefont {G.}~\bibnamefont {Zhou}},\ and\ \bibinfo {author} {\bibfnamefont {J.}~\bibnamefont {Wu}},\ }\bibfield  {title} {\bibinfo {title} {Bioinspired design of high vibration-damping supramolecular elastomers based on multiple energy-dissipation mechanisms},\ }\href@noop {} {\bibfield  {journal} {\bibinfo  {journal} {ACS Applied Materials \& Interfaces}\ }\textbf {\bibinfo {volume} {14}},\ \bibinfo {pages} {35097} (\bibinfo {year} {2022})}\BibitemShut {NoStop}%
\bibitem [{\citenamefont {Zhang}\ \emph {et~al.}(2024)\citenamefont {Zhang}, \citenamefont {Wang}, \citenamefont {Zhu}, \citenamefont {Zhang},\ and\ \citenamefont {Zhu}}]{zhang2024dielectric}%
  \BibitemOpen
  \bibfield  {author} {\bibinfo {author} {\bibfnamefont {C.}~\bibnamefont {Zhang}}, \bibinfo {author} {\bibfnamefont {Z.}~\bibnamefont {Wang}}, \bibinfo {author} {\bibfnamefont {H.}~\bibnamefont {Zhu}}, \bibinfo {author} {\bibfnamefont {Q.}~\bibnamefont {Zhang}},\ and\ \bibinfo {author} {\bibfnamefont {S.}~\bibnamefont {Zhu}},\ }\bibfield  {title} {\bibinfo {title} {Dielectric gels with microphase separation for wide-range and self-damping pressure sensing},\ }\href@noop {} {\bibfield  {journal} {\bibinfo  {journal} {Advanced Materials}\ }\textbf {\bibinfo {volume} {36}},\ \bibinfo {pages} {2308520} (\bibinfo {year} {2024})}\BibitemShut {NoStop}%
\bibitem [{\citenamefont {Ohzono}\ \emph {et~al.}(2019)\citenamefont {Ohzono}, \citenamefont {Saed},\ and\ \citenamefont {Terentjev}}]{ohzono2019enhanced}%
  \BibitemOpen
  \bibfield  {author} {\bibinfo {author} {\bibfnamefont {T.}~\bibnamefont {Ohzono}}, \bibinfo {author} {\bibfnamefont {M.~O.}\ \bibnamefont {Saed}},\ and\ \bibinfo {author} {\bibfnamefont {E.~M.}\ \bibnamefont {Terentjev}},\ }\bibfield  {title} {\bibinfo {title} {Enhanced dynamic adhesion in nematic liquid crystal elastomers},\ }\href@noop {} {\bibfield  {journal} {\bibinfo  {journal} {Advanced Materials}\ }\textbf {\bibinfo {volume} {31}},\ \bibinfo {pages} {1902642} (\bibinfo {year} {2019})}\BibitemShut {NoStop}%
\bibitem [{\citenamefont {Saed}\ \emph {et~al.}(2021)\citenamefont {Saed}, \citenamefont {Elmadih}, \citenamefont {Terentjev}, \citenamefont {Chronopoulos}, \citenamefont {Williamson},\ and\ \citenamefont {Terentjev}}]{saed2021impact}%
  \BibitemOpen
  \bibfield  {author} {\bibinfo {author} {\bibfnamefont {M.~O.}\ \bibnamefont {Saed}}, \bibinfo {author} {\bibfnamefont {W.}~\bibnamefont {Elmadih}}, \bibinfo {author} {\bibfnamefont {A.}~\bibnamefont {Terentjev}}, \bibinfo {author} {\bibfnamefont {D.}~\bibnamefont {Chronopoulos}}, \bibinfo {author} {\bibfnamefont {D.}~\bibnamefont {Williamson}},\ and\ \bibinfo {author} {\bibfnamefont {E.~M.}\ \bibnamefont {Terentjev}},\ }\bibfield  {title} {\bibinfo {title} {Impact damping and vibration attenuation in nematic liquid crystal elastomers},\ }\href@noop {} {\bibfield  {journal} {\bibinfo  {journal} {Nature communications}\ }\textbf {\bibinfo {volume} {12}},\ \bibinfo {pages} {6676} (\bibinfo {year} {2021})}\BibitemShut {NoStop}%
\bibitem [{\citenamefont {Farre-Kaga}\ \emph {et~al.}(2022)\citenamefont {Farre-Kaga}, \citenamefont {Saed},\ and\ \citenamefont {Terentjev}}]{farre2022dynamic}%
  \BibitemOpen
  \bibfield  {author} {\bibinfo {author} {\bibfnamefont {H.~J.}\ \bibnamefont {Farre-Kaga}}, \bibinfo {author} {\bibfnamefont {M.~O.}\ \bibnamefont {Saed}},\ and\ \bibinfo {author} {\bibfnamefont {E.~M.}\ \bibnamefont {Terentjev}},\ }\bibfield  {title} {\bibinfo {title} {Dynamic pressure sensitive adhesion in nematic phase of liquid crystal elastomers},\ }\href@noop {} {\bibfield  {journal} {\bibinfo  {journal} {Advanced Functional Materials}\ }\textbf {\bibinfo {volume} {32}},\ \bibinfo {pages} {2110190} (\bibinfo {year} {2022})}\BibitemShut {NoStop}%
\bibitem [{\citenamefont {Zhang}\ \emph {et~al.}(2020{\natexlab{b}})\citenamefont {Zhang}, \citenamefont {Cheng}, \citenamefont {Zhao}, \citenamefont {Zhang}, \citenamefont {Zhao}, \citenamefont {Liu}, \citenamefont {Bai}, \citenamefont {Pan}, \citenamefont {Yu},\ and\ \citenamefont {Yan}}]{zhang2020muscle}%
  \BibitemOpen
  \bibfield  {author} {\bibinfo {author} {\bibfnamefont {Z.}~\bibnamefont {Zhang}}, \bibinfo {author} {\bibfnamefont {L.}~\bibnamefont {Cheng}}, \bibinfo {author} {\bibfnamefont {J.}~\bibnamefont {Zhao}}, \bibinfo {author} {\bibfnamefont {H.}~\bibnamefont {Zhang}}, \bibinfo {author} {\bibfnamefont {X.}~\bibnamefont {Zhao}}, \bibinfo {author} {\bibfnamefont {Y.}~\bibnamefont {Liu}}, \bibinfo {author} {\bibfnamefont {R.}~\bibnamefont {Bai}}, \bibinfo {author} {\bibfnamefont {H.}~\bibnamefont {Pan}}, \bibinfo {author} {\bibfnamefont {W.}~\bibnamefont {Yu}},\ and\ \bibinfo {author} {\bibfnamefont {X.}~\bibnamefont {Yan}},\ }\bibfield  {title} {\bibinfo {title} {Muscle-mimetic synergistic covalent and supramolecular polymers: phototriggered formation leads to mechanical performance boost},\ }\href@noop {} {\bibfield  {journal} {\bibinfo  {journal} {Journal of the American Chemical Society}\ }\textbf {\bibinfo {volume} {143}},\ \bibinfo {pages} {902} (\bibinfo {year} {2020}{\natexlab{b}})}\BibitemShut {NoStop}%
\bibitem [{\citenamefont {Huang}\ \emph {et~al.}(2021)\citenamefont {Huang}, \citenamefont {Xu}, \citenamefont {Qi}, \citenamefont {Zhou}, \citenamefont {Shi}, \citenamefont {Zhao},\ and\ \citenamefont {Liu}}]{huang2021ultrahigh}%
  \BibitemOpen
  \bibfield  {author} {\bibinfo {author} {\bibfnamefont {J.}~\bibnamefont {Huang}}, \bibinfo {author} {\bibfnamefont {Y.}~\bibnamefont {Xu}}, \bibinfo {author} {\bibfnamefont {S.}~\bibnamefont {Qi}}, \bibinfo {author} {\bibfnamefont {J.}~\bibnamefont {Zhou}}, \bibinfo {author} {\bibfnamefont {W.}~\bibnamefont {Shi}}, \bibinfo {author} {\bibfnamefont {T.}~\bibnamefont {Zhao}},\ and\ \bibinfo {author} {\bibfnamefont {M.}~\bibnamefont {Liu}},\ }\bibfield  {title} {\bibinfo {title} {Ultrahigh energy-dissipation elastomers by precisely tailoring the relaxation of confined polymer fluids},\ }\href@noop {} {\bibfield  {journal} {\bibinfo  {journal} {Nature Communications}\ }\textbf {\bibinfo {volume} {12}},\ \bibinfo {pages} {3610} (\bibinfo {year} {2021})}\BibitemShut {NoStop}%
\bibitem [{\citenamefont {Stukalin}\ \emph {et~al.}(2013)\citenamefont {Stukalin}, \citenamefont {Cai}, \citenamefont {Kumar}, \citenamefont {Leibler},\ and\ \citenamefont {Rubinstein}}]{stukalin2013self}%
  \BibitemOpen
  \bibfield  {author} {\bibinfo {author} {\bibfnamefont {E.~B.}\ \bibnamefont {Stukalin}}, \bibinfo {author} {\bibfnamefont {L.-H.}\ \bibnamefont {Cai}}, \bibinfo {author} {\bibfnamefont {N.~A.}\ \bibnamefont {Kumar}}, \bibinfo {author} {\bibfnamefont {L.}~\bibnamefont {Leibler}},\ and\ \bibinfo {author} {\bibfnamefont {M.}~\bibnamefont {Rubinstein}},\ }\bibfield  {title} {\bibinfo {title} {Self-healing of unentangled polymer networks with reversible bonds},\ }\href@noop {} {\bibfield  {journal} {\bibinfo  {journal} {Macromolecules}\ }\textbf {\bibinfo {volume} {46}},\ \bibinfo {pages} {7525} (\bibinfo {year} {2013})}\BibitemShut {NoStop}%
\bibitem [{\citenamefont {Kim}\ and\ \citenamefont {Wool}(1983)}]{kim1983theory}%
  \BibitemOpen
  \bibfield  {author} {\bibinfo {author} {\bibfnamefont {Y.~H.}\ \bibnamefont {Kim}}\ and\ \bibinfo {author} {\bibfnamefont {R.~P.}\ \bibnamefont {Wool}},\ }\bibfield  {title} {\bibinfo {title} {A theory of healing at a polymer-polymer interface},\ }\href@noop {} {\bibfield  {journal} {\bibinfo  {journal} {Macromolecules}\ }\textbf {\bibinfo {volume} {16}},\ \bibinfo {pages} {1115} (\bibinfo {year} {1983})}\BibitemShut {NoStop}%
\bibitem [{\citenamefont {Wool}(2008)}]{wool2008self}%
  \BibitemOpen
  \bibfield  {author} {\bibinfo {author} {\bibfnamefont {R.~P.}\ \bibnamefont {Wool}},\ }\bibfield  {title} {\bibinfo {title} {Self-healing materials: a review},\ }\href@noop {} {\bibfield  {journal} {\bibinfo  {journal} {Soft Matter}\ }\textbf {\bibinfo {volume} {4}},\ \bibinfo {pages} {400} (\bibinfo {year} {2008})}\BibitemShut {NoStop}%
\bibitem [{\citenamefont {Liu}\ \emph {et~al.}(2018)\citenamefont {Liu}, \citenamefont {Hao}, \citenamefont {Zhang}, \citenamefont {Yang}, \citenamefont {Wang}, \citenamefont {Han}, \citenamefont {Li}, \citenamefont {Xin},\ and\ \citenamefont {Zhang}}]{liu2018self}%
  \BibitemOpen
  \bibfield  {author} {\bibinfo {author} {\bibfnamefont {T.}~\bibnamefont {Liu}}, \bibinfo {author} {\bibfnamefont {C.}~\bibnamefont {Hao}}, \bibinfo {author} {\bibfnamefont {S.}~\bibnamefont {Zhang}}, \bibinfo {author} {\bibfnamefont {X.}~\bibnamefont {Yang}}, \bibinfo {author} {\bibfnamefont {L.}~\bibnamefont {Wang}}, \bibinfo {author} {\bibfnamefont {J.}~\bibnamefont {Han}}, \bibinfo {author} {\bibfnamefont {Y.}~\bibnamefont {Li}}, \bibinfo {author} {\bibfnamefont {J.}~\bibnamefont {Xin}},\ and\ \bibinfo {author} {\bibfnamefont {J.}~\bibnamefont {Zhang}},\ }\bibfield  {title} {\bibinfo {title} {A self-healable high glass transition temperature bioepoxy material based on vitrimer chemistry},\ }\href@noop {} {\bibfield  {journal} {\bibinfo  {journal} {Macromolecules}\ }\textbf {\bibinfo {volume} {51}},\ \bibinfo {pages} {5577} (\bibinfo {year} {2018})}\BibitemShut {NoStop}%
\bibitem [{\citenamefont {Han}\ \emph {et~al.}(2018)\citenamefont {Han}, \citenamefont {Liu}, \citenamefont {Hao}, \citenamefont {Zhang}, \citenamefont {Guo},\ and\ \citenamefont {Zhang}}]{han2018catalyst}%
  \BibitemOpen
  \bibfield  {author} {\bibinfo {author} {\bibfnamefont {J.}~\bibnamefont {Han}}, \bibinfo {author} {\bibfnamefont {T.}~\bibnamefont {Liu}}, \bibinfo {author} {\bibfnamefont {C.}~\bibnamefont {Hao}}, \bibinfo {author} {\bibfnamefont {S.}~\bibnamefont {Zhang}}, \bibinfo {author} {\bibfnamefont {B.}~\bibnamefont {Guo}},\ and\ \bibinfo {author} {\bibfnamefont {J.}~\bibnamefont {Zhang}},\ }\bibfield  {title} {\bibinfo {title} {A catalyst-free epoxy vitrimer system based on multifunctional hyperbranched polymer},\ }\href@noop {} {\bibfield  {journal} {\bibinfo  {journal} {Macromolecules}\ }\textbf {\bibinfo {volume} {51}},\ \bibinfo {pages} {6789} (\bibinfo {year} {2018})}\BibitemShut {NoStop}%
\bibitem [{\citenamefont {Tang}\ \emph {et~al.}(2021)\citenamefont {Tang}, \citenamefont {Zhang}, \citenamefont {Zhang}, \citenamefont {Xu}, \citenamefont {Li},\ and\ \citenamefont {Zhang}}]{tang2021bio}%
  \BibitemOpen
  \bibfield  {author} {\bibinfo {author} {\bibfnamefont {D.}~\bibnamefont {Tang}}, \bibinfo {author} {\bibfnamefont {L.}~\bibnamefont {Zhang}}, \bibinfo {author} {\bibfnamefont {X.}~\bibnamefont {Zhang}}, \bibinfo {author} {\bibfnamefont {L.}~\bibnamefont {Xu}}, \bibinfo {author} {\bibfnamefont {K.}~\bibnamefont {Li}},\ and\ \bibinfo {author} {\bibfnamefont {A.}~\bibnamefont {Zhang}},\ }\bibfield  {title} {\bibinfo {title} {Bio-mimetic actuators of a photothermal-responsive vitrimer liquid crystal elastomer with robust, self-healing, shape memory, and reconfigurable properties},\ }\href@noop {} {\bibfield  {journal} {\bibinfo  {journal} {ACS Applied Materials \& Interfaces}\ }\textbf {\bibinfo {volume} {14}},\ \bibinfo {pages} {1929} (\bibinfo {year} {2021})}\BibitemShut {NoStop}%
\bibitem [{\citenamefont {Van~Gurp}\ and\ \citenamefont {Palmen}(1998)}]{van1998time}%
  \BibitemOpen
  \bibfield  {author} {\bibinfo {author} {\bibfnamefont {M.}~\bibnamefont {Van~Gurp}}\ and\ \bibinfo {author} {\bibfnamefont {J.}~\bibnamefont {Palmen}},\ }\bibfield  {title} {\bibinfo {title} {Time-temperature superposition for polymeric blends},\ }\href@noop {} {\bibfield  {journal} {\bibinfo  {journal} {Rheol. Bull}\ }\textbf {\bibinfo {volume} {67}},\ \bibinfo {pages} {5} (\bibinfo {year} {1998})}\BibitemShut {NoStop}%

\end{thebibliography}


%

\end{document}


\title{Ultra-stretchable and Self-Healable Vitrimers with Tuneable Damping \\ and Mechanical Response}

\author{Jiaxin Zhao$^{1,2}$}
\author{Nicholas J. Warren$^{3,4}$}
\author{Richard Mandle$^{1,2}$}
\author{Peter Hine$^1$}
\author{Daniel J. Read$^5$}
\author{Andrew J Wilson$^{2,6}$}
\author{Johan Mattsson$^1$}
\email{k.j.l.mattsson@leeds.ac.uk}
\affiliation{$^1$School of Physics and Astronomy, University of Leeds, Leeds LS2\,9JT, United Kingdom}
\affiliation{$^2$School of Chemistry, University of Leeds, Leeds LS2\,9JT, United Kingdom}
\affiliation{$^3$School of Chemical and Process Engineering, University of Leeds, Leeds LS2\,9JT, United Kingdom}
\affiliation{$^4$School of Chemical, Materials and Biological Engineering, University of Sheffield, Sheffield S10\,2TN, United Kingdom}
\affiliation{$^5$School of Mathematics, University of Leeds, Leeds LS2\,9JT, United Kingdom}
\affiliation{$^6$School of Chemistry, University of Birmingham, Birmingham B15\,2TT, United Kingdom}

\maketitle

\date{\today}%

\begin{figure}[h!]
\centering
  \includegraphics[width=0.5\linewidth]{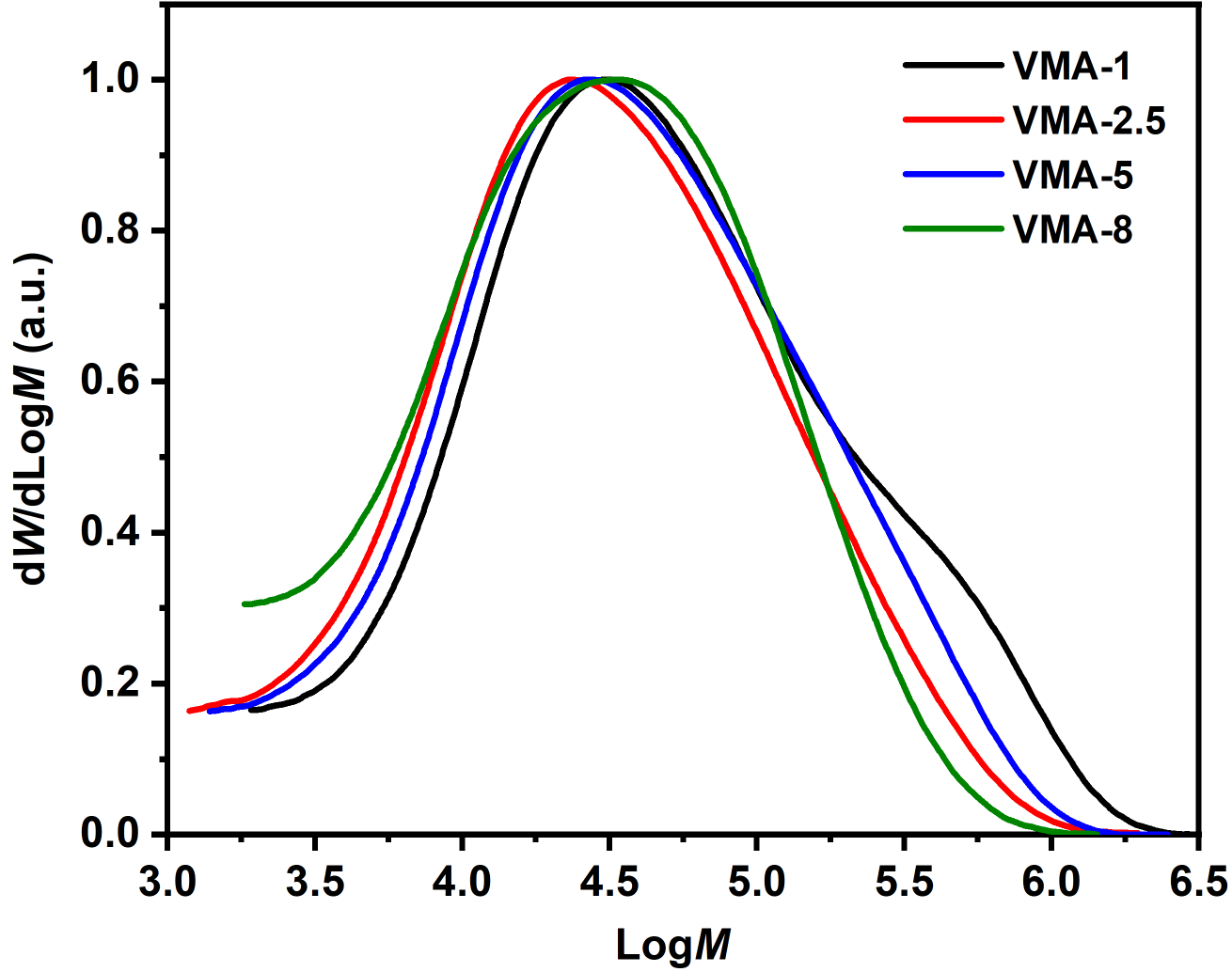}
  \caption{Gel permeation chromatography (GPC) data for the fully cured vitrimer samples.}
  \label{fig:SFigGPC-VMA}
\end{figure}

\renewcommand{\thetable}{S\arabic{table}}

\begin{table}[h!]
 \caption{GPC results for the fully cured vitrimer samples.}
 \begin{center}
  \begin{tabular}{@{}lllllllll@{}}
    \hline
    Name &VMA1 &VMA2.5&VMA5&VMA8\\
    \hline
    $M_{\textup{n}}$ (kDa) & 19.3 &12.6 & 14.9 & 13.4 \\
    $M_{\textup{w}}$  (kDa)&123.7  &69.7 &85.8  &59.7  \\
    \DJ (-) & 6.4  & 5.5 & 5.7 & 4.5   \\
    \hline
  \end{tabular}
  \end{center}
  \label{table:GPC-DATA}
\end{table}


\begin{figure}[h!]
\centering
  \includegraphics[width=0.5\linewidth]{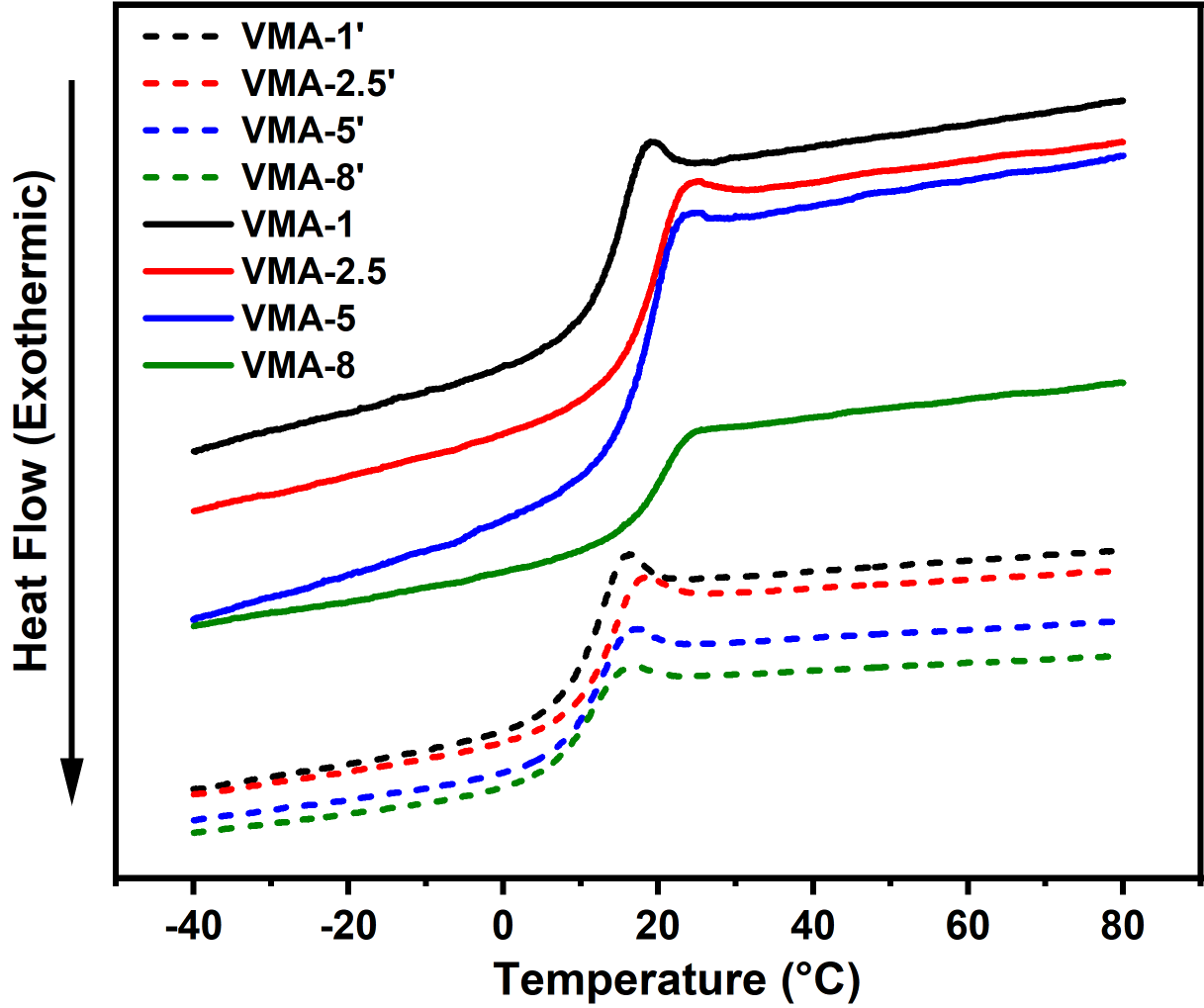}
  \caption{Differential Scanning Calorimetry (DSC) data for all investigated vitrimer samples. The shown data are recorded on heating for a heating rate of 10 K/min. See the discussion in the main paper for a detailed discussion.}
  \label{fig:SFigDSC-ALL}
\end{figure}


\begin{table}[h!]
\centering
 \caption{DSC analysis results for the onset ($T_{\textup{g}}^{\textup{onset}}$) and midpoint ($T_{\textup{g}}^{\textup{mid}}$) glass transition temperatures, together with an estimate of the glass transition breadth ($\Delta T_{\textup{g}}$), where the latter is calculated as $\Delta T_{\textup{g}}=T_{\textup{g}}^{\textup{offset}}-T_{\textup{g}}^{\textup{onset}}$, as described in the main paper.}
  \begin{tabular}{@{}lllllllll@{}}
    \hline
    Name & VMA1' &VMA1 &VMA2.5'&VMA2.5&VMA5'&VMA5&VMA8'&VMA8\\
    \hline
    $T_{\textup{g}}^{\textup{onset}}$ (°C) & 7.9 &14.0 & 9.0 & 14.5 &7.3 &14.1 & 5.5& 15.0 \\
    $T_{\textup{g}}^{\textup{mid}}$  (°C)&12.7   &19.1 &14.6  &20.4  & 12.7& 19.8&12.0 & 20.5 \\
    $\Delta T_{\textup{g}}$ (°C) & 5.7  & 5.8 & 6.9 & 7.7 &7.0 &7.7 &8.2 &8.5  \\
    \hline
  \end{tabular}
  \label{table:DSC-DATA}
\end{table}


\begin{figure}[h!]
\centering
  \includegraphics[width=0.5\linewidth]{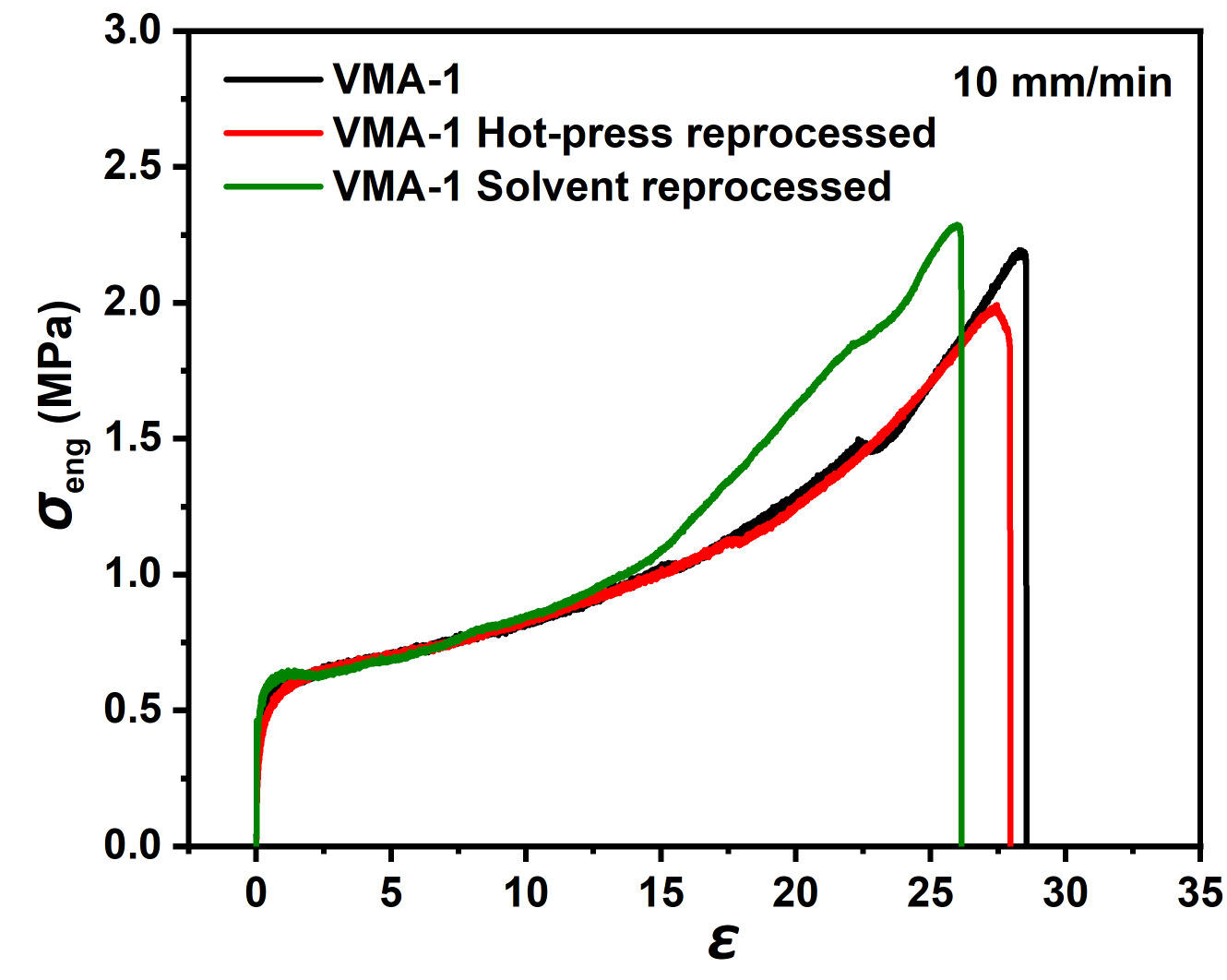}
  \caption{Tensile stress-strain data showing the engineering stress versus the strain, as measured using an extension speed of 10 mm/min. Data are shown for the fully cured VMA-1 sample, both before and after reprocessing using either a thermo-mechanical hot-press route, or an alternative solvent route.}
  \label{fig:SFigREPRO-1}
\end{figure}


\begin{figure}
\centering
  \includegraphics[width=0.5\linewidth]{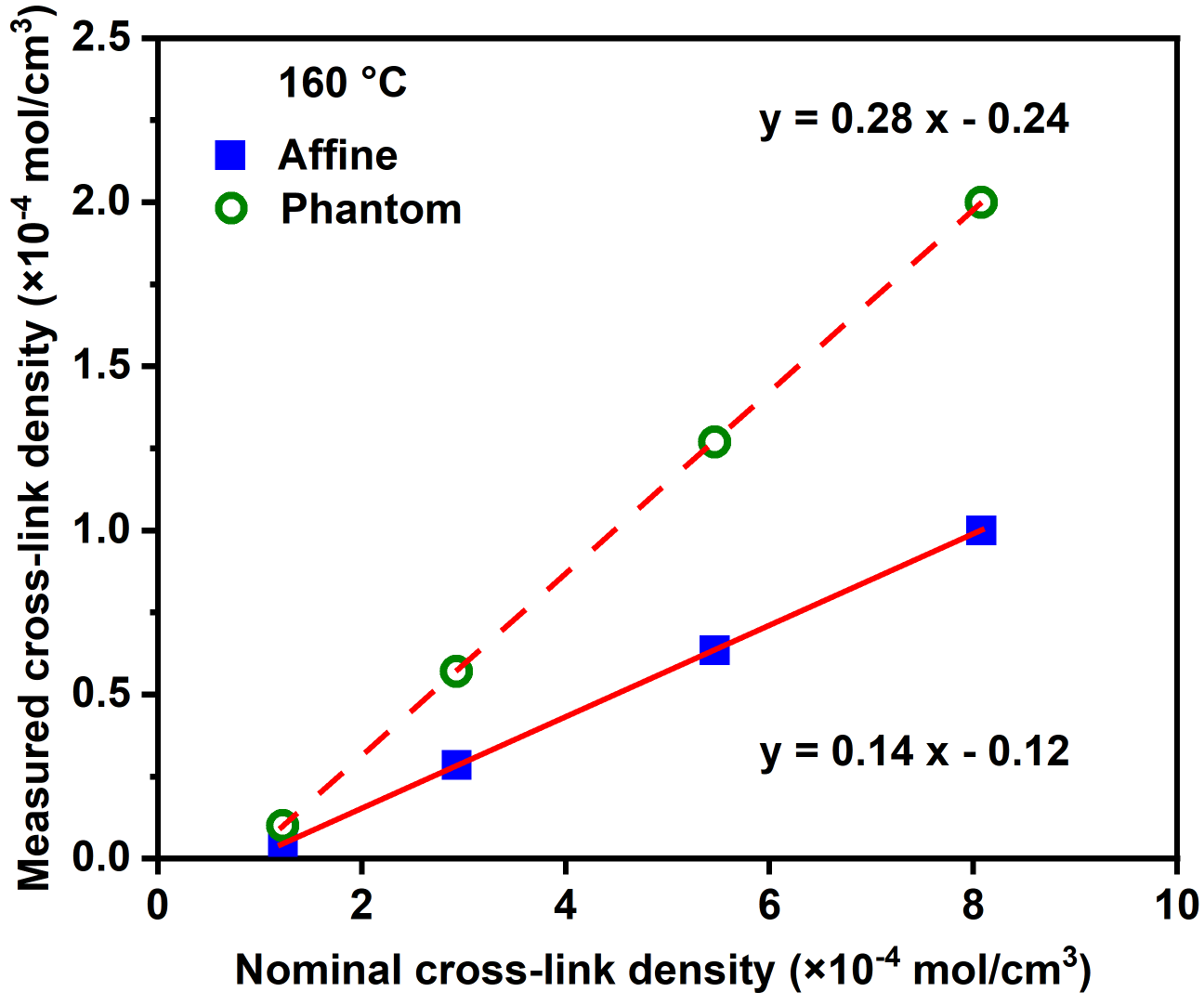}
  \caption{A plot of the estimated cross-link density for the fully cured vitrimers, as determined from the plateau moduli using either the affine network model (filled blue squares) or the phantom network model (open green circles), versus the nominal cross-link density, as determined from the sample stochiometry and the mass density of linear poly(methyl acrylate) at 160 $^\circ$C \cite{pfefferkorn2010pressure}. A detailed discussion of the calculations is found in the main paper. The solid and dashed lines are linear fits to the data for the affine (solid line) and the phantom (dashed line) network models, respectively; the fits correspond to the equations provided in the figure.}
  \label{fig:SFigCDCD}
\end{figure}


\begin{figure}
\centering
  \includegraphics[width=0.5\linewidth]{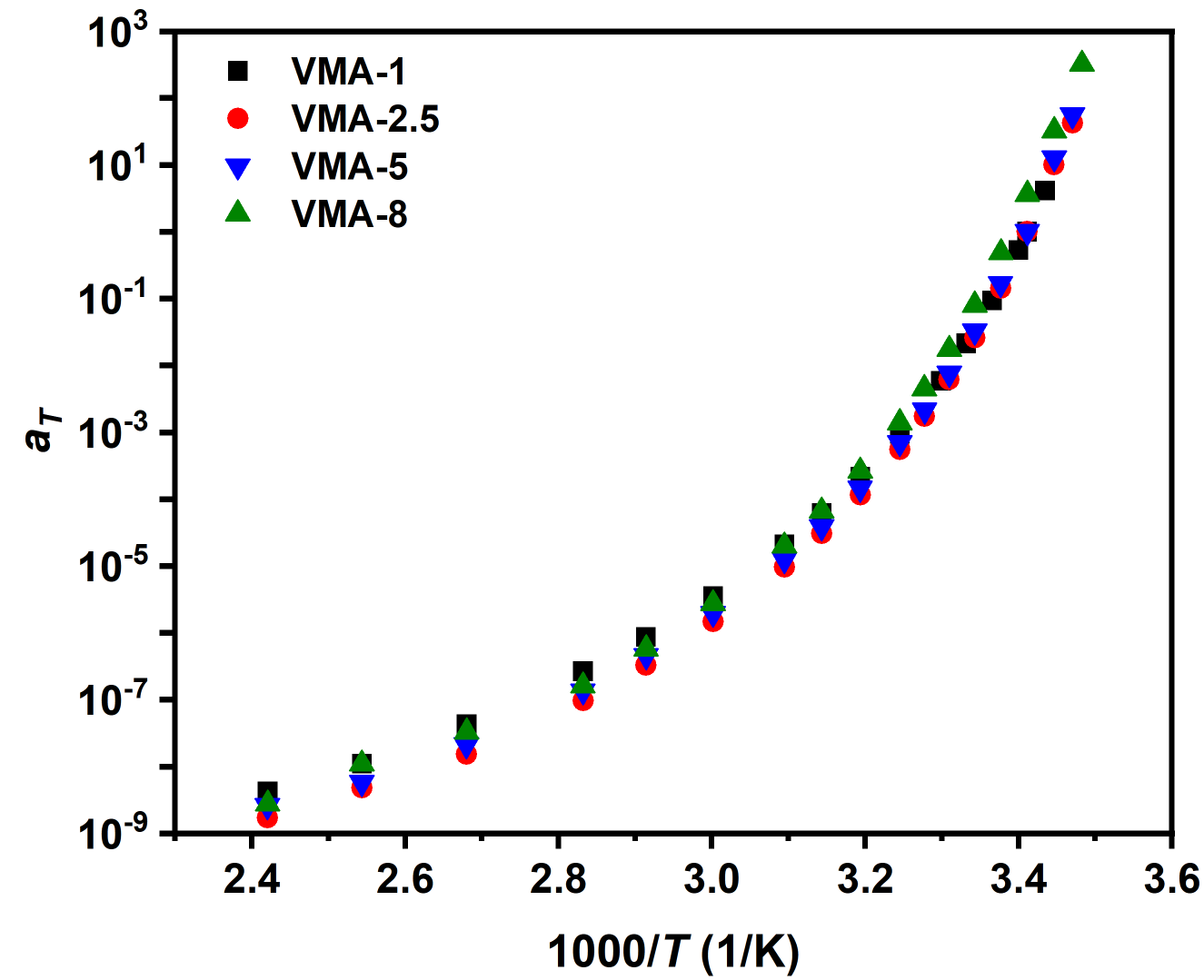}
  \caption{The $T$-dependent horizontal shift factors ($a_T(T)$) resulting from TTS analysis of SAOS data for the fully cured vitrimers, as described in the main paper.}
  \label{fig:SFigaT-VMA}
\end{figure}


\begin{figure}
\centering
  \includegraphics[width=0.5\linewidth]{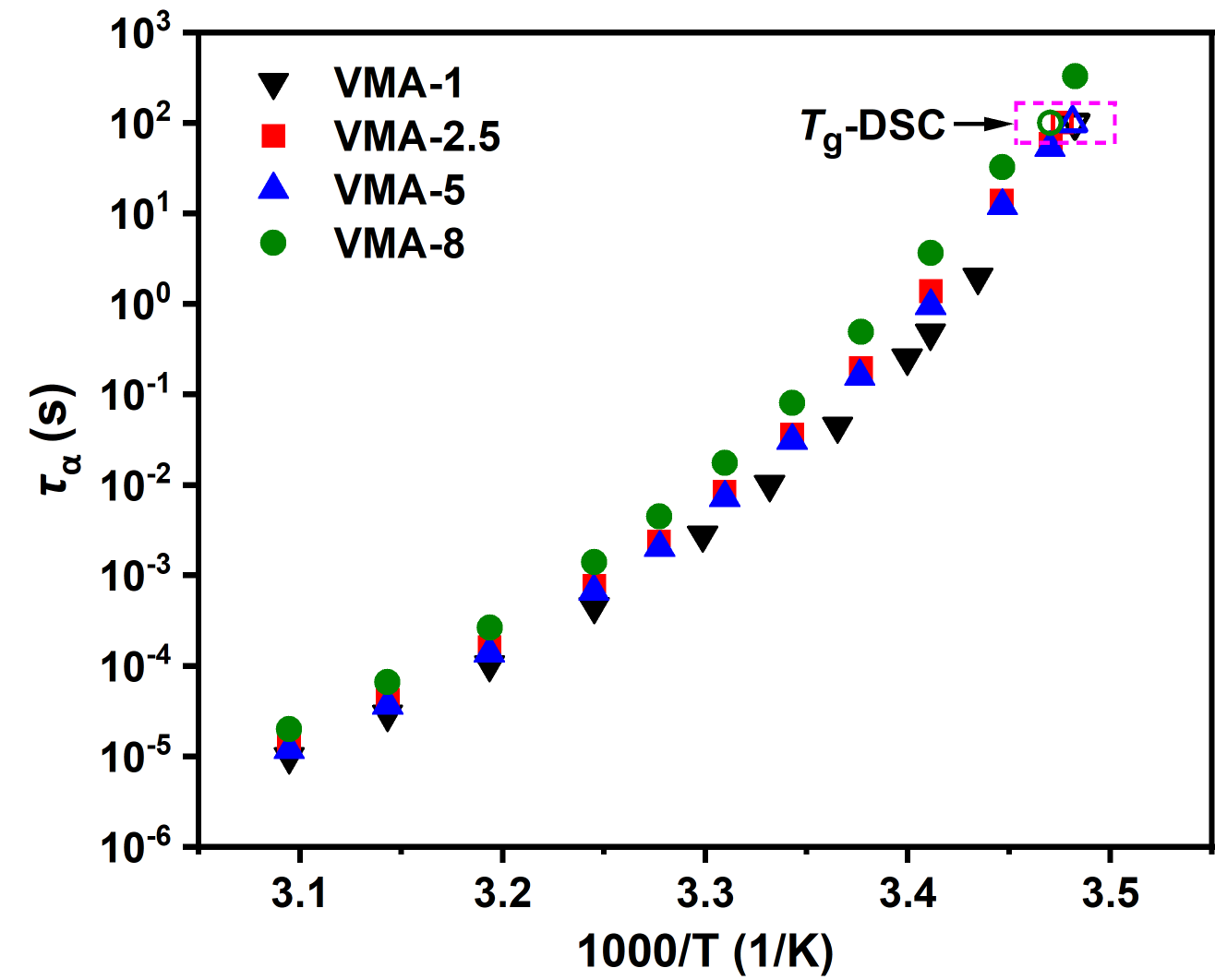}
  \caption{The $T$-dependent structural $\alpha$ relaxation time ($\tau_{\alpha}(T))$ resulting from TTS analysis of SAOS data, presented in an Arrhenius plot for the fully cured vitrimers. For comparison, DSC $\Tgval$ data are also shown for a relaxation time of 100 s.}
  \label{fig:SFigatauA}
\end{figure}


\begin{figure}
\centering
  \includegraphics[width=0.5\linewidth]{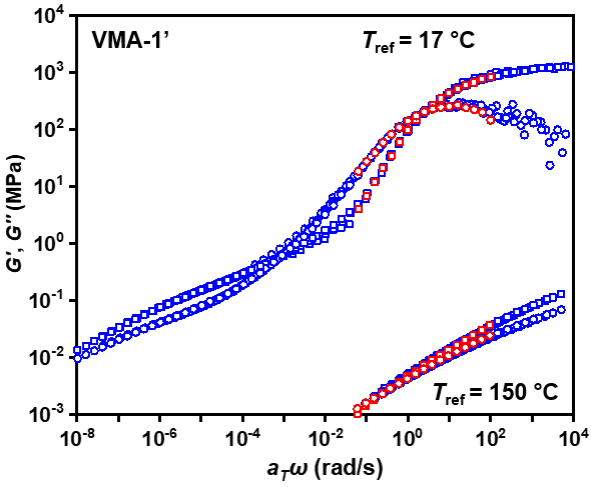}
  \caption{Master curve for $G^{'}$ and $G^{''}$ from SAOS data for the semi-cured vitrimer sample VMA-1'.}
  \label{fig:SFigMC1}
\end{figure}


\begin{figure}
\centering
  \includegraphics[width=0.5\linewidth]{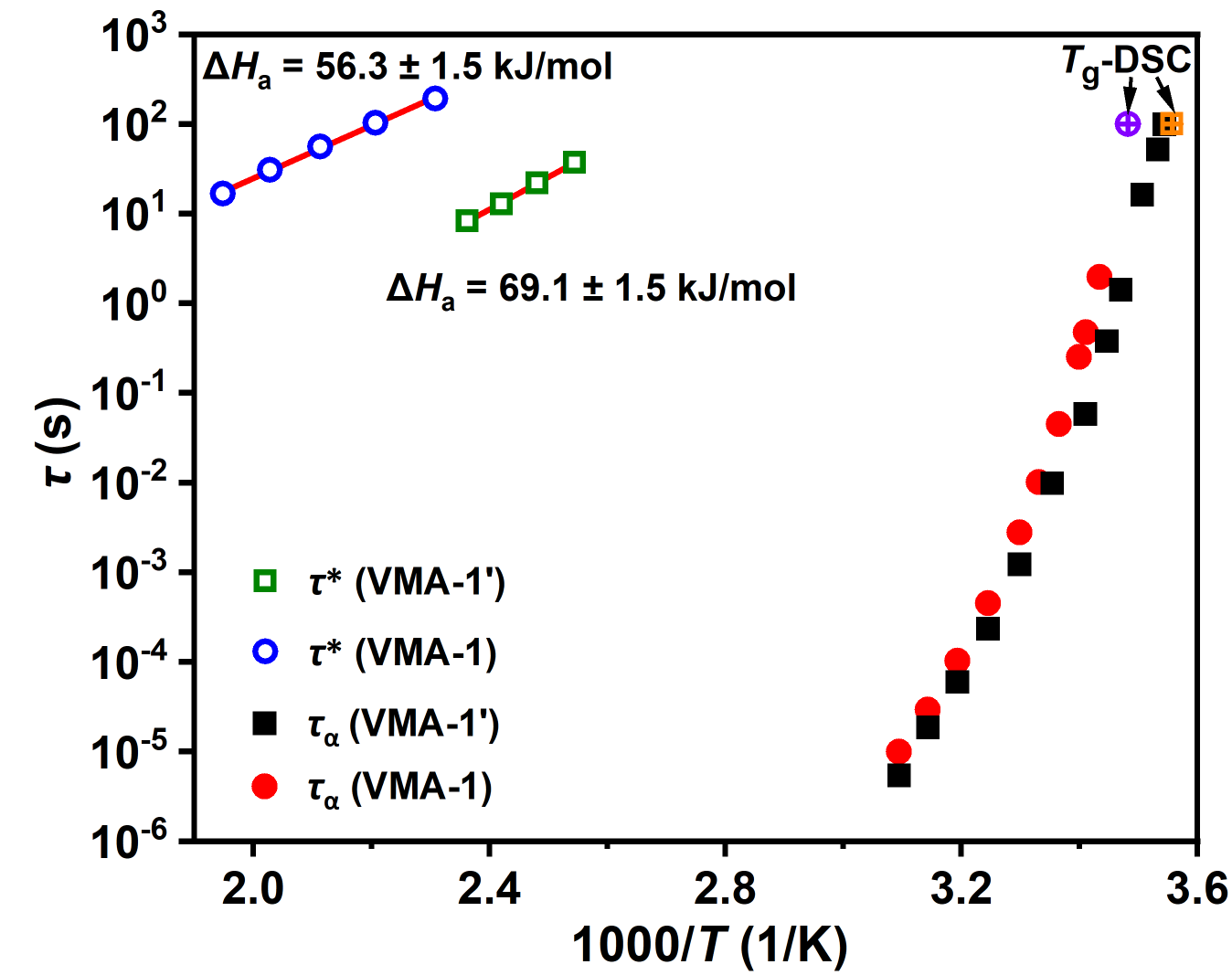}
  \caption{Arrhenius plot showing both the structural $\alpha$ relaxation time and the slow network relaxation time for the semi-cured (VMA-1') and fully cured (VMA-1) vitrimer samples, respectively. For comparison, DSC $\Tgval$ data are also shown for a relaxation time of 100 s.}
  \label{fig:SFigTau-1}
\end{figure}

\begin{figure}
\centering
  \includegraphics[width=0.5\linewidth]{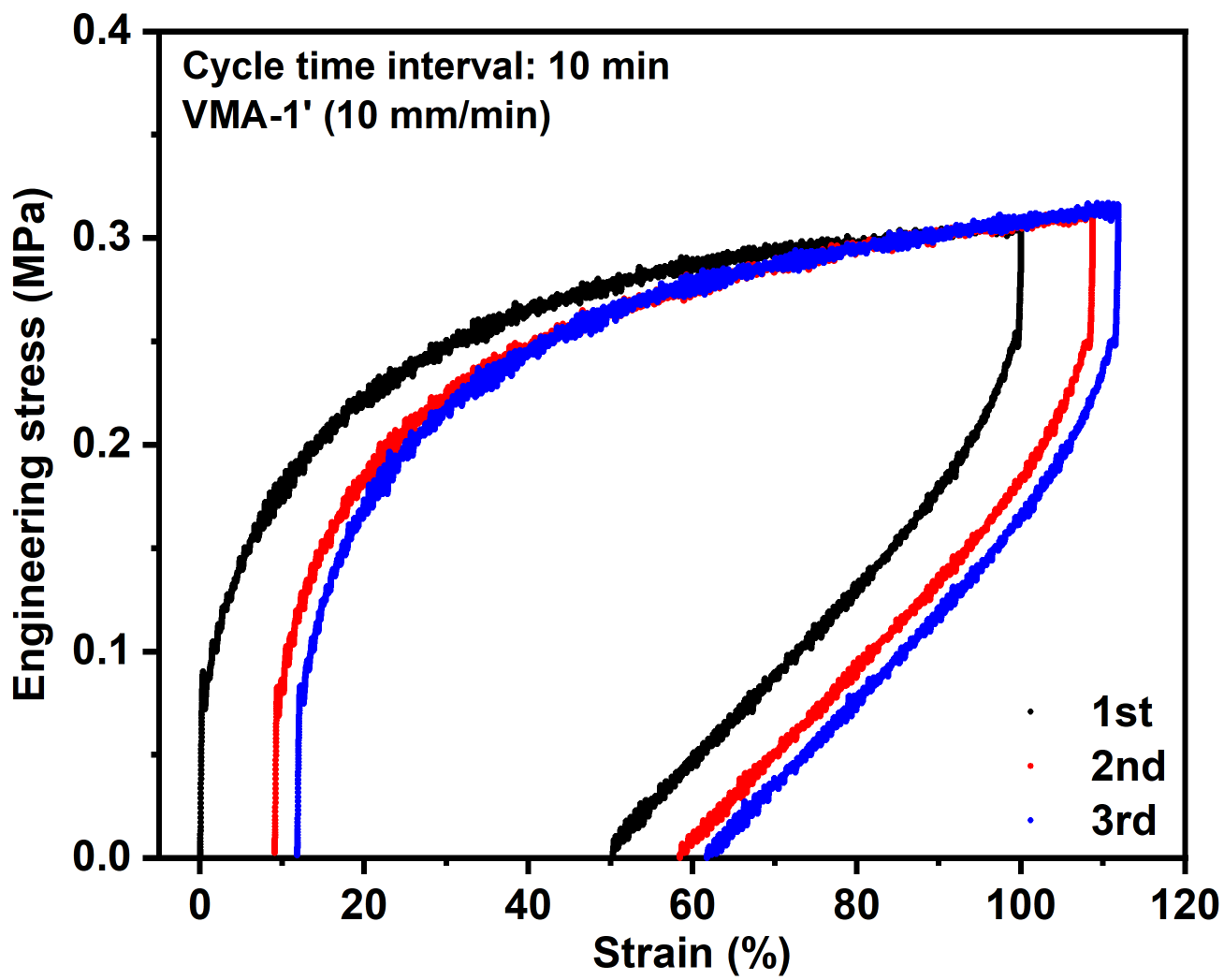}
  \caption{Tensile loading-unloading cycles, each to a maximum of 100\% strain, for the semi-cured VMA-1' sample. Three cycles are shown where the waiting time between cycles was 10 min. See the discussion in the main manuscript for detailed discussions.}
  \label{fig:SFigC100}
\end{figure}


\begin{figure}
\centering
  \includegraphics[width=0.5\linewidth]{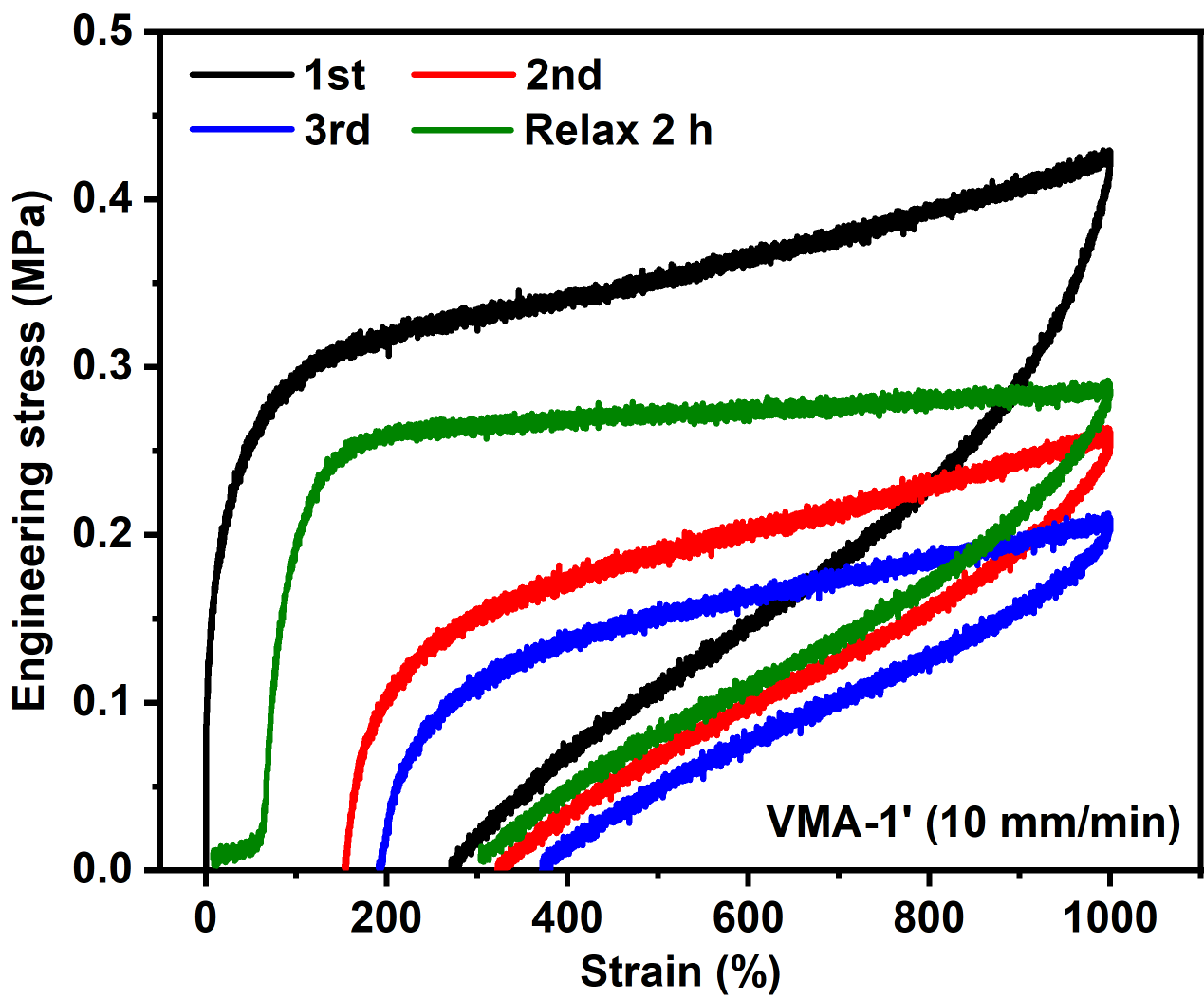}
  \caption{Tensile loading-unloading cycles, each to a maximum of 1000\% strain, for the semi-cured VMA-1' sample. Three cycles are shown where the waiting time between cycles was 10 min. Data obtained after a relaxation time of 2 hours are also shown. See the discussion in the main manuscript for detailed discussions.}
  \label{fig:SFigC1000}
\end{figure}


\begin{figure}
\centering
  \includegraphics[width=0.6\linewidth]{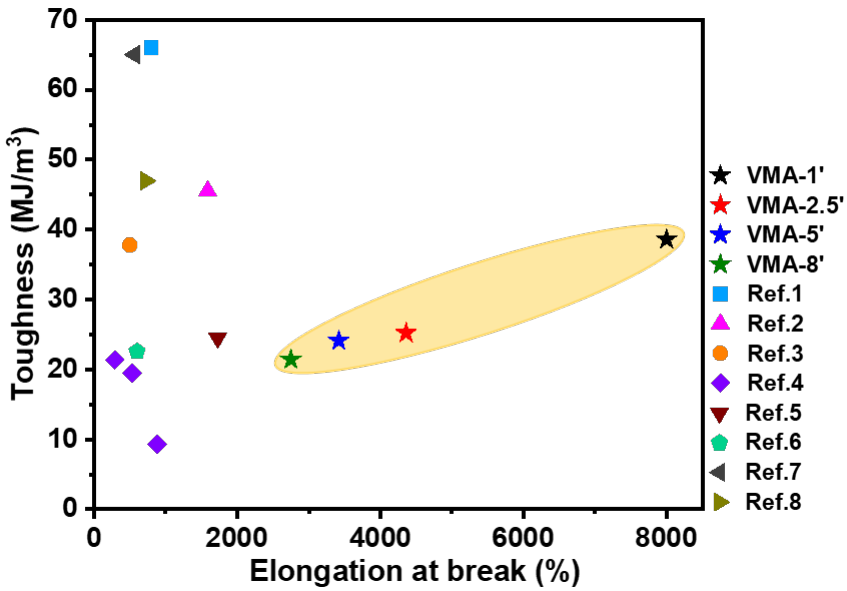}
  \caption{Comparison between the relationship between toughness and elongation at break for the semi-cured vitrimer samples and other high performance vitrimer or non-thermoset elastomers reported in the literature \cite{zhang2020synergistic,chen2023exceptionally,lu2024high,zheng2023covalently,luo2023highly,zhou2021room,leone2022dynamically,song2019ultra}.}
  \label{fig:SFigSTE}
\end{figure}


\begin{figure}
\centering
\includegraphics[width=0.45\linewidth]{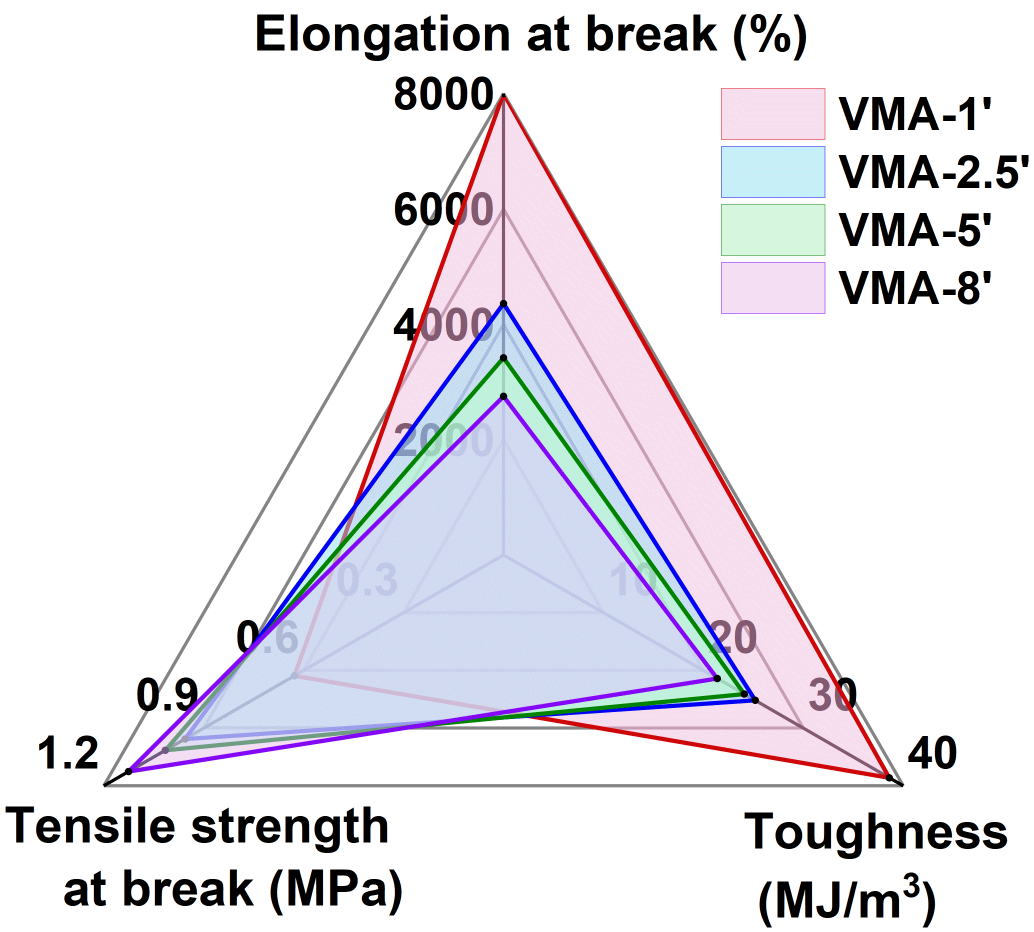}
  \caption{Radar chart of elongation at break, tensile strength at break, and toughness for the semi-cured vitrimers.}
  \label{fig:SFigRSS}
\end{figure}


\begin{figure}
\centering
  \includegraphics[width=0.3\linewidth]{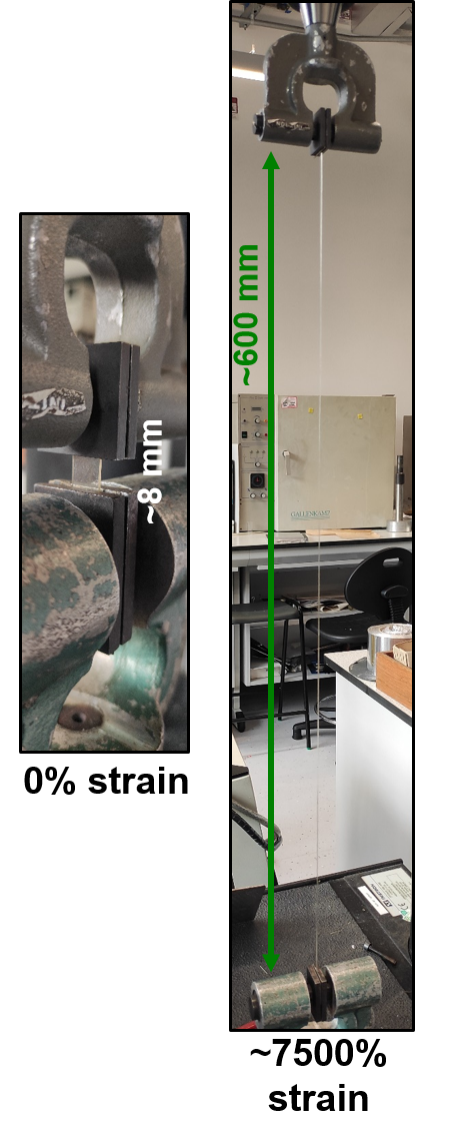}
  \caption{Photograph of the semi-cured vitrimer VMA-1' demonstrating a very high stretchability during uniaxial tensile testing at 10 mm/min.}
  \label{fig:SFigHSP}
\end{figure}


\begin{figure}
\centering
  \includegraphics[width=0.55\linewidth]{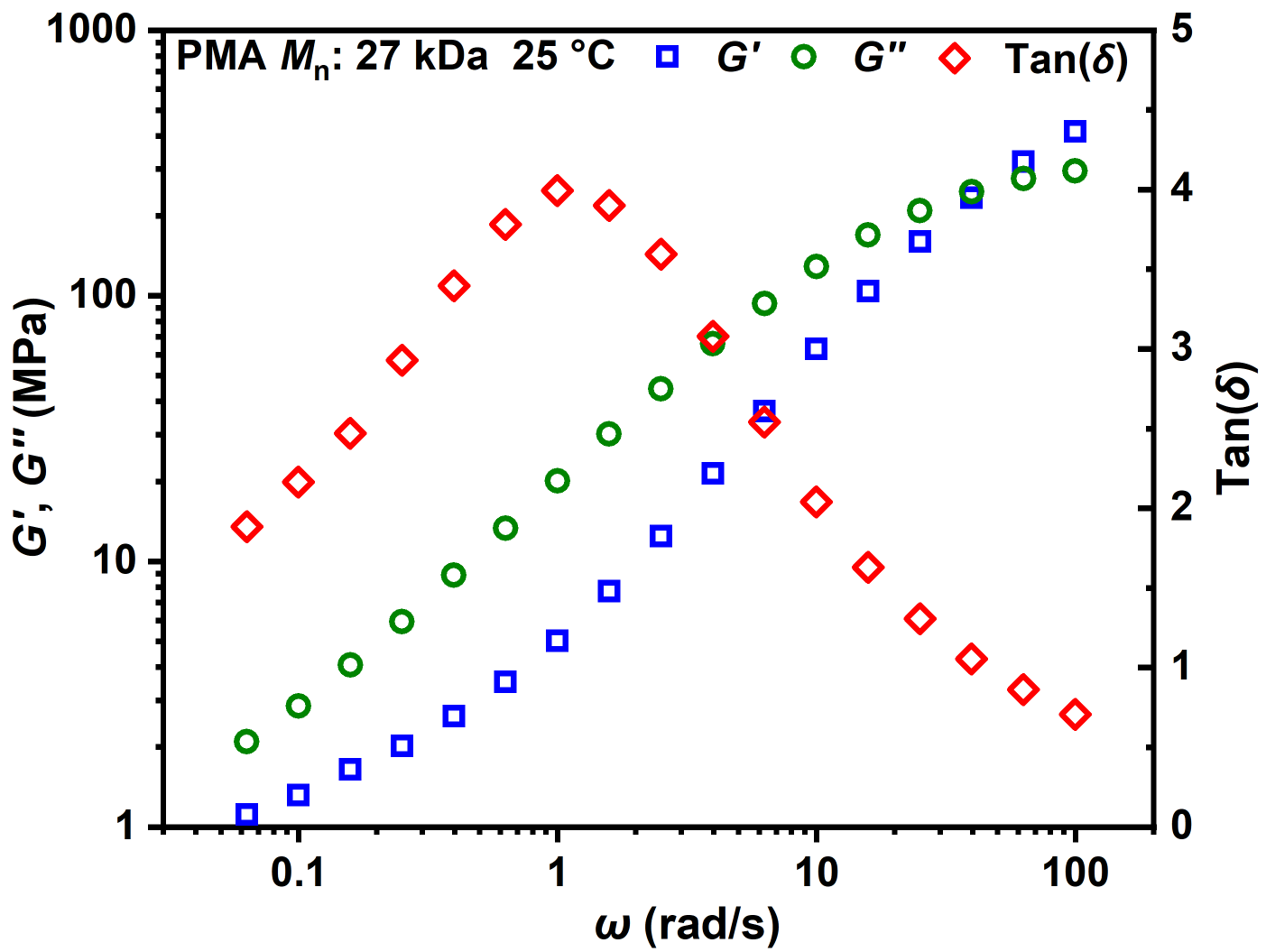}
  \caption{Small amplitude oscillatory rheology (SAOS) data for a linear poly(methyl acrylate) (PMA) ($M_n$=27 kDa) at $T=25 \degree$C. The figure shows both the storage ($G'$) and the loss ($G''$) moduli (left axis) and the loss tangent ($\tan(\delta)=G''/G'$) (right axis).}
  \label{fig:SFigPMA}
\end{figure}


\begin{figure}
\centering
  \includegraphics[width=1\linewidth]{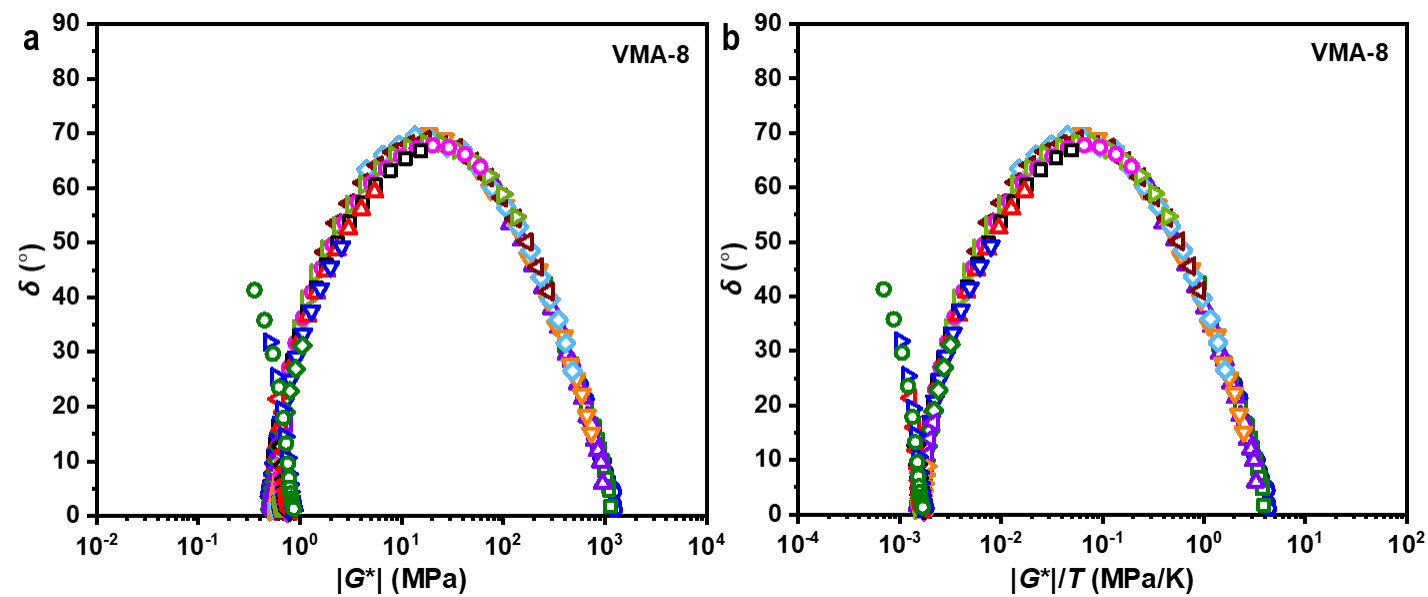}
  \caption{a) A representative so-called van Gurp Palmen plot of the phase angle versus the amplitude of the complex modulus for the fully cured VMA-8 sample. b) The same plot with a $T$-renormalized abscissa. See the discussion in the main manuscript for detailed discussions.}
  \label{fig:SFigVGP-VMA8}
\end{figure}


\renewcommand{\thetable}{S\arabic{table}}

\begin{turnpage}

\begin{table}
\centering
\caption{A comparison between the performance of our vitrimers and that of other representative previously reported high performance elastomers.} 

\centering 
\resizebox{\linewidth}{!}
{
\begin{tabular}{c c c c c c c c c c c} 
\hline\hline
 \makecell[c]{Materials}
  & \makecell[c]{{Max.stength}\\(MPa)}

   & \makecell[c]{{Max.strain}\\(\%)}

 & \makecell[c]{{Toughness}\\(MJ/m$^{3}$)} 

  & \makecell[c]{Tan($\delta$)\\max}

 & \makecell[c]{$T$-range of \\tan($\delta$)$>$0.5 (°C)}  
 
& \makecell[c]{{Self-healing}\\$T$ and $t$}

&\makecell[c]{Stength recovery\\ (MPa)}

 &\makecell[c]{Elongation\\ recovery (\%)}

 & \makecell[c]{Reprocessability}

 & \makecell[c]{Ref.} 

 \\ [0.5ex] 
\hline 

VMA (Vitrimers) &	0.6-6.5&	1000-8000&	21-54&	2.0-3.0 
&	30-50&	20 °C, 1 h/24 h&	1.5/2.0 (74\%/96\%)&	2500/3600&	Yes&this work\\

Supramolecular rubber &	3.3&	600&	$<$10&	N/A 
&	N/A&	40 °C, 3 h&	2.85 (86\%)&	530
&	Yes&\cite{cordier2008self}\\

AP Hydrogel &	0.16&	2300&	$<$3&	$<$0.1
&	None&	N/A&	N/A&	N/A&	N/A&\cite{sun2012highly}\\

Supramolecular elastomers&	0.9-3.77&	310-1570&	$<$12&	N/A 
&	N/A&	25 °C, 1 h/24 h&	0.9/1.7 (47\%/90\%)
&	390/720 &	Yes&\cite{chen2012multiphase}\\

Fe-Hpdca-PDMS &	0.23&	1860&	$<$5&	N/A 
&	N/A&	25 °C, 48 h&	0.22 (95\%)
&	1700 &	Yes&\cite{li2016highly}\\

UPyHCBA &	0.003-0.008&	2700-10000&	$<$1&	N/A 
&	N/A&	r.t., 5 min&	0.004 ($\sim$100\%)&	10000 
&	Yes&\cite{jeon2016extremely}\\

CB[8]-based network &	0.5-2.0&	10700&	$<$100&	$\sim$0.7 
&	$\sim$30&	25 °C, 1 h/12 h&	0.2/0.5 (40\%/100\%)
&	2800/4500 &	Yes&\cite{liu2017tough}\\

IP–SS (TPU) &	6.8&	923&	26.9&	N/A 
&	N/A&	25 °C, 2 h&	6.0 ($\sim$88\%)
&	920 &	Yes&\cite{kim2018superior}\\

PDMS–MPU–IU &	0.3-1.5&	1700-3200&	$<$20&	N/A 
&	N/A&	r.t., 48 h or 60 °C, 1.5 h
&	1.3/1.5 (78\%/88\%)
&	$\sim$1600 &	Yes&\cite{kang2018tough}\\

U-PDMS-Es &	0.06-1&	984-5600&	1.5-7.2&	0.6 
&	$\sim$44&	25 °C, 2 h&	0.19 (85\%)
&	2049 &	Yes&\cite{cao2018superstretchable}\\

Cu–DOU–CPU &	14.8&	1200&	87&	0.8 
&	$\sim$40&	25 °C, 130 h&	13.8 (93\%)& $\sim$1100 &	N/A&\cite{zhang2019highly}\\

LCE40 &	0.5&	100&	N/A&	1.6 
&	$\sim$42&	N/A&	N/A&	N/A&	No&\cite{saed2021impact}\\

MINs &	10/15.5&	$\sim$110&	7.4/12.9&	1.2/1.0
&	$\sim$24&	N/A&	N/A&	N/A&	N/A&\cite{zhao2021mortise}\\

Ionogels &	0.1-0.7&	1100-2100&	$<$5&	0.6 
&	$\sim$35&	20 °C, 24 h or 30 °C, 12 h
&	0.5 ($\sim$100\%)&	1800 &	Yes&\cite{xu2021transparent}\\

PTFEA-co-PFOEA &	0.6-4&	1200-2168&	5.4&	1.3 
&	N/A&	20 °C, 20 min&	0.14 (82\%)&	2060 
&	Yes&\cite{xiang2023highly}\\

Dielectric gels &	1.2 (compressed)&	70&	N/A&	1.0 
&	55&	N/A&	N/A&	N/A&	No&\cite{zhang2024dielectric}\\
\hline 
\end{tabular}
}

\label{table:ALL-DATA} 
\end{table}

\end{turnpage}

\clearpage




%